\documentclass[aps,prd,twocolumn,preprintnumbers,nofootinbib,amsmath,amssymb]{revtex4-2}

\usepackage{slashbox,latexsym,amssymb,amsfonts,amsmath,slashed,amsthm,bm,bbm}
\usepackage{graphicx}
\usepackage{mathrsfs}

\usepackage[pdftex, pdftitle={Dirac spectrum in the chirally symmetric phase of a gauge
  theory. I}]{hyperref}

\usepackage{xcolor}
\usepackage[normalem]{ulem}

\newcommand{\f}[2]{\frac{#1}{#2}}
\newcommand{\tf}[2]{{\textstyle\f{#1}{#2}}}

\newcommand{\de}{\partial}

\newcommand{\la}{\langle}
\newcommand{\ra}{\rangle}

\newcommand{\Oc}{\mathcal{O}}

\newcommand{\susc}{\mathcal{A}}

\newcommand{\coeff}{\mathcal{C}}
\newcommand{\coeffi}{\mathcal{C}^{\scriptscriptstyle\infty}}
\newcommand{\wiv}{\mathcal{W}_{\!\scriptscriptstyle\infty}}
\newcommand{\whiv}{\hat{\mathcal{W}}_{\!\scriptscriptstyle\infty}}
\newcommand{\wtiv}{\tilde{\mathcal{W}}_{\!\scriptscriptstyle\infty}}
\newcommand{\wivG}{\mathcal{W}_{G{\scriptscriptstyle\infty}}}

\newcommand{\ff}{f}
\newcommand{\fft}{\tilde{f}}
\newcommand{\ffh}{\hat{f}}

\newcommand{\topoa}{\mathcal{T}_1}
\newcommand{\topob}{\mathcal{T}_2}
\newcommand{\topoc}{\mathcal{T}_3}
\newcommand{\Iu}{I^{(1)}}
\newcommand{\Id}{I^{(2)}}
\newcommand{\Iq}{I_{Q^2}^{(1)}}
\newcommand{\Iqiv}{I_{Q^2{\scriptscriptstyle\infty}}^{(1)}}
\newcommand{\In}{I_{N_0}^{(1)}}
\newcommand{\Iniv}{I_{N_0{\scriptscriptstyle\infty}}^{(1)}}
\newcommand{\bbnn}{b_{N_0^2{\scriptscriptstyle\infty}}}
\newcommand{\bbqq}{\kappa_{4t}}

\newcommand{\chit}{\chi_t}
\newcommand{\lvol}{\mathrm{V}_4}
\newcommand{\svol}{\mathrm{V}_3}

\newcommand{\sous}{V}
\newcommand{\soup}{W}

\newcommand{\Wc}{\mathcal{W}}

\newcommand{\ssi}{j_S}
\newcommand{\seta}{j_P}

\newcommand{\spi}{\vec{\jmath}_P}
\newcommand{\sde}{\vec{\jmath}_S}

\newcommand{\des}{K}
\newcommand{\U}{\mathcal{U}}
\newcommand{\Rc}{\mathcal{R}}
\newcommand{\CP}{\U_\mathcal{CP}}

\newcommand{\Mc}{\mathcal{M}}

\renewcommand{\Re}{{\rm Re}\,}
\renewcommand{\Im}{{\rm Im}\,}

\newcommand{\tr}{{\rm tr}\,}

\newcommand{\tu}{\tilde{u}}
\newcommand{\bu}{\bar{u}}

\newcommand{\sti}{S} 
\newcommand{\mnz}{M}
\newcommand{\dcl}{\Delta}

\newcommand{\bt}{\mathrm{b}_2}

\newcommand{\mdens}{\mathcal{N}}

\makeatletter
\def\ignorecitefornumbering#1{%
     \begingroup
         \@fileswfalse
         #1
    \endgroup
}
\makeatother

\begin{document}

\title{Dirac spectrum in the chirally symmetric phase of a gauge
  theory. I}

\author{Matteo Giordano} \email{giordano@bodri.elte.hu}
\affiliation{Institute of Physics and Astronomy, ELTE E\"otv\"os
  Lor\'and University, P\'azm\'any P\'eter s\'et\'any 1/A, H-1117,
  Budapest, Hungary}

\date{\today}

\begin{abstract}
  
  I study the consequences of chiral symmetry restoration for the
  Dirac spectrum in finite-temperature gauge theories in the
  two-flavor chiral limit, using Ginsparg--Wilson fermions on the
  lattice. I prove that chiral symmetry is restored at the level of
  the susceptibilities of scalar and pseudoscalar bilinears if and
  only if all these susceptibilities do not diverge in the chiral
  limit $m\to 0$, with $m$ the common mass of the light fermions.
  This implies in turn that they are infinitely differentiable
  functions of $m^2$ at $m=0$, or $m$ times such a function, depending
  on whether they contain an even or odd number of isosinglet
  bilinears. Expressing scalar and pseudoscalar susceptibilities in
  terms of the Dirac spectrum, I use their finiteness in the symmetric
  phase to derive constraints on the spectrum, and discuss the
  resulting implications for the fate of the anomalous
  $\mathrm{U}(1)_A$ symmetry in the chiral limit. I also discuss the
  differentiability properties of spectral quantities with respect to
  $m^2$, and show from first principles that the topological
  properties of the theory in the chiral limit are characterized by an
  instanton gas-like behavior if $\mathrm{U}(1)_A$ remains effectively
  broken.
\end{abstract}

\maketitle

\section{{Introduction}}
\label{sec:intro}

Studies of the chiral limit of QCD, where $N_f$ of the quark masses
are taken to zero, have provided considerable insight into its
real-world version describing the physics of strong interactions. In
the chiral limit the theory has a classical
$\mathrm{U}(N_f)_L\times \mathrm{U}(N_f)_R= \mathrm{SU}(N_f)_L\times
\mathrm{SU}(N_f)_R\times \mathrm{U}(1)_V\times \mathrm{U}(1)_A$
continuous chiral symmetry, anomalously broken by quantum effects to
$\mathrm{SU}(N_f)_L\times \mathrm{SU}(N_f)_R\times \mathrm{U}(1)_V$
and, at low temperatures, spontaneously broken to its diagonal
$\mathrm{SU}(N_f)_V\times \mathrm{U}(1)_V$ component. The chiral
$\mathrm{SU}(N_f)_L\times \mathrm{SU}(N_f)_R$ symmetry, and the way it
is realized, largely control the low-energy behavior of QCD at
physical quark masses. Due to the lightness of the up and down quarks,
the case $N_f=2$ is particularly relevant from the phenomenological
point of view. At zero temperature, the spontaneous breaking of
$\mathrm{SU}(2)_L\times \mathrm{SU}(2)_R$ in the chiral limit explains
the lightness of the pions and the absence of parity partners in the
hadronic spectrum at the physical point, and its restoration at higher
temperatures is behind the approximate chiral symmetry of QCD at
temperatures above the crossover to the quark-gluon
plasma~\cite{Aoki:2006we,Borsanyi:2010bp,Bazavov:2011nk,
  Bhattacharya:2014ara,Bazavov:2016uvm}.

Particularly important questions concerning the chiral limit are the
nature of the $N_f=2$ chiral transition, and the fate of the anomalous
$\mathrm{U}(1)_A$ symmetry in the chirally symmetric phase. These are
not only of considerable theoretical interest, but can have
phenomenological impact on important and diverse aspects of strongly
interacting matter as described by QCD at the physical point,
including heavy-ion collisions~\cite{Pisarski:1983ms,Rajagopal:1992qz,
  Rajagopal:1993ah,Shuryak:1993ee,Kapusta:1995ww,Huang:1995fc,
  Bass:2018xmz,Pisarski:2024esv} and axion
cosmology~\cite{Bonati:2015vqz,Petreczky:2016vrs,Frison:2016vuc,
  Borsanyi:2016ksw,Lombardo:2020bvn,Athenodorou:2022aay}. In spite of
extensive theoretical and numerical studies, however, these questions
remain currently open.

According to the effective-Lagrangian analysis of the seminal paper
Ref.~\cite{Pisarski:1983ms}, for $N_f=2$ the nature of the transition
and the fate of $\mathrm{U}(1)_A$ are closely related, with the
transition being second or first order depending on whether
$\mathrm{U}(1)_A$ remains effectively broken or is effectively
restored at the transition. Here ``effective breaking'' and
``effective restoration'' mean respectively that even though
$\mathrm{U}(1)_A$ symmetry is, strictly speaking, always broken due to
its anomalous nature, symmetry-breaking effects at a given nonzero
temperature may be of comparable magnitude to those found at zero
temperature, or strongly suppressed with respect to them.  However,
since Ref.~\cite{Pisarski:1983ms}, a number of additions to the
original analysis~\cite{Butti:2003nu,Basile:2004wa,
  Pelissetto:2013hqa,Pisarski:2024esv,Giacosa:2024orp}, and the use of
alternative techniques such as chiral
Lagrangians~\cite{Meggiolaro:1992wc,Marchi:2003wq,Meggiolaro:2013swa,
  Meggiolaro:2019rkl}, the functional renormalization
group~\cite{Fejos:2022mso,Fejos:2024bgl}, and Schwinger--Dyson
equations~\cite{Bernhardt:2023hpr}, have led to a larger set of
possibilities, and so to a less tight (and less clear) relation
between chiral symmetry restoration and the fate of $\mathrm{U}(1)_A$.

Numerical calculations on the lattice, that could have in principle
settled these issues, have resulted instead in contradictory claims.
Using staggered fermions, the HotQCD collaboration concludes that
$\mathrm{U}(1)_A$ remains effectively broken at the transition, and
that this is second order in the $\mathrm{O}(4)$
class~\cite{HotQCD:2019xnw}, in agreement with earlier
studies~\cite{Chandrasekharan:1998yx}. This conclusion is further
supported by dedicated studies of the effects of the
anomaly~\cite{Dick:2015twa,Ding:2020xlj,Kaczmarek:2021ser,
  Kaczmarek:2023bxb}, including with domain-wall and M\"obius
domain-wall fermions~\cite{HotQCD:2012vvd,Buchoff:2013nra,
  Bhattacharya:2014ara,Gavai:2024mcj}. On the other hand, using
M\"obius domain-wall fermions reweighted to overlap, the JLQCD
collaboration concludes that $\mathrm{U}(1)_A$ is restored in the
symmetric phase~\cite{Aoki:2020noz}, including close to the critical
temperature~\cite{JLQCD:2024xey}, confirming earlier results using the
same discretization~\cite{Cossu:2013uua,Tomiya:2016jwr}, as well as
using improved Wilson fermions~\cite{Brandt:2016daq}. It is worth
mentioning in passing that for $N_f=3$ no sign of a first-order region
has been found in the continuum limit, suggesting that the transition
is second order~\cite{Cuteri:2021ikv, Dini:2021hug}, a result in
contrast with the analysis of Ref.~\cite{Pisarski:1983ms}, but
compatible with other studies~\cite{Fejos:2022mso,Fejos:2024bgl,
  Bernhardt:2023hpr}.

The approximate analytical methods discussed above are unfortunately
affected by serious theoretical uncertainties. Studies based on the
Ginzburg--Landau effective-Lagrangian approach typically ignore the
role of gauge symmetries, which is still not fully understood (see
Refs.~\cite{Pelissetto:2017sfd,Bonati:2020elf}). Both Ginzburg--Landau
and chiral Lagrangians require the temperature dependence of the
various coefficients in the Lagrangian as an input. Studies in the
framework of the functional renormalization group or of
Schwinger--Dyson equations suffer from the effects of hardly
controllable approximations required to make the calculations
tractable. Lattice studies have more solid foundations, as they are
based on first principles and have fully controllable uncertainties,
but are nonetheless affected by serious drawbacks. Dealing with chiral
symmetry on the lattice is in fact notoriously
difficult~\cite{Nielsen:1981hk,Nielsen:1980rz,Nielsen:1981xu,
  Friedan:1982nk}, and while fermion discretizations with good chiral
properties (``Ginsparg--Wilson fermions'')~\cite{Ginsparg:1981bj,
  Hasenfratz:1998ri,Hasenfratz:1998jp,Luscher:1998pqa,Kikukawa:1998py,
  Neuberger:1998wv,Horvath:1999bk} have been
found~\cite{Kaplan:1992bt,Shamir:1993zy,Furman:1994ky,
  Borici:1999zw,Chiu:2002ir,Brower:2005qw,Brower:2012vk,
  Narayanan:1993sk,Narayanan:1993ss,Neuberger:1997fp,Neuberger:1998wv,
  Hasenfratz:1993sp,Bietenholz:1995cy,DeGrand:1995ji,Hasenfratz:1998ri,
  Hasenfratz:1998jp,Hasenfratz:2002rp}, they are very expensive from
the numerical point of view, strongly limiting the lattice size that
one can afford, and so how close one can reliably get to the
thermodynamic and chiral limits.

A solution from first principles to the problem of the fate of
$\mathrm{U}(1)_A$ in the $N_f=2$ chiral limit was proposed by Cohen in
Ref.~\cite{Cohen:1996ng} (partly elaborating on the arguments of
Ref.~\cite{Shuryak:1993ee}).  There he argued, using the formal
continuum functional integral, that $\mathrm{U}(1)_A$ must be
effectively restored if $\mathrm{SU}(2)_L\times \mathrm{SU}(2)_R$ is.
Here ``effectively restored'' takes a stronger meaning than in
Ref.~\cite{Pisarski:1983ms}, and indicates that the effects of the
anomaly become invisible in (at least the simplest) physical
observables. This term will be understood in this sense in what
follows. However, a loophole in the argument of
Ref.~\cite{Cohen:1996ng} was pointed out in
Refs.~\cite{Evans:1996wf,Lee:1996zy}, namely the incomplete treatment
of the contributions of topologically nontrivial configurations to the
path integral. While these configurations form a set of zero measure
in the chiral limit, and were neglected in Ref.~\cite{Cohen:1996ng}
for this reason, the contribution of the corresponding zero modes of
the Dirac operator to the difference of correlators related by a
$\mathrm{U}(1)_A$ transformation can be non-vanishing in the chiral
limit. The conclusion of Refs.~\cite{Evans:1996wf,Lee:1996zy} was that
$\mathrm{SU}(2)_L\times \mathrm{SU}(2)_R$ restoration does not
necessarily imply $\mathrm{U}(1)_A$ restoration, and that, in fact,
$\mathrm{U}(1)_A$ most likely remains effectively broken in the chiral
limit also in the symmetric phase. A key assumption of
Refs.~\cite{Evans:1996wf,Lee:1996zy} is that in this phase the order
in which one takes the thermodynamic and chiral limits should not
matter, and so they can be interchanged.

A new strategy to study the relation between chiral symmetry
restoration and the fate of $\mathrm{U}(1)_A$ from first principles
was proposed by Cohen in Ref.~\cite{Cohen:1997hz}, and developed in
full depth by Aoki, Fukaya, and Taniguchi in
Ref.~\cite{Aoki:2012yj}. This strategy is to determine how chiral
symmetry restoration constrains the behavior of the spectrum of the
Dirac operator, and in turn what the resulting constraints imply for
$\mathrm{U}(1)_A$ in the symmetric phase. This strategy was also
exploited in Refs.~\cite{Kanazawa:2015xna,Azcoiti:2023xvu}, and has
been recently revisited by myself in
Refs.~\cite{Giordano:2024jnc,Giordano:2024awb} (see also
Ref.~\cite{Carabba:2021xmc}). Using Ginsparg--Wilson fermions on the
lattice, as in Refs.~\cite{Aoki:2012yj,
  Azcoiti:2023xvu,Giordano:2024jnc,Giordano:2024awb}, this approach
combines analytic (and mathematically sound) methods with the
first-principle and nonperturbative nature of the functional integral,
to extract information from such a key object as the spectrum of the
Dirac operator. In fact, the Dirac spectrum and the corresponding
eigenvectors entirely encode the interactions of quarks with the gauge
fields, and so in particular they should reflect the status of the
various symmetries. (In this context, it is worth mentioning
Ref.~\cite{Ding:2023oxy} about the relation between the spectrum and
the scaling with mass and temperature of the cumulants of the chiral
condensate.) While at zero and low temperatures the spontaneous
breaking of chiral symmetry and the appearance of massless Goldstone
bosons in the chiral limit allow one to exploit effective theories to
gain considerable insight into the Dirac
spectrum~\cite{Leutwyler:1992yt,Smilga:1993in,
  Verbaarschot:1994qf,Osborn:1998qb,Damgaard:2008zs,Giusti:2008vb,
  Verbaarschot:2000dy,Toublan:2000dn}, such a luxury is generally not
available in the symmetric phase, and one should content oneself with
constraining the spectrum.

The central assumption of Ref.~\cite{Aoki:2012yj} is that in the
symmetric phase the relevant observables (that can be expressed as
expectation values of mass-independent functionals of gauge fields
only) are analytic functions of $m^2$, with $m$ the common fermion
mass, reflecting the analyticity and symmetry properties expected in
the $N_f=2$ chiral limit. Together with certain technical assumptions
on the spectrum, this led the authors to conclude that in the
symmetric phase the spectral density, $\rho(\lambda;m)$, as a function
of the eigenvalue, $\lambda$, must vanish faster than $\lambda^2$ in
the chiral limit, i.e., $\rho(\lambda;0)=o(\lambda^2)$, and vanish at
$\lambda=0$ identically for small enough mass, i.e., $\rho(0;m)=0$ for
$|m|<m_0$; and that the topological susceptibility, $\chit$, must
vanish identically for small enough mass. This results in the
conclusion that $\mathrm{U}(1)_A$ must be effectively restored in the
chiral limit in scalar and pseudoscalar correlation functions, partly
supporting the original claim of Ref.~\cite{Cohen:1996ng}. Under
assumptions similar to (but technically weaker than) those of
Ref.~\cite{Aoki:2012yj}, Ref.~\cite{Kanazawa:2015xna} proved with a
simpler argument that $\mathrm{U}(1)_A$ is effectively restored in the
chiral limit in the symmetric phase at the level of the simplest order
parameter, i.e., the difference $\chi_\pi -\chi_\delta$ of the usual
pion and delta susceptibilities.

The predictions of Ref.~\cite{Aoki:2012yj} are supported by the
numerical results of Refs.~\cite{Cossu:2013uua,Tomiya:2016jwr,
  Brandt:2016daq,Aoki:2020noz,JLQCD:2024xey}, but are in disagreement
with the results of other
studies~\cite{Chandrasekharan:1998yx,Buchoff:2013nra,Dick:2015twa,
  HotQCD:2019xnw,HotQCD:2012vvd,Ding:2020xlj,Kaczmarek:2021ser,
  Kaczmarek:2023bxb}. While the chiral fermions used in
Refs.~\cite{Cossu:2013uua,Tomiya:2016jwr,Aoki:2020noz,JLQCD:2024xey}
(but also in Ref.~\cite{Buchoff:2013nra}) give one a better control of
theoretical uncertainties, the staggered fermions used in
Refs.~\cite{Chandrasekharan:1998yx,Dick:2015twa,
  HotQCD:2019xnw,Ding:2020xlj,Kaczmarek:2021ser,Kaczmarek:2023bxb} are
computationally cheaper, and give one a better control of the
finite-volume and finite-spacing systematics and of the statistical
uncertainties of the numerical results. Stating that the issue of the
fate of $\mathrm{U}(1)_A$ in the symmetric phase remains unsettled
seems a fair and balanced conclusion.

From the theoretical point of view, in order to avoid the conclusions
of Refs.~\cite{Aoki:2012yj,Kanazawa:2015xna} one needs to abandon at
least one of their assumptions, the easiest choice being the technical
assumptions on the spectral density. As a matter of fact, ensembles of
sparse random matrices, of which the Dirac operator in the background
of fluctuating gauge fields constitutes an example, display a wide
variety of properties concerning the dependence of the spectral
density on $\lambda$. Technical assumptions on this dependence reflect
more the experience with concrete models, mostly based on numerical
simulations, than the results of rigorous theorems. This seems to
leave some room for the possibility of effective $\mathrm{U}(1)_A$
breaking in the symmetric phase. On the other hand, while the
$m^2$-analyticity assumption is certainly reasonable, it is arguably
not more nor less reasonable \textit{a priori} than commutativity of
the thermodynamic and chiral limits, that leads to opposite
conclusions concerning the fate of
$\mathrm{U}(1)_A$~\cite{Evans:1996wf,Lee:1996zy,Carabba:2021xmc}, and
it seems in fact quite reasonable to make both assumptions at
once. Assuming $m^2$-analyticity of the free energy density and
commutativity of limits leads to severe restrictions on the functional
form of the spectral density if $\mathrm{U}(1)_A$ remains effectively
broken~\cite{Azcoiti:2023xvu}. These restrictions led
Ref.~\cite{Azcoiti:2023xvu} to conclude that effective
$\mathrm{U}(1)_A$ breaking in the chiral limit in the symmetric phase
is possible only if the spectral density of non-zero modes develops in
the thermodynamic limit a Dirac delta at $\lambda=0$ for nonzero
fermion mass, which is quite unlikely to happen on physical grounds.
This seems to take away all the room left for effective
$\mathrm{U}(1)_A$ breaking, unless $m^2$-analyticity or commutativity
of limits fail and chiral symmetry is restored in some rather
nontrivial way in the chiral limit, possibly not fully (see
Refs.~\cite{Azcoiti:2016zbi,Azcoiti:2017jsh,Azcoiti:2019moz,
  Azcoiti:2021gst,Azcoiti:2023xvu} and Refs.~\cite{Giordano:2020twm,
  Giordano:2021nat,Giordano:2022ghy} for alternative scenarios), or
unless one can find a loophole in the analysis of
Ref.~\cite{Azcoiti:2023xvu}.

The first question to address is then whether the $m^2$-analyticity or
commutativity assumptions are just reasonable assumptions, or
necessary consequences of chiral symmetry restoration. A partial
answer was provided in Ref.~\cite{Giordano:2024jnc}, where I proved
that if chiral symmetry is restored in scalar and pseudoscalar
susceptibilities, then these are $C^\infty$ functions of $m^2$ at
$m=0$, i.e., functions of $m^2$ infinitely differentiable at zero
(``$m^2$-differentiable'', for the sake of brevity), if they involve
an even number of isosinglet scalar and pseudoscalar bilinears, and
$m$ times such a function if this number is odd.  Moreover, if chiral
symmetry is restored also in susceptibilities involving scalar and
pseudoscalar bilinears and general (including nonlocal) functionals of
the gauge fields only, $m^2$-differentiability extends to the spectral
density as well. This essentially implies the $m^2$-analyticity
assumptions of
Refs.~\cite{Aoki:2012yj,Kanazawa:2015xna,Azcoiti:2023xvu}.  In fact,
while different from analyticity, infinite differentiability actually
suffices for (most of) their arguments, since they use only the
existence of $m^2$-derivatives at $m=0$. On the other hand, although
commutativity of the thermodynamic and chiral limits is supported by
reasonable arguments (see footnote 5 in Ref.~\cite{Evans:1996wf}), I
am not aware of a proof that it necessarily follows from symmetry
restoration, and one might have to abandon it.

The next question is whether the technical assumptions of
Refs.~\cite{Aoki:2012yj,Kanazawa:2015xna} on the spectral density are
not too restrictive, excluding reasonable functional forms (different
from a Dirac delta at the origin) that allow for effective
$\mathrm{U}(1)_A$ breaking (which, if the conclusions of
Ref.~\cite{Azcoiti:2023xvu} are correct, would require abandoning the
commutativity of limits). These assumptions are that $\rho(\lambda;m)$
is an analytic function of $m^2$, and admits a power-law expansion in
$|\lambda|$ sufficiently close to zero. In Ref.~\cite{Aoki:2012yj} a
relaxed form of the second assumption was also considered, allowing
the presence of a term $C|\lambda|^\alpha$ in the spectral density,
with a mass-independent non-integer exponent $\alpha> 0$ and a
mass-independent prefactor $C$.\footnote{In Ref.~\cite{Aoki:2012yj}
  these assumptions are actually made on the spectral density computed
  on individual gauge configurations of infinite size, assumed to be a
  well-defined ordinary function.} Analyticity (or rather infinite
differentiability) in $m^2$ is a consequence of symmetry restoration
(in the extended sense), as discussed above, but the assumption of a
regular behavior of the spectral density at the origin is called into
question by recent (and less recent~\cite{Edwards:1999zm}) numerical
results, indicating the presence of a possibly singular near-zero peak
in the high-temperature
phase~\cite{HotQCD:2012vvd,Buchoff:2013nra,Cossu:2013uua,Dick:2015twa,
  Alexandru:2015fxa,Tomiya:2016jwr,Kovacs:2017uiz,Aoki:2020noz,
  Alexandru:2019gdm,Ding:2020xlj,Kaczmarek:2021ser,Vig:2021oyt,
  Kovacs:2021fwq,Meng:2023nxf,Kaczmarek:2023bxb,Alexandru:2024tel,
  JLQCD:2024xey,Fodor:2025yuj}.\footnote{In the chirally broken phase
  at low temperature, partially quenched chiral perturbation theory
  predicts a logarithmic divergence in $\rho$ at
  $m\neq 0$~\cite{Osborn:1998qb,Damgaard:2008zs,Giusti:2008vb}.} It is
possible that this peak goes away in the chiral limit without any
visible effect and can be ignored. In fact,
Refs.~\cite{Tomiya:2016jwr, Aoki:2020noz} claim that it disappears
entirely at a nonzero value of the quark mass, based on results
obtained with chiral discretizations of the Dirac operator, and that
its persistence observed in Refs.~\cite{Dick:2015twa,
  Kaczmarek:2021ser} is an artefact due to the use of a mixed action,
with overlap spectra computed on staggered backgrounds. On the other
hand, Ref.~\cite{Ding:2020xlj} found persistent
$\mathrm{U}(1)_A$-breaking effects in the chiral limit, originating in
a near-zero peak, using exclusively staggered fermions. For physical
values of the quark mass, Ref.~\cite{Alexandru:2024tel} showed
directly how a near-zero peak emerges in the staggered spectrum in the
continuum limit, which supports the conclusion that it is a physical
feature of the Dirac spectrum at nonzero quark mass. If a singular
near-zero peak is indeed present, the questions are then how fast it
has to disappear in the chiral limit in order to be compatible with
chiral symmetry restoration, and if and how it can affect the fate of
$\mathrm{U}(1)_A$. Since a singular, mass-dependent peak was not
explored in Refs.~\cite{Aoki:2012yj,Kanazawa:2015xna}, this behavior
remained an interesting possibility to investigate.

In Ref.~\cite{Giordano:2024jnc} I showed that a singular peak in the
spectral density complying with chiral symmetry restoration and at the
same time effectively breaking $\mathrm{U}(1)_A$ is indeed technically
possible. I also showed that in this case effective $\mathrm{U}(1)_A$
breaking requires further peculiar features of the spectrum, including
a close relation between the peak modes and topology, and the
delocalization of these modes over the whole system; and that the
first nontrivial cumulant of the topological charge is the same as in
an ideal gas of instantons and anti-instantons, to leading order in
$m$. This was shown to hold for all cumulants in
Ref.~\cite{Kanazawa:2014cua} using an effective approach based on
considerations of analyticity and symmetry.  Of course, the
theoretical possibility of a singular peak with just the right
features may simply be an unlikely edge case. It is then reassuring
for its physical viability that an explicit mechanism leading to the
right kind of behavior is provided by a very simple instanton-based
random matrix model~\cite{Kovacs:2023vzi}.

The purpose of this series of papers is to systematize the approach to
the problem of chiral symmetry restoration and the fate of
$\mathrm{U}(1)_A$ based on the study of the Dirac
spectrum~\cite{Cohen:1997hz,Aoki:2012yj,Kanazawa:2015xna,Azcoiti:2023xvu},
expanding the analysis of Ref.~\cite{Giordano:2024jnc}. This requires
first of all to work on the foundations of the approach, looking for a
characterization of the chirally symmetric phase based on first
principles rather than on plausible but unproven assumptions. This
allows one to better disentangle conclusions that are fully justified
by the nature of the symmetric phase from the consequences of more
technical (and less controllable) assumptions. The conditions
resulting from the request of chiral symmetry restoration are then
translated into constraints on the Dirac spectrum, of very general
nature. Characterizing the symmetric phase and deriving these
constraints is the scope of the present paper. The main arguments,
already outlined in Ref.~\cite{Giordano:2024jnc}, are discussed here
in greater detail. The next step of the program is to work on the
technical assumptions on the Dirac spectrum, extending and refining
the analysis of Ref.~\cite{Giordano:2024jnc}, and generalizing some of
its results. This includes carefully scrutinizing the conclusions of
Ref.~\cite{Azcoiti:2023xvu} concerning the consequences of
commutativity of the thermodynamic and chiral limits. This will be
discussed in a separate paper.

As in Ref.~\cite{Giordano:2024jnc}, I deal with the $N_f=2$ chiral
limit of general gauge theories on the lattice using Ginsparg--Wilson
fermions.  No particular restriction is made on the theory, besides
assuming that it has a symmetric phase where
$\mathrm{SU}(2)_L\times\mathrm{SU}(2)_R$ is fully realized. Results
are obtained on the lattice, but there is no obstacle in extending
them to the continuum, assuming that a continuum limit exists (which
generally requires further restrictions on the theory). I work in the
sector of the theory generated by scalar and pseudoscalar fermion
bilinears, both flavor-singlet and flavor-triplet, referred to as
the ``scalar and pseudoscalar sector'' for brevity, where
susceptibilities can be expressed entirely in terms of Dirac
eigenvalues only. I briefly summarize here the main results, of
general nature, obtained in this paper.

Starting from the basic properties expected of the symmetric phase of
a quantum field theory, I prove that chiral symmetry restoration at
the level of scalar and pseudoscalar susceptibilities is equivalent to
finiteness of these quantities in the chiral limit. In this context,
by finite quantity I always mean a quantity that is not divergent, but
possibly vanishing, in the chiral limit.  This result implies in turn
that in the symmetric phase the scalar and pseudoscalar
susceptibilities must be $m^2$-differentiable functions, or $m$ times
an $m^2$-differentiable function. The discussion of the basic
assumptions is more detailed than in Ref.~\cite{Giordano:2024jnc}, and
the proof is simplified. As already pointed out above, this result
essentially proves that the analyticity assumptions of
Refs.~\cite{Cohen:1997hz,Aoki:2012yj,Kanazawa:2015xna,Azcoiti:2023xvu}
(and of Ref.~\cite{Kanazawa:2014cua}, see below) on susceptibilities
and on the free energy density are necessary conditions for symmetry
restoration, since the distinction between analytic and infinitely
differentiable functions is of very limited practical relevance in
this context.

These symmetry-restoration conditions are the starting point for the
derivation of constraints on the Dirac spectrum, using an explicit
expression for the generating function of scalar and pseudoscalar
susceptibilities in terms of Dirac eigenvalues, reported in
Ref.~\cite{Giordano:2024jnc} and derived in detail here. This
expression shows in particular how exact zero modes generally cannot
be ignored even in the chiral limit, further supporting the criticism
of this assumption of Ref.~\cite{Cohen:1996ng} made in
Refs.~\cite{Evans:1996wf,Lee:1996zy}.

The approach of this paper is more general than, and subsumes that of
Ref.~\cite{Aoki:2012yj}, providing constraints not only on the
spectral density and the topological susceptibility, but on eigenvalue
correlations as well. The constraints derived here do not require any
detailed assumption on the spectral density and other spectral
quantities. At this stage, full restoration of
$\mathrm{SU}(2)_L\times\mathrm{SU}(2)_R$ is compatible both with the
effective breaking and with the effective restoration of
$\mathrm{U}(1)_A$. To obtain any insight into this issue, a more
in-depth study is required, involving a detailed analysis of the
properties of eigenvalue correlators, and requiring additional
technical assumptions. This is discussed in the next paper of this
series.

Under additional assumptions on how chiral symmetry restoration
manifests in susceptibilities involving also nonlocal operators built
entirely out of gauge fields, or in the presence of external fermion
fields, I show that also the spectral density and other spectral
quantities are $m^2$-differentiable. This essentially justifies the
$m^2$-analyticity assumption on the spectral density made in
Refs.~\cite{Aoki:2012yj,Kanazawa:2015xna} starting from more
fundamental assumptions on how chiral symmetry is realized, and
provides further restrictions that can be exploited in constraining
the behavior of the Dirac spectrum.

Finally, as anticipated in Ref.~\cite{Giordano:2024jnc}, using the
formalism developed here I rederive the conclusion of
Ref.~\cite{Kanazawa:2014cua} that in the chiral limit in the symmetric
phase the cumulants of the topological charge are identical to those
found in an ideal gas of (anti)instanton-like objects, to leading
order in the fermion mass, if $\mathrm{U}(1)_A$ is effectively
broken. This result is put here on first-principles ground, justifying
the assumptions of Ref.~\cite{Kanazawa:2014cua}, and allowing for the
study of corrections to the ideal-gas behavior.

The plan of this paper is the following.  In Sec.~\ref{sec:FTLGT} I
summarize the relevant aspects of finite-temperature gauge theories on
the lattice and of Ginsparg--Wilson fermions, including the spectral
properties of the corresponding discretized Dirac operator. In
Sec.~\ref{sec:chisymrest} I discuss chiral symmetry restoration and
its consequences, and derive a set of necessary and sufficient
conditions for chiral symmetry restoration at the level of scalar and
pseudoscalar susceptibilities, and their extension to susceptibilities
involving also functionals of gauge fields. In
Sec.~\ref{sec:det_sources} I derive an explicit expression for the
generating function of scalar and pseudoscalar susceptibilities in
terms of the eigenvalues of a Ginsparg--Wilson Dirac operator. In
Sec.~\ref{sec:Dirac_constr} I use the conditions of
Sec.~\ref{sec:chisymrest} and the results of
Sec.~\ref{sec:det_sources} to obtain constraints on the Dirac spectrum
in the symmetric phase. In Sec.~\ref{sec:top_iig} I show that in the
chiral limit in the symmetric phase the topological charge behaves as
in an ideal instanton gas if $\mathrm{U}(1)_A$ remains effectively
broken. In Sec.~\ref{sec:concl1} I draw my conclusions. Technical
details are discussed in Appendices~\ref{sec:conncorrfunc},
\ref{sec:Trefl}, \ref{sec:rhochirest}, \ref{sec:ansusc},
\ref{sec:WTI_geo}, \ref{sec:invchilim}, and \ref{sec:compl_det}.

\section{Finite-temperature lattice gauge theories with Ginsparg--Wilson
  fermions}
\label{sec:FTLGT}

In this series of papers I consider 3+1-dimensional finite-temperature
lattice gauge theories based on some compact gauge group, with two
flavors of dynamical light (eventually massless) fermions of mass $m$
transforming in some irreducible $N_c$-dimensional representation of
the gauge group, and any number of additional fermions, transforming
in possibly different irreducible representations of the gauge group,
that remain massive as $m\to 0$.

It is assumed that the gauge group, fermion content, and gauge group
representations are such that a phase of the theory exists where in
the limit $m\to 0$ the $\mathrm{SU}(2)_L\times \mathrm{SU}(2)_R$
chiral symmetry of the theory (see below Sec.~\ref{sec:GWchi}) is
fully realized. Further restrictions are generally needed to ensure
the existence of a continuum limit, but this does not affect the
validity of the results as long as one works at finite lattice
spacing.

The theory is discretized on a hypercubic lattice of linear spatial
size $\mathrm{L}$ and temporal size $1/\mathrm{T}$. Here and below I
use lattice units. The spatial and spacetime lattice volumes are
denoted by $\svol=\mathrm{L}^3$ and $\lvol=\svol/\mathrm{T}$,
respectively.  The thermodynamic limit $\lvol\to\infty$ is taken by
sending $\svol\to\infty$ while keeping $1/\mathrm{T}$
fixed.\footnote{The formalism remains unchanged at $\mathrm{T}=0$, the
  only difference being that the thermodynamic limit is taken by
  setting the temporal extension to $\mathrm{L}$ and sending
  $\lvol=\mathrm{L}^4\to \infty$.} It is understood that when taking
the chiral limit $m\to 0$, this is done after taking the thermodynamic
limit, unless specified otherwise.

Gauge link variables taking values in the gauge group, denoted
collectively by $U$, are associated with the lattice edges. Periodic
boundary conditions both in time and in space are imposed on them
by including additional edges that make the hypercubic lattice into a
four-dimensional torus. The detailed form of the discretized gauge
action does not play any role in this work; it is only assumed that it
is invariant under the usual lattice spacetime symmetries (lattice
translations, rotations, and reflections).

Two sets of Grassmann variables $\Psi_{x\alpha c f}$ and
$\bar{\Psi}_{x\alpha c f}$, representing the two light fermions, are
associated with the lattice sites. These variables carry spacetime
indices, including the lattice coordinates $x$ and a discrete Dirac
index $\alpha=1,\ldots,4$, a color (i.e., gauge-group representation)
index, $c=1,\ldots,N_c$, and a flavor index, $f=1,2$, that will all
be suppressed in the following. Boundary conditions periodic in space,
and antiperiodic in time, are imposed on the fermionic variables.

Light fermions are coupled to the gauge links via a discretized
massless Dirac operator $D$ satisfying the Ginsparg--Wilson (GW)
relation $\{D,\gamma_5\} = 2DR\gamma_5 D$, where $\gamma_5$ is the
usual Dirac matrix and $R$ is a local operator that commutes with
$\gamma_5$~\cite{Ginsparg:1981bj,Hasenfratz:1998ri,Hasenfratz:1998jp,
  Neuberger:1998wv,Luscher:1998pqa,Horvath:1999bk}.\footnote{The
  general form of the GW relation is
  $\{D,\gamma_5\} = D\{R',\gamma_5\} D$, with $R'$
  local~\cite{Ginsparg:1981bj,Horvath:1999bk}.  Without loss of
  generality one can replace
  $R'\to R=\f{1}{2}\left(R' + \gamma_5 R'\gamma_5\right)$, with
  $[R,\gamma_5]=0$ and $\{R',\gamma_5\}=2R\gamma_5$.} $D$ carries only
spacetime and color indices. It is again assumed that invariance
under the usual lattice spacetime symmetries holds.  Examples of
operators obeying the GW relation are the domain wall
operator~\cite{Kaplan:1992bt,Shamir:1993zy,Furman:1994ky,Borici:1999zw,
  Chiu:2002ir,Brower:2005qw,Brower:2012vk}, the overlap
operator~\cite{Narayanan:1993sk,Narayanan:1993ss,
  Neuberger:1997fp,Neuberger:1998wv}, and the fixed-point
action~\cite{Hasenfratz:1993sp,Bietenholz:1995cy,DeGrand:1995ji,
  Hasenfratz:1998ri,Hasenfratz:1998jp,Hasenfratz:2002rp}. Massless
Dirac operators obeying the GW relation possess an exact lattice
chiral symmetry~\cite{Neuberger:1997fp,Neuberger:1998wv,
  Hasenfratz:1998ri,Luscher:1998pqa,Kikukawa:1998py,Horvath:1999bk},
that reduces to the usual chiral symmetry of continuum fermions in the
continuum limit. The associated massive operator $D_m$
is~\cite{Chandrasekharan:1998wg, Niedermayer:1998bi}
\begin{equation}
  \label{eq:GW_massive}
  D_m = D + m\left(\mathbf{1}-DR\right)\,,
\end{equation}
where $\mathbf{1}$ is the identity in 
spacetime (including Dirac) and color space.  The mass is coupled
here to the simplest proper order parameter for chiral symmetry
breaking~\cite{Hasenfratz:1998jp,Chandrasekharan:1998wg,Niedermayer:1998bi}.

Massive fermions are similarly represented by pairs of sets of
Grassmann variables, and are coupled to gauge fields via
discretizations of the Dirac operator, possibly not of GW type. Again,
their detailed form plays no role; invariance under the usual lattice
spacetime symmetries is assumed, and the usual boundary conditions for
fermions are imposed.

After integrating out the massive fermion fields, the partition
function reads
\begin{equation}
  \label{eq:general_pf2}
  \begin{aligned}
    Z & = \int DU\int D\Psi D\bar{\Psi}\, e^{-S_{\mathrm{eff}}(U)}
        e^{-\bar{\Psi} D_m(U)\mathbf{1}_{\mathrm{f}}\Psi}\\
      & = \int DU  \, e^{-S_{\mathrm{eff}}(U)}\left[\det D_m(U)\right]^2 \,,
  \end{aligned}
\end{equation}
where $S_{\mathrm{eff}}$ includes the gauge action and the
contribution of massive fermion fields, $\mathbf{1}_{\mathrm{f}}$ is
the identity in flavor space, $DU$ denotes the product of the Haar
measures associated with the link variables, and $D\Psi D\bar{\Psi}$
the product of the Berezin measures associated with the fermion
fields. The measure is invariant under gauge transformations and
lattice symmetry transformations. Expectation values of observables
depending only on gauge and light-fermion fields read
\begin{equation}
  \label{eq:general_pf3}
  \begin{aligned}
    \la \Oc\ra
    &= Z^{-1}   \int DU \int D\Psi D\bar{\Psi}\,
      e^{-S_{\mathrm{eff}}(U)}e^{-\bar{\Psi} D_m(U)\mathbf{1}_{\mathrm{f}}\Psi}\\
    & \hphantom{=} \times\Oc(\Psi,\bar{\Psi},U)\,. 
  \end{aligned}
\end{equation}
These expectation values are understood to be evaluated in a finite
spatial volume $\svol$; their dependence on $\mathrm{T}$ and $\svol$
is left implicit for notational simplicity.

Finally, I assume that the theory is invariant under a discrete ``$CP$
transformation'' of the form
\begin{equation}
  \label{eq:Trelf1}
  U\to U_{\mathcal{CP}}\,, \quad \Psi \to \CP \mathbf{1}_{\mathrm{f}}\Psi\,, \quad
  \bar{\Psi}\to \bar{\Psi}\mathbf{1}_{\mathrm{f}}\CP^\dag\,,
\end{equation}
where $\CP$ is a suitable unitary matrix (with spacetime and color
indices only), and
\begin{equation}
  \label{eq:Trelf1_bis}
  \begin{aligned}
    S_{\mathrm{eff}}(U_{\mathcal{CP}})  &= S_{\mathrm{eff}}(U)\,, 
    &&&    &   \\  
    \CP^\dag D(U_{\mathcal{CP}}) \CP   &  = D(U)\,,
    &&& \CP^\dag\gamma_5   \CP       &= -\gamma_5\,.
  \end{aligned}
\end{equation}
The integration measure is assumed to be invariant under the
transformation Eq.~\eqref{eq:Trelf1}. Calling this a $CP$
transformation is a bit of a misnomer, although common in the
literature: in fact, viable choices satisfying these requirements are
the temporal reflection, reflections through a plane perpendicular to
one of the spatial directions, or the spatial parity
transformation. $CP$ invariance is then guaranteed for all the most
common discretizations of gauge and fermion actions. Notice, however,
that the discussion in Secs.~\ref{sec:chisymrest} and
\ref{sec:det_sources} makes no use of $CP$ invariance, and the results
obtained there hold also if one includes a $CP$-violating topological
term in the action (except of course for results where the use of $CP$
is explicitly mentioned).

\subsection{Chiral symmetry of the GW Dirac operator}
\label{sec:GWchi}

Thanks to the chiral properties of GW Dirac operators, the system
under consideration has at the classical level an exact
$\mathrm{U}(2)_L\times \mathrm{U}(2)_R$ symmetry in the chiral
limit~\cite{Luscher:1998pqa,Kikukawa:1998py,Horvath:1999bk}. Let
$\vec{\sigma}=(\sigma_1,\sigma_2,\sigma_3)$ denote the usual Pauli
matrices acting in flavor space, and let
$\hat{\gamma}_5 \equiv (\mathbf{1}-2DR)\gamma_5$. The GW relation can
be recast as $D\gamma_5 + \hat{\gamma}_5D =0$, and implies
$\hat{\gamma}_5^2=\mathbf{1}$.  Flavor non-singlet,
$\mathrm{SU}(2)_L\times \mathrm{SU}(2)_R$ chiral transformations,
denoted by $\U(\vec{\alpha}_L,\vec{\alpha}_R)$, are defined by
\begin{equation}
  \label{eq:ch_transf1}
  \begin{aligned}
    \Psi
    &\to \Psi_{\U} = \left( U(\vec{\alpha}_L)  \tf{\mathbf{1}- \gamma_5}{2}
      + U(\vec{\alpha}_R) \tf{\mathbf{1}+ \gamma_5}{2}
      \vphantom{U(\vec{\alpha}_L)^\dag \tf{\mathbf{1}+ \hat{\gamma}_5}{2}}\right)
      \Psi\,, \\
    \bar{\Psi}
    &\to \bar{\Psi}_{\U}=
      \bar{\Psi}
      \left(U(\vec{\alpha}_L)^\dag \tf{\mathbf{1}+ \hat{\gamma}_5}{2}
      + U(\vec{\alpha}_R)^\dag \tf{\mathbf{1}- \hat{\gamma}_5}{2} \right)
      \,, 
  \end{aligned}
\end{equation}
where $\vec{\alpha}_{L,R}\in\mathbb{R}^3$ and
$ U(\vec{\alpha})\equiv
e^{i\vec{\alpha}\cdot\f{\vec{\sigma}}{2}}\in\mathrm{SU}(2)$. Flavor-singlet,
$\mathrm{U}(1)_L\times \mathrm{U}(1)_R$ chiral transformations,
denoted by $\U^{(0)}(\alpha_L,\alpha_R)$, are similarly defined by
\begin{equation}
  \label{eq:ch_transf3}
  \begin{aligned}
    \Psi
    &\to \Psi_{\U^{(0)}}=
      \left(   e^{i\alpha_L}\tf{\mathbf{1}- \gamma_5}{2} +
      e^{i\alpha_R}\tf{\mathbf{1}+ \gamma_5}{2}
      \vphantom{\tf{\mathbf{1}+ \hat{\gamma}_5}{2}}\right)
      \Psi\,,    \\
    \bar{\Psi}
    &\to \bar{\Psi}_{\U^{(0)}} =
      \bar{\Psi}
      \left(e^{-i\alpha_L}\tf{\mathbf{1}+ \hat{\gamma}_5}{2}+
      e^{-i\alpha_R}\tf{\mathbf{1}- \hat{\gamma}_5}{2} \right)\,.
  \end{aligned}
\end{equation}
Nonsinglet vector and axial transformations, $\mathrm{SU}(2)_V$ and
$\mathrm{SU}(2)_A$, are defined respectively as
$\U_V(\vec{\alpha}) \equiv \U(\vec{\alpha},\vec{\alpha}) $ and
$\U_A(\vec{\alpha}) \equiv \U(-\vec{\alpha},\vec{\alpha}) $, and
singlet vector and axial transformations, $\mathrm{U}(1)_V$ and
$\mathrm{U}(1)_A$, are defined respectively as
$\U_{V}^{(0)}(\alpha) \equiv \U^{(0)}(\alpha,\alpha)$ and
$\U_{A}^{(0)}(\alpha) \equiv\U^{(0)}(-\alpha,\alpha)$, with
$\mathrm{U}(1)_L\times \mathrm{U}(1)_R = \mathrm{U}(1)_V\times
\mathrm{U}(1)_A$.

It is straightforward to show that
$\bar{\Psi}D\mathbf{1}_{\mathrm{f}}\Psi$ is invariant under the chiral
$\mathrm{U}(2)_L\times \mathrm{U}(2)_R$ transformations
Eqs.~\eqref{eq:ch_transf1} and \eqref{eq:ch_transf3}. The Berezin
integration measure is invariant under
$\mathrm{SU}(2)_L\times \mathrm{SU}(2)_R$ and $\mathrm{U}(1)_V$
transformations, but not under $\mathrm{U}(1)_A$ transformations,
under which~\cite{Luscher:1998pqa}
\begin{equation}
  \label{eq:U1Ameas}
  D\Psi D\bar{\Psi} \to  D\Psi D\bar{\Psi}\, e^{-i 4\alpha Q}\,,  
\end{equation}
where $Q$ is the topological charge [see under
Eq.~\eqref{eq:GWspec4}].\footnote{The proof of Eq.~\eqref{eq:U1Ameas}
  in Ref.~\cite{Luscher:1998pqa} is for the case $2R=\mathbf{1}$, but
  can be extended to general $R$ without difficulty.} This makes
$\mathrm{U}(1)_A$ an anomalous symmetry already on the
lattice~\cite{Ginsparg:1981bj,Hasenfratz:1998ri,Luscher:1998pqa}. The
full chiral symmetry of the classical lattice action for massless
fermions is then broken by quantum effects to
$\mathrm{SU}(2)_L\times \mathrm{SU}(2)_R\times \mathrm{U}(1)_V$; a
nonzero fermion mass $m$ breaks this explicitly further, down to
$\mathrm{SU}(2)_V\times \mathrm{U}(1)_V$.  In the chiral limit
$m\to 0$, $\mathrm{SU}(2)_L\times \mathrm{SU}(2)_R$ may not be
recovered, but instead break down spontaneously to $\mathrm{SU}(2)_V$:
this is the case, e.g., for the two-flavor chiral limit of QCD and
QCD-like theories based on the gauge group $\mathrm{SU}(N_c)$ at zero
and low temperatures. At higher temperatures, above a
symmetry-restoring phase transition, chiral symmetry is instead fully
realized (see, e.g., Refs.~\cite{Tomiya:2016jwr,HotQCD:2019xnw}). The
results of this work apply to gauge theories in such an
$\mathrm{SU}(2)_L\times \mathrm{SU}(2)_R$-symmetric phase.

Since the two-flavor chiral limit is taken along the line of
mass-degenerate light fermions, $m_1=m_2=m$, $\mathrm{SU}(2)_V$
symmetry is exactly realized for any $m$, although in principle the
system may not be in a pure phase in the thermodynamic limit. However,
this is guaranteed by the fact that $\mathrm{SU}(2)_V$ cannot break
down spontaneously if the integration measure is positive-definite:
even if one breaks it explicitly by setting $m_1\neq m_2$, one always
recovers it when $m_{1,2}\to m$ (at least for nonzero $m$). The
impossibility of spontaneous breaking of $\mathrm{SU}(N_f)_V$ was
shown in the continuum in Ref.~\cite{Vafa:1983tf}, and on the lattice
for various discretizations including GW fermions in
Ref.~\cite{Giordano:2023spj} (see also
Refs.~\cite{Aloisio:2000rb,Azcoiti:2010ns}).

\subsection{Scalar and pseudoscalar bilinears}
\label{sec:sps_dens}

The quantities of interest in this work are the connected correlation
functions of the following fermion bilinears,
\begin{equation}
  \label{eq:densities}
  \begin{aligned}
    S &\equiv \bar{\Psi} (\mathbf{1}-DR)\mathbf{1}_{\mathrm{f}} \Psi \,, &&& P &\equiv \bar{\Psi}
    (\mathbf{1}-DR)\mathbf{1}_{\mathrm{f}} \gamma_5\Psi\,, \\
    \vec{P} &\equiv \bar{\Psi} (\mathbf{1}-DR) \vec{\sigma}\gamma_5\Psi \,, &&& \vec{S}
    &\equiv \bar{\Psi} (\mathbf{1}-DR) \vec{\sigma}\Psi \,,
  \end{aligned}
\end{equation}
i.e., the spacetime integral of scalar and pseudoscalar isosinglet and
isotriplet densities, including a suitable subtraction
term~\cite{Chandrasekharan:1998wg,Niedermayer:1998bi,Hasenfratz:2002rp}.
Under infinitesimal chiral transformations of the fermion fields, the
following four-component vectors of fermionic bilinears,
\begin{equation}
  \label{eq:OVW_def}
  O_V    \equiv    \begin{pmatrix}
    S \\ i\vec{P}
  \end{pmatrix}\,,
  \qquad
  O_W  \equiv   \begin{pmatrix}
    iP \\ -\vec{S}
    \end{pmatrix}\,, 
\end{equation}
transform by an infinitesimal $\mathrm{SO}(4)$
rotation~\cite{Hasenfratz:2002rp}. Using the exponential map, under a
finite non-singlet chiral transformation $\U$,
Eq.~\eqref{eq:ch_transf1}, one finds then
\begin{equation} 
  \label{eq:O_VW_transf}
  O_{V,W}   \to (  O_{V,W})_{\U}=   \Rc(\U)  O_{V,W}  \,,
\end{equation}
for some $\Rc(\U) \in \mathrm{SO}(4)$, where $( O_{V,W})_{\U}$ denotes
the chirally transformed scalar and pseudoscalar bilinears, i.e., the
bilinears defined in Eq.~\eqref{eq:densities} with $\Psi$ and
$\bar{\Psi}$ replaced by $\Psi_{\U}$ and $\bar{\Psi}_{\U}$.  Since the
exponential map is surjective for compact connected groups (see
Ref.~\cite{BrockerDieck1985}, ch.~IV, theorem 2.2), the mapping
$\U \to \Rc(\U)$ provides a representation of
$\mathrm{SU}(2)_L\times \mathrm{SU}(2)_R$ that is onto
$\mathrm{SO}(4)$.

Concerning flavor-singlet transformations, Eq.~\eqref{eq:ch_transf3},
the bilinears in Eq.~\eqref{eq:densities} are manifestly invariant
under vector transformations, while $O_V$ and $O_W$ get mixed under
axial transformations: under the transformation
$\U_{A}^{(0)}\!(\tf{\alpha}{2})\in\mathrm{U}(1)_A$ one finds
\begin{equation}
  \label{eq:OVW_transf_singl}
  \begin{pmatrix}
    O_V\\  O_W
  \end{pmatrix} 
  \to
  \begin{pmatrix}
    \cos\alpha &  \sin\alpha\\
    -\sin\alpha & \cos\alpha
  \end{pmatrix}
  \begin{pmatrix}
    O_V\\  O_W
  \end{pmatrix}\,.
\end{equation}
Finally, under the $CP$ transformation, Eqs.~\eqref{eq:Trelf1} and
\eqref{eq:Trelf1_bis}, scalar and pseudoscalar bilinears transform as
\begin{equation}
  \label{eq:Trelf4}
  S\to S\,,  \quad \vec{P}\to -\vec{P}\,, \quad  P\to -P\,, \quad \vec{S}\to\vec{S}\,. 
\end{equation}

\subsection{Generating function}
\label{sec:genfunc}

The correlation functions of scalar and pseudoscalar bilinears are
conveniently handled through generating functions. Let
\begin{equation}
  \label{eq:sources}
  \begin{aligned}
    \des( \Psi,\bar{\Psi},U; \sous,\soup)
    &\equiv \ssi S + i\spi\cdot
      \vec{P} + i \seta P - \sde\cdot \vec{S}\\
    &= \sous\cdot O_V + \soup\cdot O_W\,,
  \end{aligned}
\end{equation}
with
\begin{equation}
  \label{eq:sources_bis}
  \sous\equiv
  \begin{pmatrix}
    \ssi \\
    \spi 
  \end{pmatrix}\,, \qquad
  \soup\equiv
  \begin{pmatrix}
  \seta\\ \sde  
  \end{pmatrix}\,,
\end{equation}
where $\ssi$ and $\seta$ are isosinglet scalar and pseudoscalar
external sources, and similarly $j_{Sa}$ and $j_{Pa}$, $a=1,2,3$, are
isotriplet scalar and pseudoscalar external sources, collected in the
vectors $\sde$ and $\spi$. One defines the generating functions
$\mathcal{Z}$ and $\mathcal{W}$ of full and connected correlation
functions, respectively, as
\begin{equation}
  \label{eq:partfunc}
  \begin{aligned}
    \mathcal{Z}(\sous,\soup;m)
    &\equiv \int DU    \int D\Psi D\bar{\Psi} \, e^{-S_{\mathrm{eff}}(U)}\\
    &\phantom{=}\times e^{- \bar{\Psi} D_m (U)\mathbf{1}_{\mathrm{f}}  \Psi
      - \des( \Psi,\bar{\Psi},U; \sous,\soup) }\\
    &\equiv \exp\left\{ \lvol      \mathcal{W}(\sous,\soup;m) \right\} \,.
  \end{aligned}
\end{equation}
For notational simplicity, the dependence of $\mathcal{Z}$ and
$\mathcal{W}$ on $\svol$ and $\mathrm{T}$ is omitted.  Up to constant
factors, the derivatives of $\mathcal{Z}$ with respect to the sources
evaluated at vanishing sources are the full correlation functions of
the scalar and pseudoscalar bilinears defined in
Eq.~\eqref{eq:densities} [see Eqs.~\eqref{eq:general_pf2} and
\eqref{eq:general_pf3}],
\begin{equation}
  \label{eq:largeV_cu4}
  \begin{aligned}
    &   \left\la S^{n_S} (iP)^{n_P} 
      (i\vec{P})^{\vec{n}_{P}} 
      \vec{S}^{\,\vec{n}_{S}}\right\ra (-1)^{n_S+n_P+\sum_{a=1}^3 n_{Pa}}\\
    &  =          Z^{-1}
      \left(\de_{j_S}\right)^{n_S}
      \left(\de_{j_P}\right)^{n_P}
      \prod_{a=1}^3\left[
      \left(\de_{j_{Pa}}\right)^{n_{P_a}}
      \left(\de_{j_{Sa}}\right)^{n_{S_a}}\right]
    \\ &\phantom{=}\times       \mathcal{Z}(\sous,\soup;m)|_0  \,,
  \end{aligned}
\end{equation}
where $\de_x \equiv \de/\de x$, $|_0$ denotes setting $V=W=0$, and I
have used the shorthand
$\vec{X}^{\vec{n}_X}\equiv\prod_{a=1}^3 X_a^{n_{X a}}$. 
Similarly,
\begin{equation}
  \label{eq:genfuncdef2}
  \begin{aligned}
    &  \f{1}{\lvol}\!
      \left\la S^{n_S}
      (iP)^{n_P}(i\vec{P})^{\vec{n}_{P}} \vec{S}^{\vec{n}_{S}}\right\ra_c\!(-1)^{n_S+n_P+\sum_{a=1}^3 n_{Pa}}\\
    &= (\de_{j_S})^{n_S}
      (\de_{j_P})^{n_P}\prod_{a=1}^3\left[(\de_{j_{Pa}})^{n_{Pa}}(\de_{j_{Sa}})^{n_{Sa}}\right]
    \\ &\phantom{=}\times      \mathcal{W}(\sous,\soup;m)|_0  \,,
  \end{aligned}
\end{equation}
where $\la\ldots\ra_c$ denotes connected correlation functions,
defined recursively in the usual way (see
Appendix~\ref{sec:ccf_rec}). Up to the same constant factors as above,
in the thermodynamic limit the derivatives of $\mathcal{W}$ at zero
sources yield the scalar and pseudoscalar susceptibilities,
$\chi\left(S^{n_S} (i\vec{P})^{\vec{n}_P} (iP)^{n_P}
  \vec{S}^{\vec{n}_S}\right)$, where
\begin{equation}
  \label{eq:genfuncdef2_s}
  \chi\left( \textstyle\prod_iO_i \right)
  \equiv \lim_{\lvol\to\infty}\f{1}{\lvol}        \left\la  \textstyle\prod_iO_i\right\ra_c
  \,.
\end{equation}
For practical purposes, it is convenient to treat $\mathcal{W}$ as a
formal power series in the sources, with the normalized connected
correlators of scalar and pseudoscalar bilinears in
Eq.~\eqref{eq:genfuncdef2} as coefficients, that can be truncated at
any sufficiently high but finite order $n$ if one is interested only
in correlators involving no more than $n$ bilinears. This guarantees
the possibility to exchange taking derivatives with respect to the
sources at zero sources with any other operation, in particular taking
derivatives with respect to $m$, and taking the thermodynamic or
chiral limit. Moreover, exchanging the thermodynamic limit with
derivatives with respect to $m$ is expected to be allowed at any
nonzero $m$, where the finite correlation length of the system
guarantees that the thermodynamic limit is uniform in $m$ in any range
of nonzero masses.\footnote{\label{foot:der}Notice that in numerical
  simulations of lattice theories, the derivative of the thermodynamic
  limit of a quantity with respect to some parameter (such as the mass
  or the $\theta$ angle) is by practical necessity \textit{defined} as
  the thermodynamic limit of the derivative. Lacking a rigorous proof
  of the properties of the relevant quantities in the thermodynamic
  limit, the other order of operations is currently unattainable, and
  one is forced to assume that derivatives and thermodynamic limit
  commute at $m\neq 0$ to make any progress.}

With this in mind, it is convenient to denote with
$ \wiv = \lim_{\lvol\to\infty}\mathcal{W}$ the formal power series
collecting the thermodynamic limit of the relevant correlators, i.e.,
the generating function of the susceptibilities, and write
\begin{equation}
  \label{eq:genfuncdef2_ss}
  \begin{aligned}
    &  \chi\left(S^{n_S}
      (i\vec{P})^{\vec{n}_P} (iP)^{n_P}
      \vec{S}^{\vec{n}_S}\right)(-1)^{n_S+n_P+\sum_{a=1}^3 n_{Pa}}\\
    &= 
      (\de_{j_S})^{n_S}
      (\de_{j_P})^{n_P}\prod_{a=1}^3\left[(\de_{j_{Pa}})^{n_{Pa}}(\de_{j_{Sa}})^{n_{Sa}}\right]
    \\ &\phantom{=}\times      \wiv(\sous,\soup;m) |_0 \,.
  \end{aligned}
\end{equation}
It follows from the discussion in Sec.~\ref{sec:sps_dens} that for a
generic nonsinglet chiral transformation, Eq.~\eqref{eq:ch_transf1},
one has
\begin{equation}
  \label{eq:sources2}
  \des(\Psi_{\U},\bar{\Psi}_{\U},U; \sous,\soup)
  = \des( \Psi,\bar{\Psi},U; \Rc(\U)^T\sous,\Rc(\U)^T\soup)\,,
\end{equation}
with $\Rc(\U)\in \mathrm{SO}(4)$ the rotation matrix associated with
the chiral transformation, see Eq.~\eqref{eq:O_VW_transf}. The
generating function of the susceptibilities of the chirally
transformed scalar and pseudoscalar bilinears, $( O_{V,W})_{\U}$, is
then simply $\wiv(\Rc(\U)^T\sous,\Rc(\U)^T\soup;m)$.

The transformation properties under $CP$, Eq.~\eqref{eq:Trelf4},
imply also that
\begin{equation}
  \label{eq:sources3}
  \des( \CP\Psi,\bar{\Psi}\CP^\dag,U_{\mathcal{CP}}; \sous,\soup)
  =
  \des(\Psi,\bar{\Psi},U; \mathcal{C}\sous,-\mathcal{C}\soup)\,,
\end{equation}
where $\mathcal{C}\equiv\mathrm{diag}(1,-1,-1,-1)$. Together with the
invariance of the integration measure this implies that
\begin{equation}
  \label{eq:sources4}
    \mathcal{Z}(\mathcal{C}\sous,-\mathcal{C}\soup;m)=\mathcal{Z}(\sous,\soup;m)\,,
\end{equation}
and similarly for $\mathcal{W}$ and $\wiv$. In the case of
$\gamma_5$-Hermitean GW operators with $2R=\mathbf{1}$ (see below
Sec.~\ref{sec:GWspec}), Eq.~\eqref{eq:sources4} implies that
$\mathcal{Z}$ is real, see Appendix~\ref{sec:Trefl}, and therefore the
derivatives of $\wiv$ at zero sources, i.e., the susceptibilities in
Eq.~\eqref{eq:genfuncdef2_ss}, are real. (Since $\mathcal{Z}|_0=Z$ is
positive at zero sources, the free energy density $-\wiv|_0$ is real
as well.)

\subsection{Spectrum of the GW Dirac operator}
\label{sec:GWspec}

The main purpose of this work is to investigate the consequences of
chiral symmetry restoration for the Dirac spectrum, more precisely for
the eigenvalues of the massless GW operator,
\begin{equation}
  \label{eq:GWspec1}
  D(U)  \psi_n(U) = \mu_n(U) \psi_n(U)\,.
\end{equation}
The eigenvalues are generally complex, $\mu_n(U)\in\mathbb{C}$. The
eigenvectors $\psi_n(U)$ carry spacetime and color indices, and the
index $n$ ranges over $N_{\mathrm{tot}} = 4 N_c \lvol$ values. Both
eigenvalues and eigenvectors depend on the gauge configuration; this
dependence will be often omitted for notational simplicity.

Common realizations of GW fermions, such as domain
wall~\cite{Kaplan:1992bt,Shamir:1993zy,Furman:1994ky,Borici:1999zw,
  Chiu:2002ir,Brower:2005qw,Brower:2012vk} or overlap
fermions~\cite{Narayanan:1993sk,Narayanan:1993ss,
  Neuberger:1997fp,Neuberger:1998wv}, have the additional properties
that $2R=\mathbf{1}$ and that $D$ is $\gamma_5$-Hermitean,
$\gamma_5 D \gamma_5 = D^\dag$.  In this case $\mathbf{1}-D$ is
unitary, so $D$ and $D^\dag$ have a common basis of orthonormal
eigenvectors, with $D\psi_n=\mu_n\psi_n$, $\mu_n = 1-e^{-i\varphi_n}$,
$\varphi_n\in (-\pi,
\pi]$, and $D^\dag\psi_n=\mu_n^*\psi_n$, which implies that also
$ \gamma_5 \psi_n$ is an eigenvector of $D$ with eigenvalue
$\mu_n^*$. Complex modes, $\mu_n\neq \mu_n^*$, come then in conjugate
pairs, with $\psi_n$ and $ \gamma_5 \psi_n$ the corresponding
orthogonal eigenvectors. If $\mu_n=\mu_n^*$ is real, it must be either
$\mu_n=0$ ($\varphi_n=0$), or $\mu_n=2$ ($\varphi_n=\pi$).  These
modes can be, and usually are, chosen to be chiral,
\begin{equation}
  \label{eq:GWspec4}
  D\psi_n^{(r)} = r \psi_n^{(r)}\,, \qquad
  \gamma_5\psi_n^{(r)} = \xi^{(r)}_n \psi_n^{(r)}
  \,, 
\end{equation}
where $r=0,2$ and $\xi^{(0,2)}_n = \pm 1$. I denote the number of zero
modes with chirality $\pm 1$ in configuration $U$ by $N_\pm(U)$; the
total number of zero modes by $N_0(U)\equiv N_+(U)+N_-(U)$; and the
topological charge by $Q(U)\equiv N_+(U) - N_-(U)$.  The
identification of the topological charge with the index of $D$ is
justified by the lattice index theorem of
Ref.~\cite{Hasenfratz:1998ri}. One similarly has $N_\pm'(U)$
``doubler'' modes with eigenvalue 2 and chirality $\pm 1$, and
$N_2(U) \equiv N_+'(U) + N_-'(U)$ doubler modes in total, with
$Q=N_+-N_- = N_-'-N_+'$ (since $\gamma_5$ is traceless).

The transformation property Eq.~\eqref{eq:Trelf1_bis} implies that
$\CP\psi_n(U)$ are eigenvectors of $D(U_{\mathcal{CP}})$ with
eigenvalue $\mu_n(U)$.  The gauge action and the Dirac spectrum are
then the same for the gauge configurations $U$ and $U_{\mathcal{CP}}$,
that have therefore the same weight in the partition function. On the
other hand, for the real chiral modes $\psi_n^{(r)}(U)$ one finds that
$\CP \psi_n^{(r)}(U) $ has chirality opposite to that of
$\psi_n^{(r)}(U) $, and so $Q(U_{\mathcal{CP}})=-Q(U)$, implying in
particular that $\la Q^{2k+1}\ra=0$ for any nonnegative integer $k$
(this of course is not true anymore if a topological term is added to
the action).

The density of complex modes and their correlation functions are
conveniently expressed in terms of
$\lambda_n \equiv 2\sin\f{\varphi_n}{2}$, with
$\lambda_n^2=|\mu_n|^2$. Denoting with
$N(U)\equiv\f{1}{2}[N_{\mathrm{tot}}-N_0(U)-N_2(U)]$ the number of
pairs of complex modes, it is convenient to label as $\mu_n$,
$n=1,\ldots,N$, the modes with $\Im\mu_n>0$ [$\varphi_n\in (0,\pi)$],
and with $\mu_{-n}\equiv \mu_n^*$ their complex conjugates, so
$\Im\mu_{-n}<0$, $\varphi_{-n}=-\varphi_n\in (-\pi,0)$, and
$\lambda_{-n}=-\lambda_n$.  I define the spectral density in a given
gauge configuration as
\begin{equation}
  \label{eq:rho_def1}
  \rho_U(\lambda)  \equiv
  \sum_n \delta \left(\lambda - \lambda_n(U)\right)\,,
\end{equation}
where the sum runs over $n=\pm 1,\ldots,\pm N(U)$.  This is a
distribution supported in $(-2,0)\cup (0,2)$, symmetric about the
origin, and normalized as
\begin{equation}
  \label{eq:rho_def2}
  \int_{-2}^2 d\lambda\,\rho_U(\lambda) =
  2  \int_0^2 d\lambda\,\rho_U(\lambda)  =  2N(U)
  \,.  
\end{equation}
(This differs by a factor $\lvol$ from the normalization used in
Ref.~\cite{Giordano:2024jnc}.) While $\rho_U$ is a highly singular
object, one expects that the (normalized) spectral density,
\begin{equation}
  \label{eq:rho_def3}
  \rho^{(1)}_{c}(\lambda;m)\equiv 
  \f{1}{\lvol}\la \rho_U(\lambda) \ra\,,
\end{equation}
obtained after averaging over gauge configurations, is an ordinary
function (although it may still develop distributional contributions
in the thermodynamic limit).\footnote{The spectral density
  $\rho^{(1)}_{c}$ may still contain Dirac-delta terms if there are
  eigenvalues that appear repeatedly on a set of configurations of
  nonzero measure.  In the system under consideration this should
  apply only to $\lambda_n=0,2$, that are excluded in
  Eq.~\eqref{eq:rho_def1}.}  In analogy with $\rho_U$ one defines also
higher-point eigenvalue correlation functions in a fixed
configuration, removing contact terms for coinciding arguments,
\begin{equation}
  \label{eq:rho_def4}
  \begin{aligned}
    &  \rho_U^{(k)}(\lambda_1,\ldots,\lambda_k)\\
    & \equiv \sum_{\substack{n_1,\ldots,n_k\\ n_i\neq
    \pm  n_j,\, \forall i\neq j}} \delta \left(\lambda_1 -
    \lambda_{n_1}(U)\right)\ldots
    \delta \left(\lambda_k - \lambda_{n_k}(U)\right)\,,
  \end{aligned}
\end{equation}
where of course $ \rho_U^{(1)}= \rho_{U}$.  By construction
$\rho_U^{(k)}$ is symmetric under permutations of its arguments; and
is symmetric under reflection of any of its arguments,
$\lambda_i\to -\lambda_i$, thanks to the symmetry of the spectrum.
One then defines suitably normalized connected $k$-point eigenvalue
correlators,
\begin{equation}
  \label{eq:rho_def5_0}
  \rho_{c}^{(k)}  (\lambda_1,\ldots,\lambda_k;m)
  \equiv   \f{1}{\lvol} 
  \la
  \rho_U^{(k)}(\lambda_1,\ldots,\lambda_k)
  \ra_c\,,
\end{equation}
according to the usual recursive procedure, see
Appendix~\ref{sec:ccf_rec}. The quantities $\rho_{c}^{(k)}$ are
obviously symmetric under permutations and reflections of their
arguments.  One similarly defines normalized connected correlation
functions involving the complex Dirac modes and other quantities $O$
(see Appendix~\ref{sec:ccf_rec}),
\begin{equation}
  \label{eq:rho_def8_gen}
  \rho_{O\,c}^{(k)}(\lambda_1,\ldots,\lambda_k;m)
  \equiv \f{1}{\lvol}
\left\la O\, \rho_U^{(k)}(\lambda_1,\ldots,\lambda_k)\right\ra_c\,. 
\end{equation}
For $O=1$ one gets back $\rho_{1\,c}^{(k)}=\rho_{c}^{(k)}$; in this
case the subscript $O$ will be omitted. For $k=0$ one gets the
cumulants
\begin{equation}
  \label{eq:generic_cumulant}
 b_O \equiv \rho_{O\,c}^{(0)}=
  \f{\la O\ra_c}{\lvol}\,. 
\end{equation}
Examples relevant in the following are the correlation functions
involving both zero and complex modes, i.e., $O=N_+^{k_+} N_-^{k_-}$,
or $O=N_0^{k_0}Q^{k_1}$. Since $N_+,N_-$ and $N_0,Q$ are linearly
related, their connected correlation functions are related by the same
linear transformation that relates their full correlation functions
(see Appendix~\ref{sec:ccf_cr}).  Relevant examples for $k=1$ are
\begin{equation}
  \label{eq:rho_def12_bis}
  \begin{aligned}
    \rho_{N_0\,c}^{(1)}(\lambda;m) 
    &= \f{\la N_0 \rho_U(\lambda)\ra_c}{\lvol}=
      \f{     \la      N_0 \rho_U(\lambda)      \ra 
      - \la      N_0\ra \la \rho_U(\lambda)      \ra}{\lvol}
      \,,    \\
    \rho_{Q^2\,c}^{(1)}(\lambda;m) 
    &= \f{\la Q^2 \rho_U(\lambda)\ra_c}{\lvol}=
      \f{      \la      Q^2 \rho_U(\lambda)      \ra 
      -\la      Q^2\ra
      \la \rho_U(\lambda)      \ra
      }{\lvol}\,,
  \end{aligned}
\end{equation}
where the last passage on the second line applies if $CP$ invariance
holds.  The case $k=0$ reduces to the cumulants of $N_+$ and $N_-$, or
those of $N_0$ and $Q$,
\begin{equation}
  \label{eq:Ncum_def}
  b_{N_0^{k_0}Q^{k_1}} =  \rho^{(0)}_{N_0^{k_0}Q^{k_1}\,c}
  =    \f{\la N_0^{k_{0}}Q^{k_{1}}\ra_c}{\lvol}\,.
\end{equation}
For future utility, I define the following integrals associated with
$\rho^{(k)}_{O\,c}$,
\begin{equation}
  \label{eq:cum_sys_3_0_again}
  I^{(k)}_{O}[g_1,\ldots, g_k]   
  \equiv    \left[\prod_{i=1}^k\int_0^2 d\lambda_i\,
    g_i(\lambda_i)\right]\!
  \rho^{(k)}_{O\,c}(\lambda_1,\ldots,\lambda_k;m)\,.
\end{equation}
Their dependence on $\svol$, $\mathrm{T}$, and $m$ is left implicit.
The subscript $O$ will again be omitted if $O=1$. Of course
$I^{(0)}_{O} = \rho^{(0)}_{O\,c}=b_O$.

The spectral correlators $\rho_{c}^{(k)}$ measure the correlations
between the number of modes in infinitesimal spectral intervals, and
are expected to have a well-defined thermodynamic limit in the light
of the typical behavior of random-matrix
systems~\cite{mehta2004random,Pastur2011},
\begin{equation}
  \label{eq:rho_def5_0_th}
  \rho_{c\,{\scriptscriptstyle\infty}}^{(k)}(\lambda_1,\ldots,\lambda_k;m)
  \equiv \lim_{\lvol\to\infty}\rho_{c}^{(k)}(\lambda_1,\ldots,\lambda_k;m)
  \,.
\end{equation}
For $k=1$ one obtains the spectral density in infinite volume (see
Ref.~\cite{van_Hemmen_1982}),
\begin{equation}
  \label{eq:rho_def3_th}
  \rho(\lambda;m) \equiv \rho^{(1)}_{c\,\scriptscriptstyle\infty}(\lambda;m)=
  \lim_{\lvol\to\infty}\rho^{(1)}_{c}(\lambda;m)\,,
\end{equation}
and for $k=2$ the connected two-point function,
\begin{equation}
  \label{eq:rho_def7}
  \begin{aligned}
    & \rho_{c\,{\scriptscriptstyle\infty}}^{(2)}(\lambda_1,\lambda_2;m)    \\
    &=\left[\lim_{\lvol\to\infty}
      \f{1}{\lvol}\left(
      \la
      \rho_U(\lambda_1)\rho_U(\lambda_2)\ra -
      \la\rho_U(\lambda_1)\ra \la\rho_U(\lambda_2)\ra
      \right)\right]\\
    &\phantom{=}  -
      \left[\delta(\lambda_1-\lambda_2)+
      \delta(\lambda_1+\lambda_2)\right]\rho(\lambda_1;m)\,.
  \end{aligned}
\end{equation}
These quantities are expected to have at most integrable
singularities.\footnote{\label{foot:rho}The limit
  Eq.~\eqref{eq:rho_def5_0_th} should be understood in the
  distributional sense. For the spectral density,
  Eq.~\eqref{eq:rho_def3_th}, one starts from the normalized mode
  number in a spectral interval $[\lambda_0,\lambda]$,
  $$
  \mdens(\lambda_0,\lambda;m) \equiv \lim_{\lvol\to\infty}
  \f{1}{\lvol} \left\la\int_{\lambda_0}^\lambda d\lambda'\,
    \rho_U(\lambda') \right\ra \,,
  $$
  and obtains the spectral density as the function (or more generally
  distribution) obeying
  $\rho(\lambda;m) = \de_\lambda \mdens(\lambda_0,\lambda;m)$. By
  definition, integrating $\rho(\lambda;m)$ one gets back the mode
  density, so $\rho(\lambda;m)$ is integrable. Similar constructions
  are used to define precisely the quantities in
  Eq.~\eqref{eq:rho_def5_0_th}.} One
similarly writes 
\begin{equation}
  \label{eq:rho_def8_gen_th}
  \rho_{O\,c\,{\scriptscriptstyle\infty}}^{(k)}(\lambda_1,\ldots,\lambda_k;m)\equiv
  \lim_{\lvol\to\infty}  \rho_{O\,c}^{(k)}(\lambda_1,\ldots,\lambda_k;m)
\,.
\end{equation}
These quantities may generally have non-integrable singularities, and
the thermodynamic limit of the integrals
Eq.~\eqref{eq:cum_sys_3_0_again},
\begin{equation}
  \label{eq:cum_sys_3_0_again_iv}
  I^{(k)}_{O\,{\scriptscriptstyle\infty}}[g_1,\ldots, g_k] \equiv
  \lim_{\lvol\to\infty}  I^{(k)}_{O}[g_1,\ldots, g_k]\,,
\end{equation}
may be divergent, depending on the choice of the functions
$g_i$.\footnote{More precisely, one obtains
  $ \rho_{O\,c\,{\scriptscriptstyle\infty}}^{(k)}$ from
  $I^{(k)}_{O\,{\scriptscriptstyle\infty}}$ by taking for $g_i$ the
  indicator functions of the spectral intervals
  $[\lambda_{0i},\lambda_i]$, and then taking derivatives (in the
  distributional sense) with respect to $\lambda_i$ (see
  footnote~\ref{foot:rho}), as long as
  $I^{(k)}_{O\,{\scriptscriptstyle\infty}}$ is finite for the given
  $\lambda_{0i}$ and for $\lambda_i$ in some small range.  A divergent
  $I^{(k)}_{O\,{\scriptscriptstyle\infty}}$ signals instead the
  presence of a non-integrable singularity of
  $\rho_{O\,c\,{\scriptscriptstyle\infty}}^{(k)}$ in at least one of
  the spectral intervals.} Similarly, the thermodynamic limit of the
cumulants $b_O$ may be divergent. A finite thermodynamic limit is
certainly expected for
\begin{equation}
  \label{eq:bdef}
  \begin{aligned}
    n_0 &\equiv \lim_{\lvol\to\infty} b_{N_0}= \lim_{\lvol\to\infty}\f{\la N_0\ra}{\lvol}\,,\\
    \chit &\equiv \lim_{\lvol\to\infty} b_{Q^2} = \lim_{\lvol\to\infty}\f{\la Q^2\ra}{\lvol}\,,
  \end{aligned}
\end{equation}
which are the zero-mode density and the topological susceptibility,
respectively (and I made again use of $CP$ invariance in the last
step). More generally, the existence of a thermodynamic limit is
expected for the cumulants of the topological charge, $b_{Q^{k}}$, as
$Q$ is the sum of local quantities, $Q=\sum_x q(x)$ with
$q(x)=-\tr \{ (DR)_{xx}\gamma_5\}$, with the trace running over Dirac
and color indices
only~\cite{Hasenfratz:1998ri,Luscher:2004fu,Vicari:2008jw}; for more
general $b_{N_0^{k_0}Q^{k_1}}$ this should be checked on a
case-by-case basis.

It is argued, and well supported by numerical results, that
$ N_+(U) N_-(U)= 0$ almost everywhere (a.e.)\ in the space of gauge
configurations. The argument is that a non-minimal realization of the
index theorem, i.e., $Q(U) = N_+(U)-N_-(U)$ with
$N_+(U) N_-(U)\neq 0$, requires that the gauge configuration be finely
tuned to provide more zero modes than strictly required, and the set
of such gauge configurations is of zero measure~\cite{Blum:2001qg}.
This implies in particular that $ \la N_0^2\ra=\la Q^2\ra$, and so
$n_0=0$ in the $CP$-invariant case (using
$\la N_0 \ra^2\le \la Q^2\ra$ and finiteness of $\chit$). Moreover,
since in this case $N_0=|Q|$ a.e., one expects the cumulants
$b_{N_0^k}$ to have a well-defined thermodynamic limit. By the same
argument, one expects $ N_+'(U) N_-'(U)= 0$ a.e., and so $N_2=|Q|=N_0$
a.e., and the density of doubler modes to vanish,
$n_2 \equiv \lim_{\lvol\to \infty}\f{\la N_2\ra}{\lvol}=0$, in the
$CP$-invariant case.

\section{Chiral symmetry restoration}
\label{sec:chisymrest}

The fundamental requirement for symmetry restoration in a local
quantum field theory is that correlation functions of local operators
that are related by a symmetry transformation become equal in the
symmetric limit, i.e., when all the terms breaking the symmetry
explicitly are removed from the theory. In the case at hand, this is
the chiral limit in which the common mass $m$ of the two flavors of
light fermions is sent to zero, and chiral symmetry becomes exact at
the classical level.

To obtain information on the Dirac spectrum, however, one would rather
work with susceptibilities, specifically with the scalar and
pseudoscalar susceptibilities in Eq.~\eqref{eq:genfuncdef2_ss}, that
can be expressed solely in terms of the Dirac eigenvalues (see
Sec.~\ref{sec:det_sources} below). Restoration of chiral symmetry at
the level of the local correlators implies restoration at the level of
the susceptibilities (i.e., that susceptibilities related by a chiral
transformation become equal as $m\to 0$), if the zero-momentum limit
corresponding to integrating over the whole spacetime volume commutes
with the chiral limit. This is the case if the correlation length of
the system remains finite in the chiral limit, as one generally
expects in the symmetric phase due to the expected absence of massless
excitations, resulting in finite susceptibilities that are manifestly
symmetric. A notable exception is the critical point of a continuous
transition, where the correlation length diverges, and so do one or
more of the susceptibilities. Symmetry at the level of the
susceptibilities is not guaranteed in this case, although it may be
possible in principle that the difference of symmetry-related
susceptibilities still vanishes in the chiral limit, even if they
separately diverge. Another notable exception are free continuum
fermions at $\mathrm{T}=0$, for which chiral symmetry is restored in
the chiral limit but (most of the) susceptibilities diverge due to the
divergent correlation length, and the difference of symmetry-related
susceptibilities does not always vanish in the chiral
limit.\footnote{\label{foot:contfree}The continuum massless fermion
  propagator decays only algebraically at large distances,
  $\slashed{\de}^{-1}\propto \slashed{x}/|x|^4$.  For susceptibilities
  involving $b$ bilinears, integrating in the range
  $1/\Lambda \le |x| \le \mathrm{L}$, this leads by power counting to
  a dependence $\Lambda^{4-b}$ on the UV cutoff, and
  $\mathrm{L}^{b-4}$ on the IR cutoff, and so to divergent
  susceptibilities for $b\ge 4$ as $\mathrm{L}\to\infty$ (at fixed
  $\Lambda$). In particular, $\chi(S^5)$ diverges as $m\to 0$ while
  $\chi((iP_a)^5)=0$ identically, so that chiral symmetry is not
  realized at the level of susceptibilities.} Conversely, it may be
mathematically possible that restoration of chiral symmetry at the
level of susceptibilities does not reflect the actual restoration of
the symmetry at the fundamental level of local correlators, and comes
about only due to cancelations in the spacetime integrals of these
correlators. This possibility, however, seems physically very
unlikely, and will be ignored in the following.\footnote{As will
  become clear below, in this case the consequences of the present
  analysis would still apply, even though chiral symmetry would not be
  restored in the sense of local quantum field theory.}

In this work chiral symmetry restoration will be understood to mean
restoration at the level of the susceptibilities, i.e., that for any
chiral transformation $\U$ [see Eq.~\eqref{eq:genfuncdef2_s} and under
Eqs.~\eqref{eq:O_VW_transf} and \eqref{eq:largeV_cu4} for notation]
\begin{equation}
  \label{eq:susc_condition}
  \begin{aligned}
    \lim_{m\to 0}
    &\left[\chi\left(S^{n_S}  (i\vec{P})^{\vec{n}_P}
      (iP)^{n_P}  \vec{S}^{\vec{n}_S}\right)  \right. \\
    & \left. - \chi\left(S_{\U}^{n_S}  (i\vec{P}_{\U})^{\vec{n}_P}
      (iP_{\U})^{n_P}  \vec{S}_{\U}^{\vec{n}_S}\right)\right] =0\,. 
  \end{aligned}
\end{equation}
As formulated, the request of symmetry restoration,
Eq.~\eqref{eq:susc_condition}, allows in principle for divergent
susceptibilities in the chiral limit, as long as the divergences of
susceptibilities related by a chiral transformation cancel out.  On
the other hand, while symmetric (and finite) local correlators and a
finite correlation length in the chiral limit imply both finiteness
and symmetry of the susceptibilities, finiteness of the
susceptibilities in the chiral limit alone does not guarantee
\textit{a priori} that they will also be symmetric. (As already
mentioned in Sec.~\ref{sec:intro}, in this context a finite quantity
is a quantity that is non-divergent -- including vanishing -- in the
chiral limit.)  In principle, then, in the chiral limit
susceptibilities could diverge yet be symmetric; or remain finite yet
not be symmetric. I show in Sec.~\ref{sec:chisymrest_nsc} that in the
scalar and pseudoscalar sector this is not possible: chiral symmetry
is restored at the level of scalar and pseudoscalar susceptibilities
if and only if these susceptibilities are finite. Finiteness of the
susceptibilities then fully characterizes symmetry restoration in the
scalar and pseudoscalar sector except when the correlation length
diverges (and barring ``accidental'' restoration at the level of the
susceptibilities but not at the level of local correlators). In
particular, this characterization should apply within a
finite-temperature symmetric phase.

As I will also show in Sec.~\ref{sec:chisymrest_nsc}, finiteness of
the susceptibilities has the corollary that the ``even''
(respectively, ``odd'') susceptibilities, i.e., those involving an
even (respectively, odd) number of the isosinglet bilinears $S$ and
$P$, must be $C^\infty$ (i.e., infinitely differentiable) functions of
$m^2$ at $m=0$ (respectively $m$ times a $C^\infty$ function), a
property that I will refer to as ``$m^2$-differentiability'' for
short. This is an equivalent characterization of symmetry restoration
at the level of susceptibilities.

Thanks to the transformation properties of the relevant bilinears
under chiral transformations, Eq.~\eqref{eq:O_VW_transf}, the symmetry
restoration condition Eq.~\eqref{eq:susc_condition} can be expressed
in compact form in terms of the generating function $\wiv$ [see
Eqs.~\eqref{eq:genfuncdef2_ss} and \eqref{eq:sources2} and the
following discussion in Sec.~\ref{sec:genfunc}],
\begin{equation}
  \label{eq:symrest1}
  \begin{aligned}
    \lim_{m\to 0}
    &\left[\wiv(\sous,\soup;m)\right. \\
    & \left.  - \wiv(R\sous,R\soup;m)\right] = 0  \,, ~
      \forall R\in \mathrm{SO}(4)\,.
  \end{aligned}
\end{equation}
This is the starting point of the analysis carried out in this work.

In a symmetric phase with a finite correlation length, the symmetry
restoration condition Eq.~\eqref{eq:susc_condition} naturally extends
to susceptibilities involving the spacetime integrals
$G_i=\sum_x G_i(x)$ of local operators $G_i(x)$ built out entirely of
gauge fields. Within the symmetric phase one has then
\begin{equation}
  \label{eq:susc_condition_g2}
  \begin{aligned}
    \lim_{m\to 0}
    & \left[ \chi\left(S^{n_S}  (i\vec{P})^{\vec{n}_P}
      (iP)^{n_P}  \vec{S}^{\vec{n}_S}{\textstyle\prod_i} G_i\right) \right.\\
    &  \left. - \chi\left(S_{\U}^{n_S}  (i\vec{P}_{\U})^{\vec{n}_P}
      (iP_{\U})^{n_P}  \vec{S}_{\U}^{\vec{n}_S} {\textstyle\prod_i} G_i\right)  \right] =0\,.
  \end{aligned}
\end{equation}
As shown in Sec.~\ref{sec:gauge_ext}, also in this case
Eq.~\eqref{eq:susc_condition_g2} is satisfied if and only if all these
susceptibilities are finite in the chiral limit, or equivalently if
they are $m^2$-differentiable or $m$ times an $m^2$-differentiable
function depending on whether they are even or odd. A relevant example
are susceptibilities involving the topological charge, since $Q$
admits a representation as the integral of a local density [see under
Eq.~\eqref{eq:bdef}].  From this result follows in particular the
$m^2$-differentiability of its cumulants.
Equation~\eqref{eq:susc_condition_g2} can be reformulated in a manner
similar to Eq.~\eqref{eq:symrest1} by defining the augmented
generating functions $\mathcal{Z}_G$, $\mathcal{W}_G$, and $\wivG$
[see Eq.~\eqref{eq:partfunc_G}], adding sources $J_{G_i}$ for the
gauge operators $G_i$ and replacing
$\des \to \des - \sum_i J_{G_i} G_i$ in the definition of
$\mathcal{Z}$ and $\mathcal{W}$ [see Eqs.~\eqref{eq:sources} and
\eqref{eq:partfunc}]. Since $G_i$ are unaffected by chiral
transformations, the symmetry restoration condition reads
\begin{equation}
  \label{eq:symrest1_G}
  \begin{aligned}
    \lim_{m\to 0}
    & \left[\wivG(\sous,\soup;J_G;m)\right. \\
    & \left.- \wivG(R\sous,R\soup;J_G;m)\right] = 0    \,,  ~
      \forall R\in \mathrm{SO}(4)\,,
  \end{aligned}
\end{equation}
where $J_G$ denotes collectively the gauge-operator sources.

The fact that gauge fields are unaffected by chiral transformations
suggests that one could reasonably expect that symmetry restoration is
manifest also in susceptibilities involving scalar and pseudoscalar
bilinears and general nonlocal operators built out of gauge fields
only.  In other words, Eq.~\eqref{eq:susc_condition_g2} is expected to
hold also if one includes (translation-invariant) nonlocal operators
in the set $\{G_i\}$, leading to the same conclusions as above
concerning finiteness and $m^2$-differentiability of the
susceptibilities. Of course, if one adopts a strictly local point of
view, then whether or not symmetry is manifest in this kind of
susceptibilities has no bearing on its being physically realized.  On
the other hand, nothing prevents one to use also these nonlocal
functionals for a more detailed characterization of the phases of the
theory. I will then treat this as an additional assumption, logically
quite independent from the (essentially) local symmetry-restoration
assumptions discussed above, and refer to it as ``nonlocal
restoration'' when invoked.

For the purposes of this work, the main consequence of nonlocal
restoration is the resulting $m^2$-differentiability of spectral
observables, such as the spectral density or the two-point function of
Dirac eigenvalues, that are precisely of the relevant type --
susceptibilities of operators that are (highly) nonlocal but involve
only the gauge fields [see Eqs.~\eqref{eq:rho_def4} and
\eqref{eq:rho_def5_0}]. However, while sensible, the additional
assumption of nonlocal restoration does not follow directly from the
basic request of symmetry restoration for local correlators (and from
the finiteness of the correlation length). For the spectral quantities
of interest a perhaps more palatable argument can be obtained by
making use of partially quenched theories.  It is reasonable to expect
that if one probes the system with external fields, coupled in such a
way as not to break chiral symmetry explicitly, then in the symmetric
phase chiral symmetry will remain manifest in local correlation
functions involving both dynamical and external fields, as well as in
the corresponding susceptibilities if the correlation length of the
system is finite.  In a partially quenched setup where both fermion
and pseudofermion fields of the same mass $M$ are added to the theory,
canceling exactly each other's contribution to the partition
function, one expects then that scalar and pseudoscalar
susceptibilities involving both bilinears built out of the dynamical
fermion fields, Eq.~\eqref{eq:densities}, and their counterparts built
out of the external fermion fields, will still display exact chiral
symmetry in the chiral limit. This leads again to the same conclusions
about $m^2$-differentiability of (all) susceptibilities as in the
original theory.  As this should hold for arbitrary complex mass $M$,
including when this approaches purely imaginary values where
discontinuities appear; and since the spectral density, the two-point
eigenvalue correlation function, and similar quantities can be
obtained from these discontinuities (with no insertion of bilinears
built with dynamical fields), it would then follow that they are
$m^2$-differentiable. While this approach still requires an additional
assumption, it has the advantage of involving only susceptibilities of
local operators in its formulation. This is discussed in
Appendix~\ref{sec:rhochirest} for $\gamma_5$-Hermitean GW operators
with $2R=\mathbf{1}$.

\subsection{Functional form of the generating function} 
\label{sec:chisymrest_W}

It is obvious from their definition, Eq.~\eqref{eq:partfunc}, that
$\mathcal{Z}$ and $\mathcal{W}$ depend on the scalar isosinglet source
$j_S$ and on the fermion mass $m$ only through the combination
$j_S+m$,
\begin{equation}
  \label{eq:mshift1}
  \mathcal{W}(\sous,\soup;m) =  \mathcal{W}(\tilde{\sous}(m),\soup;0) \,,\quad
  \tilde{\sous}(m)\equiv
 \sous + me_0 \,,
\end{equation}
where $e_0\equiv(1,\vec{0}\,)^T$, i.e., the generating functions at
nonzero $m$ equal those at $m=0$ but with a shifted, mass-dependent
source. A simple consequence of Eq.~\eqref{eq:mshift1} is the relation
\begin{equation}
  \label{eq:mdep3}
  \begin{aligned}
    \de_{j_S} \mathcal{W}(\sous,\soup;m)
    =  \de_m \mathcal{W}(\sous,\soup;m)\,,
  \end{aligned}
\end{equation}
that implies the well-known fact that the mass derivatives of a
susceptibility equal other susceptibilities involving additional
scalar densities. Indeed, taking repeated derivatives at zero sources and
taking the thermodynamic limit one finds
\begin{equation}
  \label{eq:mdep4}
  \begin{aligned}
    &    \de_m^k    \chi\left(S^{n_S}  (i\vec{P})^{\vec{n}_P} (iP)^{n_P}  \vec{S}^{\,\vec{n}_S}\right) \\ 
    &= (-1)^k\chi\left(S^{n_S+k}  (i\vec{P})^{\vec{n}_P} (iP)^{n_P}  \vec{S}^{\,\vec{n}_S}\right)\,. 
  \end{aligned}
\end{equation}
Equation~\eqref{eq:mshift1} combined with the full chiral symmetry of
the massless theory strongly restricts the functional form of
$\mathcal{W}$, and so of $\wiv$. In fact, since the integration
measure in Eq.~\eqref{eq:partfunc} is invariant under non-singlet
chiral transformations, the exactly massless theory in a finite volume
is invariant under $\mathrm{SU}(2)_L\times \mathrm{SU}(2)_R$
transformations. For arbitrary sources $\sous$ and $\soup$ one has
then $\mathcal{W}(R\sous,R\soup;0)=\mathcal{W}(\sous,\soup;0)$,
$\forall R\in\mathrm{SO}(4)$, and so
\begin{equation}
  \label{eq:chiW0_2_fv}
  \mathcal{W}(\sous,\soup;0)
  \equiv \hat{\mathcal{W}}(\sous^2,\soup^2,2\sous\cdot \soup)\,, 
\end{equation}
i.e., it depends only on $\mathrm{SO}(4)$ invariants. The factor of 2
in the third argument is purely conventional. By
Eq.~\eqref{eq:mshift1}, this implies
\begin{equation}
  \label{eq:chiW0_2_fv_bis}
  \begin{aligned}
    \mathcal{W}(\sous,\soup;m)
    &= \mathcal{W}(\tilde{\sous}(m),\soup;0)\\
    &   =  \hat{\mathcal{W}}\left(m^2 + u(V;m),w(W),\tilde{u}(V,W;m)\right)
      \,,
  \end{aligned}
\end{equation}
where
\begin{equation}
  \label{eq:chiW0_4}
  \begin{aligned}
    u(V;m) &\equiv \tilde{\sous}(m)^2-m^2 = 2mj_S + \sous^2\,,\\
    w(W) &\equiv\soup^2\,,\\
    \tilde{u}(V,W;m)&  \equiv 2\tilde{\sous}(m)\cdot W = 2(mj_P +\sous\cdot \soup)\,. 
  \end{aligned}
\end{equation}
In the following, the source and mass dependences of these quantities
will be mostly dropped for simplicity. Equations~\eqref{eq:mshift1},
\eqref{eq:mdep3}, and \eqref{eq:chiW0_2_fv_bis} remain valid also in
the thermodynamic limit, where they read
\begin{eqnarray}
  \label{eq:mshift1_iv}
  \wiv(\sous,\soup;m) &=& \wiv(\tilde{\sous}(m),\soup;0)\,,  \\
  \label{eq:mdep3_iv}
  \de_{j_S} \wiv(\sous,\soup;m) &=&  \de_m \wiv(\sous,\soup;m)\,,
\end{eqnarray}
and
\begin{equation}
  \label{eq:chiW0_3}
  \begin{aligned}
    &  \wiv(\sous,\soup;m)\\
    &  = \whiv\left(m^2 + u(V;m),w(W),\tilde{u}(V,W;m)\right) \\
    &  \equiv  \lim_{\lvol\to\infty}
      \hat{\mathcal{W}}\left(m^2 + u(V;m),w(W),\tilde{u}(V,W;m)\right)\,.
  \end{aligned}
\end{equation}
These formal relations summarize exact relations between
susceptibilities, such as Eq.~\eqref{eq:mdep4}, or the Ward-Takahashi
identities (see below). The fact that the thermodynamic limit of the
generating function at $m=0$ appears on the right-hand side of
Eq.~\eqref{eq:mshift1_iv} does not imply that the thermodynamic and
chiral limits can be exchanged: since $\tilde{\sous}$ is
$m$-dependent, the right-hand side still represents the same, massive
theory as the left-hand side.  Equation~\eqref{eq:chiW0_3} then holds
independently of the possibility of exchanging limits. Note also that
while one can in practice treat $\wiv(\sous,\soup;m)$, and so
$\whiv(m^2+u,w,\tu;m)$, as polynomials (of arbitrary order) in the
sources, Eq.~\eqref{eq:mshift1_iv} does not imply that we can treat
$\wiv(\tilde{\sous},\soup;0)$ as a polynomial in the shifted source
$\tilde{\sous}$, and so $\whiv(m^2+u,w,\tu;m)$ as a polynomial of its
arguments.  Finally, notice that in the presence of $CP$ symmetry,
Eq.~\eqref{eq:sources4} implies that $\hat{\mathcal{W}}$ and so
$\whiv$ can depend only on $\tilde{u}^2$.

The relations Eqs.~\eqref{eq:chiW0_2_fv_bis} and \eqref{eq:chiW0_3}
are exact, and hold true independently of the fate of chiral symmetry
in the chiral limit, as they follow only from the symmetries of the
exactly massless theory in a finite volume. The functional forms of
$\hat{\mathcal{W}}$ and $\whiv$ imply that only powers of the three
combinations $u$, $w$, and $\tilde{u}$ will appear in the expansion of
$\mathcal{W}$ and $\wiv$ in the sources,\footnote{Since $u$ and
  $\tilde{u}$ are not homogeneous in the sources, if one truncates the
  expansion of $\mathcal{W}$ or $\wiv$ to a fixed order in $V$ and $W$
  the terms of highest order will violate this property. However,
  since one can take the order of the expansion to be arbitrarily
  high, this causes no problem in practice.} and so relations among
susceptibilities will follow. Focussing on the thermodynamic limit and
expanding $\whiv$ in powers of $u$, $w$, $\tilde{u}$, one finds
\begin{equation}
  \label{eq:expansion1}
    \wiv(\sous,\soup;m)
    = \sum_{n_{u},n_{w},n_{\tilde{u}}=0}^\infty
      \f{
      u^{n_{u}}w^{n_{w}} \tilde{u}^{n_{\tilde{u}}} }{n_{u}!n_{w}!
      n_{\tilde{u}}!} \susc_{n_{u}n_{w} n_{\tilde{u}}}(m^2)\,,
\end{equation}
where
\begin{equation}
    \label{eq:expansion1_A}
    \susc_{n_{u}n_{w} n_{\tilde{u}}}(m^2) \equiv
    \de_u^{n_u}\de_w^{n_w}\de_{\tilde{u}}^{n_{\tilde{u}}}  \whiv(m^2 + u,w,\tilde{u})|_0\,; 
\end{equation}
for brevity I will often write
$\susc_n=\susc_{n_{u}n_{w} n_{\tilde{u}}}$ and
$ \sum_{n=0}^\infty = \sum_{n_u,n_w,n_{\tilde{u}}=0}^\infty$. The
manifest dependence on $m^2$ is only formal at this stage. Scalar and
pseudoscalar susceptibilities are then finite linear combinations of
the coefficients $\susc_n$. It is clear from Eq.~\eqref{eq:expansion1}
that all the $\susc_{n}$ can be obtained from the generating function
at vanishing isosinglet sources, $j_S=j_P=0$,
\begin{equation}
  \label{eq:expansion2}
  \begin{aligned}
    &    \wiv(\sous,\soup;m)|_{j_{S,P=0}}= \whiv(m^2 +
      \vec{\jmath}_P^{\,2},\vec{\jmath}_S^{\,2},2\vec{\jmath}_P\cdot
      \vec{\jmath}_S)\\
    &= \sum_{n=0}^\infty\f{
      (\vec{\jmath}_P^{\,2})^{n_u}(\vec{\jmath}_S^{\,2})^{n_w}(2\vec{\jmath}_P\cdot
      \vec{\jmath}_S)^{n_{\tilde{u}}} }{n_u!n_w!n_{\tilde{u}}!}
      \susc_{n}(m^2)\,,  
  \end{aligned}
\end{equation}
since $\vec{\jmath}_P^{\,2}$, $\vec{\jmath}_S^{\,2}$, and
$\vec{\jmath}_P\cdot \vec{\jmath}_S$ are independent variables.  Using
this, one can show that the $\susc_{n}$ are equivalent to a subset of
susceptibilities, involving only the bilinears $\vec{P}$ and
$\vec{S}$, from which they are obtained as finite linear combinations
with $m$-independent coefficients (see Appendix~\ref{sec:ansusc}). The
properties of the coefficients $\susc_{n}$ are then easily translated
to those of the scalar and pseudoscalar susceptibilities and vice
versa. Moreover, since $\whiv$ at $j_{S,P}=0$ depends only on
$m^2+\vec{\jmath}_P^{\,2}$, from Eq.~\eqref{eq:expansion2} one proves
that
\begin{equation}
  \label{eq:nsc_chi_new4}
  \begin{aligned}
    & \de_{m^2} \whiv(m^2 +
      \vec{\jmath}_P^{\,2},\vec{\jmath}_S^{\,2},2\vec{\jmath}_P\cdot
      \vec{\jmath}_S)\\
    &=\de_{\vec{\jmath}_P^{\,2}} \whiv(m^2 +
      \vec{\jmath}_P^{\,2},\vec{\jmath}_S^{\,2},2\vec{\jmath}_P\cdot
      \vec{\jmath}_S)\,,
  \end{aligned}
\end{equation}
which implies 
\begin{equation}
  \label{eq:nsc_chi_new6}
  \de_{m^2} \susc_{n_u n_w n_{\tilde{u}}}(m^2) = \susc_{n_u+1\, n_w n_{\tilde{u}}}(m^2)\,.
\end{equation}
All these results hold also for the finite-volume generating function,
$\mathcal{W}$, and the coefficients of its expansion in powers of $u$,
$w$, $\tilde{u}$, of which $\susc_n$ represent the thermodynamic
limit. (More precisely, the properties above are first shown to hold
in a finite volume simply as a consequence of the functional form of
$\mathcal{W}$, and hold for $\wiv$ since one can commute the
thermodynamic limit with source and mass derivatives.)

The construction leading to
Eqs.~\eqref{eq:chiW0_2_fv_bis}--\eqref{eq:chiW0_3} admits a geometric
interpretation. The resulting functional forms reflect the invariance
of $\mathcal{Z}$, $\mathcal{W}$, and $\wiv$ under the affine
transformation
\begin{equation}
  \label{eq:chiW0_7_0}
  \sous \to R\sous + m(R-\mathbf{1}_4)e_0\,,
  \quad
  \soup \to R\soup\,, 
\end{equation}
where $R\in\mathrm{SO}(4)$ and $\mathbf{1}_4$ is the four-dimensional
identity matrix. This is the leftover at nonzero $m$ of the chiral
symmetry of the exactly massless theory. For infinitesimal chiral
transformations, this invariance implies the well-known integrated
Ward-Takahashi identities. This is shown in
Appendix~\ref{sec:WTI_geo}.

\subsection{Necessary and sufficient conditions for chiral symmetry
  restoration}
\label{sec:chisymrest_nsc}

The behavior in the chiral limit of the coefficients $\susc_n$,
Eq.~\eqref{eq:expansion1_A}, allows one to fully characterize the
restoration of chiral symmetry in the scalar and pseudoscalar sector
[in the sense of Eq.~\eqref{eq:susc_condition}, or equivalently
Eq.~\eqref{eq:symrest1}].  In fact, a necessary and sufficient
condition for chiral symmetry restoration in this sector is that all
$\susc_n(m^2)$, including the free energy density
$-\susc_{000}(m^2) = - \wiv(0,0;m)$, have a finite chiral
limit~\cite{Giordano:2024jnc}, i.e., chiral symmetry is restored if
and only if
\begin{equation}
  \label{eq:def_suff}
  \susc_{n*}\equiv\lim_{m\to 0}\susc_n(m^2) 
\end{equation}
exist and are finite, $\forall n$. Here I provide a simpler and
shorter proof than that of Ref.~\cite{Giordano:2024jnc}, also filling
in some details omitted there (see also Ref.~\cite{Giordano:2024awb}).

Sufficiency is an obvious consequence of the functional form
Eq.~\eqref{eq:expansion1}: if $\susc_{n*}$ exist and are finite, then
\begin{equation}
  \label{eq:proof_suff}
    \begin{aligned}
      &\lim_{m\to 0} \wiv(\sous,\soup;m)\\
      & = \sum_{n=0}^\infty
        \f{\left(V^2\right)^{n_{u}}\left(W^2\right)^{n_{w}} \left(2V\cdot W\right)^{n_{\tilde{u}}}
        }{n_{u}!n_{w}!n_{\tilde{u}}!}\susc_{n *}\,,
  \end{aligned}
\end{equation}
which is manifestly symmetric. Since the available $\mathrm{SO}(4)$
invariants, i.e., $V^2$, $W^2$, and $2 V\cdot W$, are also
$\mathrm{O}(4)$ invariants, $\wiv$ is actually
$\mathrm{O}(4)$-symmetric in the symmetric phase.

To prove necessity, the first step is to notice that the symmetry
restoration condition Eq.~\eqref{eq:symrest1} implies that in the
chiral limit the generating function $\wiv$ depends only on
$\mathrm{SO}(4)$-invariant combinations of the sources. This would be
trivial to show if one assumed the existence and finiteness of the
chiral limit of $\wiv$, but this is precisely what one wants to prove
here. The proof of the statement above (without assuming existence and
finiteness of the chiral limit) is given in
Appendix~\ref{sec:invchilim}. In the chiral limit $\wiv$, or
equivalently $\whiv$, can then depend on the sources only through
$V^2$, $W^2$, and $2 V\cdot W$, although at this stage it is not
guaranteed that the coefficients of an expansion in powers of these
invariants have a finite limit as $m\to 0$. Nonetheless, one finds
that
\begin{equation}
  \label{eq:symrest_jsp}
  \begin{aligned}
    &  \lim_{m\to 0} \left[\de_{j_S} - 2\left(j_S \de_{V^2} + j_P \de_{2V\cdot W}\right)\right] \\
    &\times \whiv\left(m^2 + 2mj_S + \sous^2,\soup^2,2(mj_P +\sous\cdot \soup)\right)=0\,, 
  \end{aligned}
\end{equation}
and so 
\begin{equation}
  \label{eq:jsp_indep_cons}
    \lim_{m\to 0} m \de_u    \whiv(m^2+u,w,\tilde{u})  = 0\,. 
\end{equation}
The second step is to use this result to characterize the behavior of
the mass derivative of $\whiv$ in the chiral limit. Using
Eqs.~\eqref{eq:mdep3} and \eqref{eq:jsp_indep_cons} one finds
\begin{equation}
  \label{eq:nsc_chi_new2}
  \begin{aligned}
    &    \lim_{m\to 0} \de_m  \wiv(\sous,\soup;m) =
      \lim_{m\to 0} \de_{j_S} \wiv(\sous,\soup;m)\\
    & = 2\lim_{m\to 0} \left(j_S\de_u +
      j_P\de_{\tilde{u}}\right)\whiv(m^2 + u,w,\tilde{u})\,,
  \end{aligned}
\end{equation}
and so for $j_{S,P}=0$
\begin{equation}
  \label{eq:nsc_chi_new3}
  \begin{aligned}
    &  \lim_{m\to 0} \de_m \wiv(\sous,\soup;m)|_{j_{S,P}=0}
      =0\,,
  \end{aligned}
\end{equation}
that readily implies\footnote{More directly,
  Eq.~\eqref{eq:jsp_indep_cons} implies
  $ \lim_{m\to 0} 2m \susc_{n_u+1,n_w,n_{\tu}}(m^2) =0$, as one sees
  by computing the derivative, setting $j_{S,P}=0$, and using the
  independence of $\vec{\jmath}_P^{\,2}$, $\vec{\jmath}_S^{\,2}$, and
  $\vec{\jmath}_P\cdot \vec{\jmath}_S$.  By
  Eq.~\eqref{eq:nsc_chi_new6} this implies
  $\lim_{m\to 0} 2m \de_{m^2}\susc_n (m^2)=\lim_{m\to 0}\de_{m}\susc_n
  (m^2)= 0$.}
\begin{equation}
  \label{eq:nsc_chi_new3_bis}
  \lim_{m\to 0}\de_m \susc_n(m^2) =0\,,
\end{equation}
for every $n=(n_u,n_w,n_{\tilde{u}})$. Then $\susc_{n}$ in the chiral limit,
\begin{equation}
  \label{eq:nsc_chi7}
    \susc_{n*}
   = \susc_{n}(m_0^2) + \lim_{m\to
      0}\int_{m_0}^m d\bar{m} \, \de_{\bar{m}} \susc_{n}(\bar{m}^2)\,,
\end{equation}
exists finite $\forall n$, for an arbitrary choice of the integration
limit $m_0$, since the integrand is regular as $\bar{m}\to 0$ (indeed,
it vanishes there) and so the integral in Eq.~\eqref{eq:nsc_chi7} has
a finite limit as $m\to 0$.\footnote{The integrand in
  Eq.~\eqref{eq:nsc_chi7} is a regular function of $m$ for $m\neq 0$,
  since $\susc_{n}(m^2)$ are linear combinations of physical
  susceptibilities (see Appendix~\ref{sec:ansusc}), and therefore so
  are $\de_m \susc_{n} (m^2)=2m\susc_{n_u+1\, n_w n_{\tu}}(m^2)$ [see
  Eq.~\eqref{eq:nsc_chi_new6}].} This completes the proof.

The equivalence of $\susc_n$ and scalar and pseudoscalar
susceptibilities (see Appendix~\ref{sec:ansusc}) allows one to
characterize chiral symmetry restoration directly in terms of the
latter: chiral symmetry is restored if and only if all scalar and
pseudoscalar susceptibilities are finite in the chiral
limit.\footnote{More precisely, since the $\susc_n$ are equivalent to
  a subset of susceptibilities (see Appendix~\ref{sec:ansusc}), chiral
  symmetry is restored if and only if these are finite. As all the
  other susceptibilities are obtained from $\susc_n$, finiteness of
  all susceptibilities follows from that of those in the subset.}
Barring the unlikely case of symmetric susceptibilities in the absence
of chiral symmetry restoration, spontaneous chiral symmetry breaking
requires then one or more susceptibilities to diverge in the chiral
limit, in agreement with the appearance of massless excitations and of
a divergent correlation length implied by Goldstone's theorem both at
zero~\cite{Goldstone:1962es} and nonzero
temperature~\cite{Lange:1965zz,Kastler:1966wdu,Swieca:1966wna,
  Morchio:1984ph,Morchio:1987wd} (see also
Ref.~\cite{Giordano:2022ghy} for the specific case of chiral
symmetry).

It follows from the finiteness of all the $ \susc_{n *} $ and from
Eq.~\eqref{eq:nsc_chi_new6} that
\begin{equation}
  \label{eq:nsc_chi_new3_quater}
    \lim_{m\to 0}\de_{m^2} \susc_{n_u n_w n_{\tilde{u}}}(m^2)
    =  \susc_{n_u+1\,
      n_w n_{\tilde{u}}\,*}
\end{equation}
is finite, i.e., $\susc_n$ are $m^2$-differentiable. This implies that
all the odd mass-derivatives of $\susc_n$ vanish in the chiral
limit. Since $m^2$-differentiability implies finiteness, an equivalent
characterization of the symmetric phase is then the following: chiral
symmetry is restored [in the sense of Eq.~\eqref{eq:susc_condition}]
if and only if $\susc_n$ are $m^2$-differentiable for all $n$. As a
consequence of the functional form Eq.~\eqref{eq:expansion1} of the
generating function, even (respectively, odd) susceptibilities are
linear combinations of the $\susc_n(m^2)$ with coefficients that are
even (respectively, odd) polynomials in $m$. It follows that chiral
symmetry is restored in the chiral limit if and only if even
susceptibilities are $m^2$-differentiable, and odd susceptibilities
are $m$ times an $m^2$-differentiable function, which is another
characterization of the symmetric phase for what concerns the scalar
and pseudoscalar sector.

\subsection{Remarks} 
\label{sec:gauge_ext}

The extension of the proof in Sec.~\ref{sec:chisymrest_nsc} to
susceptibilities involving translation-invariant operators $G_i$ built
out of gauge fields is straightforward. Since $G_i$ are unaffected by
chiral transformations, the corresponding sources $J_{G_i}$ in the
augmented generating function, $\wivG$ [see
Eq.~\eqref{eq:partfunc_G}], act merely as spectators in the
symmetry-restoration condition Eq.~\eqref{eq:symrest1_G}, and all the
results derived above still hold. These include the functional form
Eq.~\eqref{eq:chiW0_3} for $\wivG$, that can be expanded as in
Eq.~\eqref{eq:expansion1}, with coefficients $\susc_n(m^2;J_G)$ that
are now generating functions themselves (that can be re-expressed in
terms of a subset of generating functions involving $\{G_i\}$,
$\vec{P}$, and $\vec{S}$, see Appendix~\ref{sec:ansusc}) that obey
Eq.~\eqref{eq:nsc_chi_new6}. Expanding $\susc_n(m^2;J_G)$ in powers of
$J_{G_i}$, the resulting susceptibilities must all be finite in the
chiral limit, by the same argument as in
Sec.~\ref{sec:chisymrest_nsc}, and therefore $m^2$-differentiable or
$m$ times an $m^2$-differentiable function, depending on whether they
are even or odd in the number of isosinglet bilinears. Of course,
finiteness is also a sufficient condition for symmetry restoration at
the level of susceptibilities, involving now gauge operators as well.

All the above applies also to nonlocal gauge operators if one assumes
nonlocal restoration, which in particular implies the
$m^2$-differentiability of the spectral density $\rho$,
Eq.~\eqref{eq:rho_def3_th}, in the symmetric phase.  Indeed,
$\rho=\chi\left(\rho_U\right)$ in the notation of
Eq.~\eqref{eq:genfuncdef2_s}, with $\rho_U$, Eq.~\eqref{eq:rho_def1},
a translation-invariant nonlocal functional of the gauge
fields.\footnote{To proceed rigorously, one starts from the normalized
  mode number $\mdens(\lambda_0,\lambda;m)$ (see
  footnote~\ref{foot:rho}), which is the expectation value of a
  real-valued translation-invariant nonlocal functional of the gauge
  fields, and so is $m^2$-differentiable under the assumption of
  nonlocal restoration. The spectral density is then obtained by
  differentiating $\mdens$ with respect to $\lambda$, and similarly
  its $m^2$-derivatives are obtained, by definition, by
  differentiating the $m^2$-derivatives of $\mdens$ with respect to
  $\lambda$.} As already mentioned in the introductory remarks of this
section, an alternative proof of $m^2$-differentiability of spectral
quantities under the assumption that chiral symmetry remains manifest
when probing the system with external fields is discussed in
Appendix~\ref{sec:rhochirest}.

Clearly, $m^2$-differentiability is a weaker property than the
$m^2$-analyticity assumed in
Refs.~\cite{Cohen:1997hz,Aoki:2012yj,Kanazawa:2015xna,Azcoiti:2023xvu},
as it only implies that the quantities of interest can be written as
$ \mathcal{O}= z(m^2) + \sum_{k=0}^\infty a_{k} m^{2k} $, with
$|a_{k}|<\infty$ and $z$ vanishing with all its derivatives at $m=0$,
and with no guarantee that the sum has a finite radius of
convergence. Nonetheless, $m^2$-differentiability suffices for almost
all the arguments of Refs.~\cite{Cohen:1997hz,Aoki:2012yj,
  Kanazawa:2015xna,Azcoiti:2023xvu}, and so for most practical
purposes their $m^2$-analyticity assumptions are essentially justified
as a necessary consequence of symmetry restoration.

In the discussion above the lattice regularization plays no specific
role, other than putting quantum field theories on sound mathematical
footing. If a continuum, infinite-volume generating function
$\mathcal{W}_{\mathrm{cont}}$ can be defined at nonzero light-fermion
mass after suitable renormalization, one would still require
Eq.~\eqref{eq:symrest1} to hold for chiral symmetry to be restored in
the chiral limit. The chiral symmetry of the massless theory is exact
at any nonzero lattice spacing and so it can be preserved under
renormalization (see
Refs.~\cite{Hasenfratz:1998jp,Alexandrou:2000kj}), implying in turn
that the resulting functional form Eq.~\eqref{eq:chiW0_3} for
$\mathcal{W}_{\mathrm{cont}}$ is also preserved. All the consequences
derived above will then still hold in the continuum, including the
result that chiral symmetry is restored in the scalar and pseudoscalar
sector if and only if all the (renormalized) susceptibilities remain
finite in the chiral limit, implying in turn the appropriate
$m^2$-differentiability property for even and odd susceptibilities.

\section{Generating function from the Dirac spectrum}
\label{sec:det_sources}

In this section I compute explicitly the fermionic determinant in the
presence of scalar and pseudoscalar sources for $N_f=2$ degenerate
flavors in terms of Dirac eigenvalues, and obtain an exact expression
for the generating function in terms of eigenvalue correlation
functions, in the case of $\gamma_5$-Hermitean GW Dirac operators $D$
with $2R = \mathbf{1}$. In the rest of this paper I restrict to this
case, whose special properties have been discussed in
Sec.~\ref{sec:GWspec}. The result confirms the general analysis of the
functional form of the partition function discussed in
Sec.~\ref{sec:chisymrest_W}, and is the starting point for deriving
constraints on the Dirac spectrum in the chirally symmetric phase.

\subsection{Fermionic determinant in the presence of sources}
\label{sec:comp_fdet}

After integrating out the light fermion fields in
Eq.~\eqref{eq:partfunc}, the generating function of full correlators
reads
\begin{equation}
  \label{eq:fdet0}
  \mathcal{Z}(\sous,\soup;m) = \int \!DU\, e^{-S_{\mathrm{eff}}(U)}\det\Mc(U;\sous,\soup;m)\,,
\end{equation}
where
\begin{equation}
  \label{eq:fdet1}
  \begin{aligned}
    &    \Mc(U;\sous,\soup;m)\\
    &   \equiv        D(U)\mathbf{1}_{\mathrm{f}} +
      \left(\mathbf{1}-\tf{1}{2}D(U)\right)\left( A(\sous,\soup) + iB(\sous,\soup)\gamma_5    \right)  \,,
  \end{aligned}
\end{equation}
with
\begin{equation}
  \label{eq:fdet1_bis}
  \begin{aligned}
    A(\sous,\soup) &\equiv (j_S + m)\mathbf{1}_{\mathrm{f}} -\vec{\jmath}_S\cdot\vec{\sigma}\,,\\
    B(\sous,\soup) &\equiv j_P \mathbf{1}_{\mathrm{f}}+ \vec{\jmath}_P\cdot\vec{\sigma}      \,,
  \end{aligned}
\end{equation}
is a matrix carrying spacetime (including Dirac), color, and flavor
indices. Its determinant is most easily obtained in the orthonormal
basis $\{\psi_n\phi_f\}$, where $\psi_n$ are the orthonormal
eigenvectors of $D$, and $\phi_f$, $f=1,2$, is the canonical basis of
flavor space, $(\phi_f)_i=\delta_{fi}$. Zero modes and doubler modes
are chosen with definite chirality. For pairs of conjugate complex
modes, $\mu_n$ and $\mu_n^*$, the corresponding eigenvectors are
chosen to be $\psi_n$ and $\gamma_5\psi_n$, respectively. In this
basis the $2\times 2$ blocks
$(\Mc_{n'n})_{f'\!f}=
\left(\psi_{n'}\phi_{f'},\Mc\psi_n\phi_{f}\right)$ of $\Mc$ read
\begin{equation}
  \label{eq:fdet0_1}
  \Mc_{n' n}  =  \mu_n \mathbf{1}_{\mathrm{f}}\delta_{n'n} 
  +\left(1-\tf{1}{2}\mu_{n'}\right) \left[ A    \delta_{n'n} + i B (\gamma_5)_{n'n}  \right]\,,
\end{equation}
where
$ (\gamma_5)_{n'n}\equiv \left(\psi_{n'}, \gamma_5 \psi_n\right)$, and
so $\Mc$ has a simple block structure in the Dirac eigenmode indices
$n,n'$, with nonzero matrix elements only on the diagonal for chiral
zero modes and doubler modes, and off-diagonal for $n,n'$ a pair of
conjugate complex modes. It follows then
\begin{equation}
  \label{eq:fdet3}
  \begin{aligned}
    \det \Mc &= 2^{2 N_2}\left[\det\left( A + iB \right)\right]^{N_+}
               \left[ \det\left( A -iB \right)\right]^{N_-}    \\
             &\phantom{=}\times    \prod_{n,\,\Im\mu_n>0}\det \mnz(\mu_n)\,,
  \end{aligned}
\end{equation}
where for $z\in\mathbb{C}$
\begin{equation}
  \label{eq:fdet2_2_0}
  \mnz(z) \equiv       
  \begin{pmatrix}
    z\mathbf{1}_{\mathrm{f}} + \left(1-\tf{z}{2}\right)A
    & i\left(1-\tf{z}{2}\right)B\\
    i\left(1-\tf{z^*}{2}\right)B
    & z^*\mathbf{1}_{\mathrm{f}}+
      \left(1-\tf{z^*}{2}\right)A
  \end{pmatrix}  \,. 
\end{equation}
A zero mode of chirality $\xi$ contributes a factor
\begin{equation}
  \label{eq:fdet_zero}
  \begin{aligned}
    \det\left( A + i\xi B \right)
    &= \tilde{V}^2 - W^2 + 2i \xi\tilde{V}\cdot W \\
    &= m^2+ u - w + i \xi\tilde{u}\,,
     \end{aligned}
\end{equation}
with $\tilde{V}$ defined in Eq.~\eqref{eq:mshift1}, and
$u,w,\tilde{u}$ defined in Eq.~\eqref{eq:chiW0_4}. For the
contribution of a pair of complex modes $\mu_n,\mu_n^*$, one finds
(see Appendix~\ref{sec:compl_det} for details)
\begin{equation}
  \label{eq:fdet8_shorter_VW}
  \begin{aligned}
    &    \det \mnz(\mu_n)\\
    & = \lambda_n^4 + 2\lambda_n^2h(\lambda_n)(\tilde{V}^2 + W^2 )\\
    &\phantom{=} + h(\lambda_n)^2\left( (\tilde{V}^2 -W^2 )^2 +   (2\tilde{V}\cdot W)^2  \right)\\
    & = [\lambda_n^2 + m^2 h(\lambda_n)]^2 \\
    &\phantom{=}+ 2h(\lambda_n)[ (\lambda_n^2+ m^2 h(\lambda_n))u + (\lambda_n^2- m^2h(\lambda_n))w ] \\
    &\phantom{=}+  h(\lambda_n)^2\left[ (u -w )^2 + \tilde{u}^2 \right]\,,
  \end{aligned}
\end{equation}
where $\lambda_n$ is defined above Eq.~\eqref{eq:rho_def1}, with
$\lambda_n^2=|\mu_n|^2$, and
\begin{equation}
  \label{eq:fdet_hdef}
  h(\lambda)\equiv  1-\f{\lambda^2}{4}\,. 
\end{equation}
The first expression in Eq.~\eqref{eq:fdet8_shorter_VW} shows that
$\det \mnz(\mu_n)> 0$. Equations~\eqref{eq:fdet_zero} and
\eqref{eq:fdet8_shorter_VW} show explicitly that the partition
function $\mathcal{Z}$ depends only on $\mathrm{SO}(4)$-invariant
combinations of $\tilde{V}$ and $W$, as anticipated [see
Eq.~\eqref{eq:chiW0_2_fv_bis}]. In conclusion,
\begin{equation}
  \label{eq:fdet_final}
  \begin{aligned}
    \det \Mc
    &= (\det D_m)^2\left(1+ X_0\right)^{N_+} \left(1 +  X_0^*\right)^{N_-} \\
    &\phantom{=}\times\prod_{n,\lambda_n>0} \left[1 +  X(\lambda_n)\right]\,,
  \end{aligned}
\end{equation}
where $(\det D_m)^2$ is the fermionic determinant at zero sources,
\begin{equation}
  \label{eq:fdet_final_det0}
  \det D_m =  m^{N_0}2^{N_2}    \prod_{n,\lambda_n>0}
  \left[\lambda_n^2 +    m^2h(\lambda_n)\right]  \,,
\end{equation}
and 
\begin{equation}
  \label{eq:fdet_final2}
  \begin{aligned}
    X_0&\equiv   \f{u - w + i \tilde{u}}{m^2}\,,\\
    X(\lambda)
       &\equiv 
         2\left( \ff(\lambda;m) u +  \fft(\lambda;m) w\right)
    \\ &\phantom{\equiv}
         +  \ff(\lambda;m)^2\left( ( u -w )^2
         + \tilde{u}^2 \right)  
    \\ &=   2 \ffh(\lambda;m) (u + w)
    \\ &\phantom{=}  + m^4\ff(\lambda;m)^2 ( 2 \Re X_0  +  |X_0|^2)  \,,
  \end{aligned}
\end{equation}
where
\begin{equation}
  \label{eq:fdet8_shorter_bis}
  \begin{aligned}
    \ff(\lambda;m)
    &\equiv\f{h(\lambda)}{\lambda^2+    m^2h(\lambda)}\,, \\
    \fft(\lambda;m)
    &\equiv  \ff(\lambda;m)-2m^2 \ff(\lambda;m)^2\,, \\
    \ffh(\lambda;m)
    &\equiv  \ff(\lambda;m)-m^2 \ff(\lambda;m)^2 \,.
  \end{aligned}
\end{equation}
The dependence of $X_0$ and $X(\lambda)$ on the sources and on $m$ is
left implicit for notational simplicity.

For certain choices of gauge group and gauge-group representation
[e.g., $\mathrm{SU}(N_c)$ and adjoint representation], the Dirac
spectrum has Kramers degeneracy due to the existence of an antiunitary
operator $\mathcal{T}$ that obeys $[\mathcal{T},D]=0$ and
$\mathcal{T}^2=-\mathbf{1}$ (see
Ref.~\cite{Verbaarschot:2000dy}). This implies that complex modes are
doubly degenerate, and $N_\pm$ are even. One can then replace
$1 + X(\lambda_n)\to [1 + X(\lambda_n)]^2$ in
Eq.~\eqref{eq:fdet_final}, and
$\lambda_n^2 + m^2h(\lambda_n)\to [\lambda_n^2 + m^2h(\lambda_n)]^2$
in Eq.~\eqref{eq:fdet_final_det0}, while limiting the product to the
reduced spectrum. In the rest of this paper, when explicit expressions
in terms of the Dirac spectrum are provided for the relevant
quantities, it is assumed that the Dirac spectrum is not degenerate
(see also footnote~\ref{foot:kram}).

\subsection{Cumulant expansion of the partition function}
\label{sec:cumexp}

Using the results above, labeling the eigenvalues for a given gauge
configuration $U$ so that the $N=(N_{\mathrm{tot}} - N_0 -N_2)/2$
complex eigenvalues with positive imaginary part correspond to
$n=1,\ldots, N$, one finds
\begin{equation}
  \label{eq:fdet_final3}
  \f{  \mathcal{Z}(V,W;m)  }{  \mathcal{Z}(0,0;m)} = \left\la  e^{N_+ \sti(X_0)} e^{N_- \sti(X_0^*)}
    \prod_{n=1}^N   \left[1 +  X(\lambda_n) \right]\right\ra\,,
\end{equation}
where $\sti(t) \equiv \ln(1+t)$.\footnote{\label{foot:PVL}For complex
  arguments $\ln$ denotes the principal value of the complex
  logarithm, $\ln z = \ln |z| + i \,\mathrm{arg}\, z$, with
  $\mathrm{arg}\, z\in (-\pi,\pi]$.} Writing
\begin{equation}
  \label{eq:cumexp_simpler1}
  \prod_{n=1}^N     \left[1 + X(\lambda_n) \right]    =   \sum_{k=0}^\infty \f{1}{k!}Y_k \,,
\end{equation}
where $Y_0\equiv 1$, $Y_k \equiv 0$ for $k> N$, and
\begin{equation}
  \label{eq:cumexp_simpler1_def}
  Y_k  \equiv \sum_{\substack{n_1,\ldots,n_k=1    \\ n_i\neq n_j,\forall i,j}}^N
  X(\lambda_{n_1}) \ldots X(\lambda_{n_k})\,,\quad 1\le k\le N
  \,,
\end{equation}
expanding the exponentials in Eq.~\eqref{eq:fdet_final3} in power
series, and setting $A_{\vec{k}}\equiv N_+^{k_1} N_-^{k_2} Y_{k_3}$,
where $\vec{k}=(k_1,k_2,k_3)$, one finds by standard combinatorics
(see Appendix~\ref{sec:ccf_pf})
\begin{equation}
  \label{eq:cumexp_simpler2}
  \f{  \mathcal{Z}(V,W;m)  }{  \mathcal{Z}(0,0;m)}= \exp\left\{ \sum_{\vec{k}\neq \vec{0}}
    \left(\prod_{j=1}^3\f{t_j^{k_j}}{k_j!}\right)
    \la A_{\vec{k}} \ra_{c}       \right\}  \,,
\end{equation}
where $t_1=\sti(X_0)$, $t_2=\sti(X_0^*)$, $t_3=1$. The connected
correlation functions, $\la A_{\vec{k}} \ra_{c}$, are defined in the
usual way (see Appendices~\ref{sec:ccf_rec} and \ref{sec:ccf_relob}),
and in terms of spectral quantities they read
\begin{equation}
  \label{eq:cumexp_simpler4}  
  \la A_{\vec{k}}\ra_{c} =   I^{(k_3)}_{N_+^{k_1}N_-^{k_2}}[X,\ldots,X]\,,
\end{equation}
see Eq.~\eqref{eq:cum_sys_3_0_again}. Moreover (see
Ref.~\cite{AbramowitzStegun1964}, \S 24.1.3)
\begin{equation}
  \label{eq:u_func}
  \f{\sti(x)^k}{k!} = \sum_{n=k}^\infty s(n,k) \f{x^n}{n!} \,,
\end{equation}
where $s(n,k)$ are the Stirling numbers of the first kind, and since
$s(n,0)=s(0,n)=\delta_{n0}$, one can finally write [see
Eqs.~\eqref{eq:rho_def5_0}, \eqref{eq:rho_def8_gen}, and
\eqref{eq:cum_sys_3_0_again}]
\begin{equation}
  \label{eq:cumexp_simpler5}
  \begin{aligned}
    &    \mathcal{W}(V,W;m)-  \mathcal{W}(0,0;m)
      = \f{1}{\lvol}\ln  \f{  \mathcal{Z}(V,W;m)  }{  \mathcal{Z}(0,0;m)}  \\
    &= \sum_{\vec{n}\neq  0}\f{X_0^{n_1}X_0^{*n_2}}{n_1!n_2!n_3!}\sum_{k_1=0}^{n_1}\sum_{k_2=0}^{n_2}  
      s(n_1,k_1) s(n_2,k_2)
    \\ &\phantom{=}      \times I^{(n_3)}_{N_+^{k_1}N_-^{k_2}}[X,\ldots,X]  \,.
  \end{aligned}
\end{equation}
This result provides an explicit representation of the generating
function in terms of spectral correlators of the GW Dirac operator,
involving both complex and zero modes; and therefore a spectral
representation of all scalar and pseudoscalar susceptibilities once
the thermodynamic limit is taken.  While individual terms may not have
a well-define thermodynamic limit (see end of Sec.~\ref{sec:GWspec}),
the combinations corresponding to the various susceptibilities
certainly do at $m\neq 0$ (and in the symmetric phase also as
$m\to 0$, see Sec.~\ref{sec:chisymrest_nsc}).  If desired, one can
straightforwardly re-express
$I^{(n_{\smash{3}})}_{N_{\smash{+}}^{\smash{k_{\smash{1}}}}N_{\smash{-}}^{\smash{k_{\smash{2}}}}}$
in terms of
$I^{(n_{\smash{3}})}_{N_{\smash{0}}^{\smash{k_{\smash{0}}}}Q^{\smash{k_{\smash{1}}}}}$
[see Eq.~\eqref{eq:inpnminq}].  Notice that
$I^{(n_{\smash{3}})}_{N_{\smash{+}}^{\smash{k_{\smash{1}}}}N_{\smash{-}}^{\smash{k_{\smash{2}}}}}$
generally do not vanish in the thermodynamic limit, even under the
additional assumption $N_+ N_-=0$ a.e.; and that there is \textit{a
  priori} no reason for their contribution to $\wiv$ to be negligible
in the chiral limit, even though zero modes are suppressed, as they
are multiplied by factors $X_0^{n_1}X_0^{*n_2}\propto m^{-2(n_1+n_2)}$
(with $n_i\ge k_i$).

An alternative route to a spectral representation of $\mathcal{W}$ is
to directly expand $(1+X_0)^{N_+}$ and $(1+X_0^*)^{N_-}$ in
Eq.~\eqref{eq:fdet_final} in powers of $X_0$ and $X_0^*$. Setting
\begin{equation}
  \label{eq:spol}
  s_k(t) \equiv t (t-1) \ldots (t-k+1) = \sum_{l=1}^k s(k,l) t^l\,,
\end{equation}
for $k\ge 1$, and $s_0(t)\equiv 1$, one has
\begin{equation}
  \label{eq:spol0}
  (1+x)^n = \sum_{k=0}^n \f{ x^k}{k!}s_k(n)  = \sum_{k=0}^\infty \f{ x^k}{k!}s_k(n)\,, 
\end{equation}
since $s_k(n)=0$ for $n\in \mathbb{N}_0$ if $k>n$. One still finds
Eq.~\eqref{eq:cumexp_simpler2} but this time with
$A_{\vec{k}}= s_{k_1}(N_+) s_{k_2}(N_-) Y_{k_3} $, and $t_1=X_0$,
$t_2=X_0^*$, $t_3=1$, and one obtains a more compact expression for
the generating function,
\begin{equation}
  \label{eq:cumexp_S}
  \begin{aligned}
    & \mathcal{W}(V,W;m)- \mathcal{W}(0,0;m)  \\
    &=  \sum_{\vec{n}\neq  0}\f{X_0^{n_1}X_0^{*n_2}}{n_1!n_2!n_3!}
      I^{(n_3)}_{s_{n_1}(N_+)s_{n_2}(N_-)}[X,\ldots,X]\,,
  \end{aligned}
\end{equation}
where $I^{(k)}_{s_{s_{1}(N_{+})}s_{s_{2}(N_{-})}}$ is defined
according to Eq.~\eqref{eq:cum_sys_3_0_again} in terms of the spectral
correlator
$\rho^{(k)}_{s_{s_{1}(N_{+})}s_{s_{2}(N_{-})}\,c}= \lvol^{-1}\la
s_{s_1}(N_+)s_{s_2}(N_-) \rho_U^{(k)}\ra_c$, obtained recursively by
applying Eq.~\eqref{eq:part_g1} to
$A_{\vec{s}}^{(k)}= s_{s_1}(N_+)s_{s_2}(N_-) \rho_U^{(k)}$.
Equations~\eqref{eq:cumexp_S} and \eqref{eq:cumexp_simpler5} are
equivalent by virtue of the following combinatorial result,
\begin{equation}
  \label{eq:sti_cu1}
  \begin{aligned}
    &  \sum_{\sigma_1=0}^{n_1}\sum_{\sigma_2=0}^{n_2} s(n_1,\sigma_1)  s(n_2,\sigma_2)
      \rho^{(n_3)}_{N_+^{\sigma_1}N_-^{\sigma_2}\,c}(\lambda_1,\ldots,\lambda_{n_3};m) \\
    &  = \rho^{(n_3)}_{s_{n_1}(N_+)s_{n_2}(N_-)\,c}(\lambda_1,\ldots,\lambda_{n_3};m)\,,
  \end{aligned}
\end{equation}
which is a consequence of Eq.~\eqref{eq:u_func} (see
Appendix~\ref{sec:ccf_cr}).\footnote{\label{foot:kram}For a spectrum
  with Kramers degeneracy, the final results
  Eqs.~\eqref{eq:cumexp_simpler5} and \eqref{eq:cumexp_S} should be
  modified by constructing the spectral quantities entering
  $I^{(k)}_{O}$ using $\rho_U$ defined in terms of the reduced
  spectrum, and replacing $X(\lambda)\to 2X(\lambda) + X(\lambda)^2$.}

Since $X_0$ is linear and $X(\lambda)$ is quadratic
in $u,w,\tilde{u}$, the order of the polynomial in these variables
that multiplies the correlation function
$\rho^{(n_3)}_{s_{n_1}(N_+)s_{n_2}(N_-)\,c}$ is $n_1+n_2+2n_3$, and so
this quantity appears in coefficients $\susc_n$ of order at most
$n_u + n_w + n_{\tilde{u}} = n_1+n_2+2n_3$. In particular,
$\rho^{(n_3)}_{c}$ appears at most at order $2n_3$. If $CP$ symmetry
holds, a more natural expansion is in powers of $u$, $w$, and
$\tilde{u}^2$. In such an expansion, the spectral density
$\rho=\rho^{(1)}_{c\,{\scriptscriptstyle\infty}}$ appears in
first-order coefficients, and in coefficients that are second order in
$u$ and $w$, so one cannot get any constraint on the spectral density
from coefficients of order higher than 2.  One can in fact show that
there are no new constraints even from the second-order coefficients
(see end of Sec.~\ref{sec:constr_first}).

$CP$ invariance was not used to obtain Eqs.~\eqref{eq:cumexp_simpler5}
and \eqref{eq:cumexp_S}, which remain unchanged also in the presence
of $CP$-breaking terms, such as a $\theta$-term, in the action
$S_{\mathrm{eff}}$. In a $CP$-symmetric theory the roles of $N_+$ and
$N_-$ can be interchanged, and so in Eqs.~\eqref{eq:cumexp_simpler5}
and \eqref{eq:cumexp_S} one can replace
\begin{equation}
  \label{eq:cheb1_short}
  \begin{aligned}
    X_0^{n_1}X_0^{*n_2}
    &\to  \Re\left( X_0^{n_1} X_0^{*n_2}\right) \\
    &= |X_0|^{n_1+n_2}T_{|n_1-n_2|}(\Re X_0/|X_0|)\,,
  \end{aligned}
\end{equation}
with $T_n(x)$ the Chebyshev polynomial of order $n$. Since
$T_n(-x) = (-1)^nT_n(x)$, and $n_1+n_2$ and $|n_1-n_2|$ have the same
parity, one sees that $ \Re( X_0^{n_1} X_0^{*n_2})$ depends only on
$\Re X_0$ and $|X_0|^2$, and so that $\mathcal{W}$ involves only
powers of $u+w$, $|X_0|^2= (u-w)^2 + \tilde{u}^2$ (so only
$\tilde{u}^2$ appears), and $\Re X_0 = u-w$. From
Eq.~\eqref{eq:OVW_transf_singl} follows that the generating function
of the $\mathrm{U}(1)_A$-rotated scalar and pseudoscalar
susceptibilities is obtained by replacing
$(V,W) \to (V \cos \alpha -W\sin\alpha,V\sin\alpha +W\cos \alpha)$.
In the limit $m\to 0$, both $u+w \to V^2 + W^2$ and
$ |X_0|^2\to (V^2 + W^2)^2 - 4(V^2 W^2 -(V\cdot W)^2)$ become
$\mathrm{U}(1)_A$ invariants, so $\mathrm{U}(1)_A$-breaking effects
originate in a nontrivial dependence on $\Re X_0 = u-w\to V^2 -W^2$
surviving the chiral limit, since this quantity is not
$\mathrm{U}(1)_A$-invariant.

\subsection{Lowest orders}
\label{sec:lowest}

I now obtain explicitly the lowest-order terms of $\mathcal{W}$ in the
$CP$-symmetric case. As already pointed out, $CP$ invariance implies
that $\mathcal{W}$ contains only even powers of $\tilde{u}$, and so it
is natural to treat $\tilde{u}^2$ on the same level as $u$ and $w$ by
setting
\begin{equation}
  \label{eq:def_C}
  \begin{aligned}
    \mathcal{C}(u,w,\tilde{u}^2;m)
    &\equiv \mathcal{W}(V,W;m)-  \mathcal{W}(0,0;m) \\
    &= \hat{\mathcal{W}}(m^2+u,w,\tilde{u})- \hat{\mathcal{W}}(m^2,0,0)\,,
  \end{aligned}
\end{equation}
and expanding in powers of $u$, $w$, and $\tu^2$,
\begin{equation}
  \label{eq:cumexp_simpler9_simplified2}
  \begin{aligned}
    &      \mathcal{C}(u,w,\tilde{u}^2;m) \\
    &      =  u\coeff_u(m^2) + w\coeff_w(m^2)  +\tilde{u}^2\coeff_{\tilde{u}^2}(m^2) \\
    &\hphantom{=}+ \tf{1}{2}\left[
      u^2 \coeff_{uu}(m^2) + 2uw \coeff_{uw}(m^2) + w^2 \coeff_{ww}(m^2) \right. \\
    &\hphantom{=}    \left.  + 2 u\tilde{u}^2 \coeff_{u\tilde{u}^2}(m^2) +
      2w\tilde{u}^2\coeff_{w\tilde{u}^2}(m^2)  \right. \\
    &\hphantom{=} \left. + \tilde{u}^4\coeff_{\tilde{u}^2\tilde{u}^2}(m^2)  \right]  + \ldots\,,
  \end{aligned}
\end{equation}
with omitted terms of order three or higher. The thermodynamic limit
of these quantities are denoted by
$\coeffi_X\equiv\lim_{\lvol\to\infty}\coeff_X$, and $\coeffi$ denotes
the corresponding generating function.

The coefficients $\coeffi_X$ can be expressed as linear combinations
(with $m$-independent coefficients) of a restricted set of
susceptibilities (see Appendix~\ref{sec:ansusc}), but they have
simpler expressions in terms of the full set,
Eq.~\eqref{eq:genfuncdef2_ss}, that can be obtained using the
functional form of $\wiv$, Eq.~\eqref{eq:chiW0_3}, and the
consequences of the anomalous $\mathrm{U}(1)_A$ symmetry (see
Sec.~\ref{sec:top_iig}). For the first-order coefficients one has
\begin{equation}
  \label{eq:coeff_rel1}
  \begin{aligned}
    \chi_\pi
    &\equiv   \chi\left((iP_a)^2\right)
    && = 2\coeffi_u  \,,\\
    \chi_\delta
    &\equiv \chi\left( (S_a)^2\right)
    && = 2\coeffi_w  \,,
  \end{aligned}
\end{equation}
where $\chi_\pi$ and $\chi_\delta$ are the usual pion and delta
susceptibilities, and
\begin{equation}
  \label{eq:coeff_real1_quater}
  \begin{aligned}
    \chi_{\pi\delta} \equiv  \chi\left((iP_a) S_a (iP_b)  S_b\right)|_{a\neq b}
    &= 8\coeffi_{\tilde{u}^2}  \,,    \\
    \chi\left((iP_a) (iP) S_a \right) 
    &= -8m\coeffi_{\tilde{u}^2}  \,.
  \end{aligned}
\end{equation}
Notice that for the chiral condensate one has
\begin{equation}
  \label{eq:cond}
  \Sigma \equiv -\chi(S) =   2m\coeffi_u  =  m\chi_\pi\,.
\end{equation}
For the $\eta$ susceptibility one has
\begin{equation}
  \label{eq:coeff_rel1_bis}
  \chi_\eta  \equiv  \chi   \left(( iP)^2 \right)
  = 2\coeffi_w + 8m^2\coeffi_{\tilde{u}^2}
  = \chi_\delta + m^2    \chi_{\pi\delta} 
  \,.
\end{equation}
Using the anomalous Ward-Takahashi identity
$\f{\chi_\pi - \chi_\eta}{4} = \f{\chit}{m^2}$ [see, e.g.,
Ref.~\cite{Giusti:2004qd}, and Eq.~\eqref{eq:chitop_ward2} below], the
first equation in Eq.~\eqref{eq:coeff_real1_quater} becomes
\begin{equation}
  \label{eq:coeff_real1_quater_bis}
  8\coeffi_{\tilde{u}^2} =  \chi_{\pi\delta} =  \f{\chi_\pi-\chi_\delta}{m^2}-\f{4\chit}{m^4}
  =     \f{4}{m^2}\left(\f{\coeffi_u-\coeffi_w}{2}-\f{\chit}{m^2}\right) \,.
\end{equation}
For the $\sigma$ susceptibility one finds
\begin{equation}
  \label{eq:coeff_rel4_bis}
  \chi_\sigma   \equiv  \chi\left( S^2\right)   =2\coeffi_u + 4m^2 \coeffi_{uu}    \,,
\end{equation}
involving a second-order coefficient. For the second-order
coefficients one has
\begin{equation}
  \label{eq:coeff_rel3}
  \begin{aligned}
        \chi\left(    (iP_a)^2(iP_b)^2\right)|_{a\neq b}
    &= 
         4\coeffi_{uu}
      \,,\\
        \chi\left(
    S_a^2S_b^2\right)|_{a\neq b}
    &= 
         4\coeffi_{ww}
      \,,\\
    \chi\left(
    (iP_a)^2 S_b^2\right)|_{a\neq b}
    &= 
      4\coeffi_{uw}
      \,,
  \end{aligned}
\end{equation}
and moreover
\begin{equation}
  \label{eq:coeff_rel4}
  \begin{aligned}
    \chi\left( (iP_a) S_a (iP_b) S_b (iP_c)^2\right)|_{a\neq b\neq c\neq a}
    &= 16\coeffi_{u\tilde{u}^2} \,,\\
    \chi\left( (iP_a) S_a (iP_b) S_b S_c^2\right)|_{a\neq b\neq c\neq a}
    &=  16\coeffi_{w\tilde{u}^2}  \,.
  \end{aligned}
\end{equation}
For $\coeff_{\tilde{u}^2\tilde{u}^2}$ the simplest relation is
\begin{equation}
  \label{eq:coeff_rel5}
    \chi\left( \left(\textstyle\prod_{a=1}^3(iP_a) S_a \right)  (iP)\right)
    =  192m\coeffi_{\tilde{u}^2\tilde{u}^2}  \,,
\end{equation}
where also $m$ appears (an $m$-independent expression for
$\coeff_{\tilde{u}^2\tilde{u}^2}$ can be obtained in terms of
susceptibilities involving eight fermion bilinears, using the results
in Appendix~\ref{sec:ansusc}).

Specializing now to a non-degenerate Dirac spectrum in the
$CP$-symmetric case, it is straightforward to compute the coefficients
in Eq.~\eqref{eq:cumexp_simpler9_simplified2}. Recalling
Eqs.~\eqref{eq:Ncum_def} and \eqref{eq:cum_sys_3_0_again}, one has for
the first-order coefficients
\begin{equation}
  \label{eq:first_order_expl}
  \begin{aligned}
    \coeff_u &= \f{b_{N_0}}{m^2} + 2\Iu[\ff] \,, \\
    \coeff_{w} &= -\f{b_{N_0}}{m^2} + 2\Iu[\fft] \,, \\
    \coeff_{\tilde{u}^2}  &= \f{1}{m^2}\left(\f{b_{N_0}-b_{Q^2}}{2m^2}+m^2\Iu[\ff^2]\right)  \\
             &    =     \f{1}{2m^2}\left( \f{\coeff_u - \coeff_w}{2} -  \f{b_{Q^2}}{m^2}\right)   \,,
  \end{aligned}
\end{equation}
while the second-order coefficients are
\begin{equation}
  \label{eq:second_order_expl}
  \begin{aligned}
    \coeff_{uu}    &= \topoa + 4\Id[\ff,\ff] + \f{4}{m^2}\In[\ff] \,,    \\
    \coeff_{uw}   &=  -\topoa + 4\Id[\ff,\ff]-8m^2\Id[\ff,\ff^2]  -    4\In[\ff^2]\,,  \\
    \coeff_{ww}  &= \topoa + 4\Id[\fft,\fft] - \f{4}{m^2}\In[\fft] \,,   \\
    \coeff_{u\tilde{u}^2}   &=   \topob + 2\Id[\ff,\ff^2]   +\f{1}{m^2} \In[\ff^2] \\
                   &\phantom{=} +\f{1}{m^4}\In[\ff] -\f{1}{m^4}\Iq[\ff]   \,,\\
    \coeff_{w\tilde{u}^2}  &=   - \topob + 2\Id[\fft,\ff^2]  -\f{1}{m^2} \In[\ff^2]\\
                   &\phantom{=}  +\f{1}{m^4}\In[\fft]  -\f{1}{m^4}\Iq[\fft]  \,,\\
    \coeff_{\tilde{u}^2\tilde{u}^2}  &= \topoc + \Id[\ff^2,\ff^2] +  \f{1}{m^4}\In[\ff^2]  - \f{1}{m^4}\Iq[\ff^2] \,,
\end{aligned}
\end{equation}
where
 \begin{equation}
   \label{eq:second_order_expl_bis}
   \begin{aligned}
     \topoa &\equiv \f{1}{m^4}\! \left(b_{N_0^2}-b_{N_0} +   2m^4\Iu[\ff^2]\right)\\
            &=   2\coeff_{\tu^2} + \f{1}{m^4}\! \left(b_{N_0^2}+b_{Q^2} -2b_{N_0} \right) \,,  \\
     \topob &\equiv \f{1}{m^6}\!\left(- \f{1}{2} b_{N_0 Q^2} + b_{Q^2} +  \f{1}{2}b_{N_0^2} - b_{N_0}  \right)\,,\\
     \topoc &\equiv \f{1}{m^8}\!\left(\f{b_{Q^4}}{12} - \f{b_{N_0 Q^2}}{2}+
              \f{2 b_{Q^2}}{3} +\f{b_{N_0^2}}{4}  -\f{b_{N_0}}{2} \right)\,.
  \end{aligned}
\end{equation}
The second-order coefficients $\coeff_{uu}$, $\coeff_{uw}$, and
$\coeff_{u\tilde{u}^2}$ can be written in a more compact form by
recognizing the presence of $m^2$-derivatives in their
expressions. This also allows one to show explicitly that they are the
$m^2$-derivative of $\coeff_{u}$, $\coeff_{w}$, and
$\coeff_{\tilde{u}^2}$ [see Eq.~\eqref{eq:nsc_chi_new6}]. The mass
derivative of the expectation value of any mass-independent observable
$\Oc$ reads
\begin{equation}
  \label{eq:mder1}
  \de_m \la \Oc\ra  = -\left[\la \Oc S \ra - \la \Oc\ra \la S\ra\right]\,.
\end{equation}
For $\Oc=\Oc(U)$ depending only on the gauge fields, after integrating
fermions out, and exploiting the properties of the spectrum, one finds
after a short calculation
\begin{equation}
  \label{eq:mder4}
  \begin{aligned}
    \de_m \la \Oc\ra
    &=  \f{2}{m}  \left( \left\la \Oc N_0 \right\ra
      - \la \Oc\ra \left\la N_0 \right\ra\right) \\
    &\phantom{=}+   4m\int_0^2 d\lambda\,      f(\lambda;m)  
      \left[\left\la \Oc \rho_U(\lambda) \right\ra
      -\left\la \Oc\right\ra  \left\la\rho_U(\lambda) \right\ra\right]\,.
  \end{aligned}
\end{equation}
For $\Oc=N_0/\lvol$, $\Oc=Q^2/\lvol$, and $\Oc=\rho_U/\lvol$, using
$ 2m\de_{m^2} =\de_m$, one finds [see
Eqs.~\eqref{eq:rho_def12_bis}--\eqref{eq:cum_sys_3_0_again}]
\begin{equation}
  \label{eq:mder8}
  \begin{aligned}
    \de_{m^2}\, b_{N_0}  &= \f{b_{N_0^2}}{m^2} + 2\In[f] \,,\\
    \de_{m^2}\,b_{Q^2} &= \f{b_{N_0Q^2}}{m^2} + 2\Iq[f] \,,
  \end{aligned}
\end{equation}
and
\begin{equation}
  \label{eq:mder8_nate}
  \begin{aligned}
    \de_{m^2}\, \rho^{(1)}_c(\lambda;m)
    &=    \f{1}{m^2}\rho_{N_0\,c}^{(1)}(\lambda;m) + 2  f(\lambda;m)\rho^{(1)}_c(\lambda;m)\\
    &\phantom{=}  + 2\int_0^2 d\lambda'\,      f(\lambda';m)  \rho_c^{(2)}(\lambda,\lambda';m)\,.
  \end{aligned}
\end{equation}
Moreover, since $\de_{m^2}\,f = - f^2$, one has
\begin{equation}
  \label{eq:mder9}
        \de_{m^2}\,\Iu[f^2]   =  \f{1}{m^2}\In[f^2]  + 2 \Id[f^2,f] \,.
\end{equation}
Notice that under the assumption $ N_+ N_- = 0$ a.e., in the
thermodynamic limit one finds $n_0=0$ and so $\de_m n_0=0$, and the
first equation in Eq.~\eqref{eq:mder8} results in 
\begin{equation}
  \label{eq:mder_ext0}
  \Iniv[f] =  -  \f{\bbnn}{2m^2}   \le 0\,,\qquad
  \bbnn \equiv \lim_{\lvol\to\infty}b_{N_0^2} \,.
\end{equation}
This shows that exact zero modes and complex modes generally repel
each other, independently of the status of chiral symmetry.  Note that
in this case $\bbnn$ is surely finite, as $\la N_0^2\ra$ and
$\la N_0\ra^2$ are both (no more than) $O(\lvol)$, and
$\rho_{c\,N_0\,{\scriptscriptstyle\infty}}^{(1)}(\lambda)$ is
integrable,
\begin{equation}
  \label{eq:rhon0int}
  \begin{aligned}
    &  \int_0^2d\lambda\,   \rho_{c\,N_0\,{\scriptscriptstyle\infty}}^{(1)}(\lambda) =
      \lim_{\lvol\to \infty}  \int_0^2d\lambda\,   \f{\la N_0 \rho_U(\lambda)\ra_c}{\lvol} \\
    &= -\lim_{\lvol\to \infty} \f{\la N_0^2 \ra_c}{\lvol}  = -\bbnn\,.
\end{aligned}
\end{equation}

\section{Constraints on the Dirac spectrum}
\label{sec:Dirac_constr}

As shown in Sec.~\ref{sec:chisymrest_nsc}, chiral symmetry restoration
requires the finiteness in the chiral limit of the coefficients
$\susc_n$ in Eq.~\eqref{eq:expansion1_A}, or equivalently of the
coefficients $\coeffi_X$, see
Eq.~\eqref{eq:cumexp_simpler9_simplified2} (and of the free energy
density $\susc_{000}=-\wiv|_0$).  Using Eq.~\eqref{eq:cumexp_simpler5}
or Eq.~\eqref{eq:cumexp_S}, this requirement translates into
constraints on the Dirac spectrum. In this section I discuss the
constraints obtained imposing finiteness in the chiral limit of the
lowest-order coefficients $\coeffi_X$, i.e., the thermodynamic limit
of Eqs.~\eqref{eq:first_order_expl} and
\eqref{eq:second_order_expl}. This is done from a general point of
view, without making additional technical assumptions on the spectrum.

\subsection{Constraints from first-order coefficients}
\label{sec:constr_first}

The constraints from the first-order coefficients of the generating
function, Eq.~\eqref{eq:first_order_expl}, amount to imposing
finiteness in the chiral limit of $\chi_\pi$, $\chi_\delta$, and
$\chi_{\pi\delta}$ [see Eqs.~\eqref{eq:coeff_rel1},
\eqref{eq:coeff_real1_quater}, and \eqref{eq:coeff_real1_quater_bis}],
\begin{equation}
  \label{eq:firstorder_again0_0}
  \begin{aligned}
    \f{\chi_\pi}{2}
    &= \coeffi_u  = \f{n_0}{m^2} + 2\Iu_{{\scriptscriptstyle\infty}}[\ff] \,, \\
    \f{\chi_\delta}{2}
    &= \coeffi_w  =  -\f{n_0}{m^2} + 2\Iu_{{\scriptscriptstyle\infty}}[\fft] \,, \\ 
    \f{\chi_{\pi\delta}}{8}
    &= \coeffi_{\tilde{u}^2}
      = \f{1}{2m^2}\left( \f{\chi_\pi -\chi_\delta}{4} -   \f{\chit}{m^2}\right) \,.
  \end{aligned}
\end{equation}
Independently of the request of finiteness, since $n_0\ge 0$ and
$\Iu_{{\scriptscriptstyle\infty}}[\ff^k]\ge 0$ one has
$\chi_\pi\ge 0$. Moreover, since $m^2\ff^2\le \ff$, one has
$\chi_\pi \pm \chi_\delta\ge 0$, implying
$|\chi_\delta | \le \chi_\pi$.  Finiteness of $\chi_\delta$ then
follows after imposing finiteness of $\chi_\pi$, and since both terms
in $\chi_\pi$ are positive this requires that they be separately
finite. Requiring finiteness of $\chi_\pi$ in the chiral limit is then
equivalent to requiring that
\begin{equation}
  \label{eq:firstorder_again}
  \lim_{m\to 0} \f{n_0}{m^2} <\infty\,, \qquad
  \lim_{m\to 0}    \Iu_{{\scriptscriptstyle\infty}}[\ff] <\infty\,.
\end{equation}
Finiteness of $\chi_{\pi\delta}$ in the chiral limit requires that
\begin{equation}
  \label{eq:firstorder_again1}
  \f{\chi_\pi -\chi_\delta}{4}  =   \f{\chit}{m^2} +  O(m^2)\,,
\end{equation}
which since the left-hand side must be finite requires in turn the
finiteness of $ \f{\chit}{m^2}$ in the chiral limit. Using the
explicit expression for $\chi_\pi$ and $\chi_\delta$ in terms of the
spectrum, Eq.~\eqref{eq:firstorder_again1} reads
\begin{equation}
  \label{eq:firstorder_again1_2}
  \f{n_0}{m^2} + 2m^2\Iu_{{\scriptscriptstyle\infty}}[f^2]   = 
  \f{\chit}{m^2} +  O(m^2)\,.
\end{equation}
For the usual $\mathrm{U}(1)_A$ order parameter $\dcl$,
\begin{equation}
  \label{eq:firstorder_again2_0}
  \dcl\equiv \lim_{m\to 0}\f{\chi_\pi -\chi_\delta}{4}\,, 
\end{equation}
Eq.~\eqref{eq:firstorder_again1_2} implies that
\begin{equation}
  \label{eq:firstorder_again2}
  \dcl = \lim_{m\to 0} \left(\f{n_0}{m^2} +
    2m^2\Iu_{{\scriptscriptstyle\infty}}[f^2]\right) = \lim_{m\to 0}
  \f{\chit}{m^2} \,.
\end{equation}
Equations~\eqref{eq:firstorder_again} and
\eqref{eq:firstorder_again1_2} fully summarize all the constraints
from the first-order coefficients. Under the additional assumption
$N_+ N_-=0$ a.e., one has $n_0=0$, so the first-order constraints boil
down to requiring finiteness of
$\Iu_{{\scriptscriptstyle\infty}}[\ff]$ and $\f{\chit}{m^2}$ in the
chiral limit, and that
$2m^2\Iu_{{\scriptscriptstyle\infty}}[\ff^2]-\f{\chit}{m^2}=O(m^2)$,
with Eq.~\eqref{eq:firstorder_again2} simplifying to
\begin{equation}
  \label{eq:firstorder_again2_bis}
  \dcl=  \lim_{m\to 0} 2m^2\Iu_{{\scriptscriptstyle\infty}}[f^2]=
  \lim_{m\to 0}      \f{\chit}{m^2}\,.
\end{equation}
As shown in Section~\ref{sec:chisymrest_nsc}, the coefficients of the
generating function in the thermodynamic limit, and so the even
susceptibilities, must be not only finite but also
$m^2$-differentiable in the chiral limit. From the last equation in
Eq.~\eqref{eq:firstorder_again0_0} recast as
$ \f{\chit}{m^2} = \f{1}{4}\left(\chi_\pi-\chi_\delta -
  m^2\chi_{\pi\delta}\right)$, or directly from the anomalous
Ward-Takahashi identity $\f{\chit}{m^2} = \f{\chi_\pi - \chi_\eta}{4}$
[see Eq.~\eqref{eq:chitop_ward2} below], follows then that
$\f{\chit}{m^2}$ must be $m^2$-differentiable,
a stronger result than $m^2$-differentiability of $\chit$ [see
discussion after Eq.~\eqref{eq:susc_condition_g2} in
Sec.~\ref{sec:chisymrest}], and obtained using only symmetry restoration
in the scalar and pseudoscalar sector.

Using the representation of susceptibilities in terms of the Dirac
spectrum one can obtain a lower bound on $ \chi_\pi -\chi_\delta$,
that implies that effective restoration of $\mathrm{U}(1)_A$ symmetry
in the scalar and pseudoscalar sector at a nonzero value of the quark
mass is impossible on the lattice.  Making use of the assumption
$N_+ N_-=0$ a.e., so that $n_0=0$ (as well as $n_2=0$), one finds for
the chiral condensate $\Sigma$, Eq.~\eqref{eq:cond},
\begin{equation}
  \label{eq:condensate}
  \Sigma =   4m \Iu_{{\scriptscriptstyle\infty}}[\ff]\,,
\end{equation}
and using the Cauchy--Schwarz inequality one shows that
\begin{equation}
  \label{eq:Delta2_bis}
  2 \Iu_{{\scriptscriptstyle\infty}}[\ff]^2   \le   \nu  \Iu_{{\scriptscriptstyle\infty}}[\ff^2]  \,,
\end{equation}
where $\nu\equiv 2\int_0^2 d\lambda\,\rho(\lambda;m)=4N_c$ is the
average number of complex modes per unit four-volume [see
Eq.~\eqref{eq:rho_def2}]. One has then
\begin{equation}
  \label{eq:Delta4}
  \chi_\pi -\chi_\delta\ge   \f{\Sigma^2}{\nu} \,. 
\end{equation}
The quantity on the right-hand side is strictly positive for any
nonzero mass, so $ \chi_\pi -\chi_\delta \neq 0$ at $m\neq 0$, and
$\mathrm{U}(1)_A$ cannot be effectively restored at nonzero quark mass
on the lattice. Concerning the chiral limit, in the broken phase
$\Sigma(0^\pm) = \pm \Sigma(0^+)$ with $\Sigma(0^+)>0$, $\chi_\pi$
diverges, and $ \dcl\neq 0$ (in fact one expects it to diverge, since
$\chi_\delta$ is expected to remain finite), so $\mathrm{U}(1)_A$ is
effectively broken; in the symmetric phase $\Sigma(0^\pm)=0$,
Eq.~\eqref{eq:Delta4} reduces to $\dcl \ge 0$, and so both effective
breaking (by $\dcl > 0$) and effective restoration (that requires
$\dcl =0$) are allowed. Extending Eq.~\eqref{eq:Delta4} to the
continuum limit is hampered by the additive and multiplicative UV
divergences affecting both of its sides. Note, however, that
Eq.~\eqref{eq:Delta2_bis} still holds if the integrals defining
$\Iu_{{\scriptscriptstyle\infty}}$ and $\nu$ are cut off at the same
point.  If one deals with the additive divergences of $\chi_\pi$,
$\chi_\delta$, and $\Sigma$ by cutting off the corresponding integrals
over $\lambda$, then Eq.~\eqref{eq:Delta4} still holds provided all
quantities (including $\nu$) are suitably redefined. Since the
remaining multiplicative renormalization affects both sides in the
same way, one can then extend the modified inequality to the continuum
limit, showing that $\mathrm{U}(1)_A$ cannot be effectively restored
at nonzero $m$. 

As already pointed out [see comments after Eq.~\eqref{eq:sti_cu1} in
Sec.~\ref{sec:cumexp}], since coefficients of order higher than two
involve only eigenvalue correlation functions of order at least two,
there are no further coefficients involving the spectral density other
than those in Eq.~\eqref{eq:first_order_expl} and
\eqref{eq:second_order_expl}. Moreover, in the second-order
coefficients $\rho^{(1)}_c$ appears only in $\topoa$,
Eq.~\eqref{eq:second_order_expl_bis}, where it enters through the
first-order coefficient $\coeff_{\tu^2}$. In the thermodynamic limit,
$\rho$ appears in the resulting coefficients only through
$\chi_{\pi\delta}$, which is finite if the constraints discussed in
this subsection are satisfied. It follows that no new direct
constraint on the spectral density can be obtained, other than those
coming from the first-order coefficients,
Eqs.~\eqref{eq:firstorder_again} and \eqref{eq:firstorder_again1_2}.

\subsection{Constraints from second-order coefficients}
\label{sec:constr_second}

Instead of using directly the second-order coefficients in
Eq.~\eqref{eq:second_order_expl}, to obtain constraints on the
spectrum it is more convenient to work with an equivalent set of
quantities obtained by an invertible linear transformation, namely (in
the thermodynamic limit)
\begin{widetext}
  \allowdisplaybreaks
  \begin{eqnarray}
    \label{eq:secondorder_01}
    \coeffi_{uu}
    &=& \f{1}{m^2}\left(4m^2\Id[\ff,\ff] +
        \f{\chit-\bbnn}{m^2}+ 2m^2\de_{m^2}\f{n_0}{m^2}  +
        2m^2  \coeffi_{\tilde{u}^2}\right) = \de_{m^2}\coeffi_u\,,\\
    \label{eq:secondorder_03}
    \f{1}{2}\left( \coeffi_{uu} +2 \coeffi_{uw} + \coeffi_{ww}\right)
    &=& 8\Id_{\scriptscriptstyle\infty}[\ffh,\ffh]\,,\\
    \label{eq:secondorder_02}
    \f{1}{2}\left( \coeffi_{uu} - \coeffi_{uw}\right)
    &=& \de_{m^2}\left(2m^2\coeffi_{\tilde{u}^2}
        +  \f{\chit}{m^2}\right) = \f{1}{2}\de_{m^2}\left( \coeffi_{u} - \coeffi_{w}\right)\,,\\
    \coeffi_{u\tilde{u}^2}
    &=& \f{1}{2m^2}\left[ 
        -\de_{m^2}\f{\chit}{m^2} + \f{1}{2}\left( \coeffi_{uu} -
        \coeffi_{uw}\right) -2\coeffi_{\tilde{u}^2}\right] 
        =\de_{m^2}\coeffi_{\tu^2} \notag   \\
    \label{eq:secondorder_04}
    &=& \f{1}{2m^4}\left[  \left(1-m^2\de_{m^2}\right)\f{\chit-n_0}{m^2}
        + 2m^2\Iniv[f^2]  + 4m^4 \Id_{\scriptscriptstyle\infty}[f^2,f]\right]\,,\\
    \label{eq:secondorder_05}
    \f{1}{4}\left( \coeffi_{uu}- \coeffi_{ww}\right)
    - m^2\left(\coeffi_{u\tilde{u}^2}+ \coeffi_{w\tilde{u}^2}\right)
    &=& \f{2}{m^2}\Iqiv[\ffh]\,,\\
    \coeffi_{\tilde{u}^2\tilde{u}^2}
    &=& \f{1}{m^4}\left[ 
        \f{\bbqq-\chit}{12m^4} -\f{1}{16}\left( \coeffi_{uu} -2
        \coeffi_{uw} + \coeffi_{ww}-8\coeffi_{\tilde{u}^2} \right)
        + \f{m^2}{2} \left( \coeffi_{u\tilde{u}^2}- \coeffi_{w\tilde{u}^2} \right) \right] 
        \notag\\
    \label{eq:secondorder_06}
    &=& \f{1}{12 m^4}\left[ \f{\bbqq-\chit}{m^4}
        -3\left(\de_{m^2} \f{\chit}{m^2} - \f{2}{m^2}\Iqiv[\ffh] \right)
        +3m^2 \left( \coeffi_{u\tilde{u}^2}- \coeffi_{w\tilde{u}^2} \right) \right]  \,, 
  \end{eqnarray}
\end{widetext}
where $\bbnn$ is defined in Eq.~\eqref{eq:mder_ext0}, 
\begin{equation}
  \label{eq:q4cum}
  \bbqq\equiv\lim_{\lvol\to\infty} b_{Q^4}\,,
\end{equation}
and I have made use of Eq.~\eqref{eq:first_order_expl} and of the
derivative formulas in Eqs.~\eqref{eq:mder8}--\eqref{eq:mder9}.
Imposing finiteness of the quantities in
Eqs.~\eqref{eq:secondorder_01}--\eqref{eq:secondorder_06} in the
chiral limit is of course equivalent to imposing finiteness of those
in Eq.~\eqref{eq:second_order_expl} in the thermodynamic limit
followed by the chiral limit. From a general point of view, finiteness
of the left-hand side of Eqs.~\eqref{eq:secondorder_01},
\eqref{eq:secondorder_02}, and \eqref{eq:secondorder_04} corresponds
to the existence of the first $m^2$-derivative of
$\coeffi_{u,w,\tu^2}$ at $m=0$, which is expected (for all
$m^2$-derivatives) in the symmetric phase (see
Sec.~\ref{sec:chisymrest_nsc}). Notice that
Eqs.~\eqref{eq:secondorder_02} and \eqref{eq:secondorder_04} provide
the same expression for $\de_{m^2}\f{\chit}{m^2}$ in terms of first-
and second-order coefficients.

Finiteness in the chiral limit of the quantities in
Eqs.~\eqref{eq:secondorder_01}--\eqref{eq:secondorder_06} implies a
number of constraints on the Dirac spectrum. The first two are
\begin{equation}
  \label{eq:secondorder1_1}
  \begin{aligned}
    4m^2\Id_{\scriptscriptstyle\infty}[\ff,\ff]
    &= \f{\bbnn-\chit}{m^2} - 2m^2\de_{m^2}\f{n_0}{m^2}   + O(m^2)\,,\\
    \Id_{\scriptscriptstyle\infty}[\ffh,\ffh]
    &=O(1)\,.
  \end{aligned}
\end{equation}
These follow from Eqs.~\eqref{eq:secondorder_01} and
\eqref{eq:secondorder_03}, and provide constraints on the two-point
eigenvalue correlator, $\rho^{(2)}_{c\,{\scriptscriptstyle\infty}}$,
Eq.~\eqref{eq:rho_def7}, with the first one relating it to topological
properties of the theory.  Making use of the assumption $ N_+ N_-=0$
a.e., from this constraint one finds in particular
\begin{equation}
  \label{eq:secondorder4}
  -  \lim_{m\to 0} 4m^2\Id_{\scriptscriptstyle\infty}[\ff,\ff] 
  = \lim_{m\to 0} \lim_{\lvol\to\infty}
  \f{\la N_0\ra^2}{m^2 \lvol} \equiv \Delta'\,.
\end{equation}
It is straightforward to show that $0\le \Delta'\le \Delta$, so
$\Delta'$ must be finite in the symmetric
phase.\footnote{\label{foot:deltaprime}In the large-volume limit the
  distribution of $Q$ is expected to be Gaussian, and from $N_0=|Q|$
  follows that $\Delta'=\f{2}{\pi}\Delta$.}  Since the left-hand side
should be dominated by the near-zero modes, one expects that these
predominantly repel each other if $\Delta'\neq 0$. Deeper insight
will be obtained in the second paper of this series under further
technical assumptions on $\rho^{(2)}_{c\,{\scriptscriptstyle\infty}}$.

Two more constraints are
\begin{equation}
  \label{eq:secondorder1_2}
  \begin{aligned}
    \left(1-m^2\de_{m^2}\right)\f{n_0-\chit}{m^2}
    &=  2m^2\Iniv[f^2] \\
    &\phantom{=}  + 4m^4 \Id_{\scriptscriptstyle\infty}[f^2,f] +O(m^4)\,, \\
    \de_{m^2}\f{\chit}{m^2} &=O(1)\,.
  \end{aligned}
\end{equation}
The first one follows from the request of finiteness of
$\coeffi_{u\tu^2} = \de_{m^2}\coeffi_{\tu^2}$,
Eq.~\eqref{eq:secondorder_04}, in the chiral limit. The second one
follows from finiteness of the quantity appearing in
Eq.~\eqref{eq:secondorder_02}, using the finiteness of
$\coeffi_{\tu^2}$, and the required finiteness of
$\de_{m^2}\coeffi_{\tu^2}$. This constraint is just a special case of
the all-order result for $\f{\chit}{m^2}$ already discussed above [see
under Eq.~\eqref{eq:firstorder_again2_bis}].  Using this constraint
and the assumption $N_+ N_-=0$ a.e., one finds
$(1-m^2\de_{m^2})\f{\chit}{m^2} = \Delta + O(m^4)$, which leads to
\begin{equation}
  \label{eq:secondorder4_quater}
  -\Delta = \lim_{m\to 0}\left(2m^2\Iniv[f^2]  + 4m^4 \Id_{\scriptscriptstyle\infty}[f^2,f]\right)\,,
\end{equation}
with the quantity in brackets deviating from $-\Delta$ only at order
$O(m^4)$. This constraint involves both
$\rho_{N_0\,c\,{\scriptscriptstyle\infty}}^{(1)}$ and
$\rho^{(2)}_{c\,{\scriptscriptstyle\infty}}$,
Eqs.~\eqref{eq:rho_def12_bis} and \eqref{eq:rho_def7}, or equivalently
$\rho$ and $\de_{m^2}\rho$ [see Eq.~\eqref{eq:mder8_nate}], again put
into a relation with topological properties of the theory.

Finiteness of the quantity appearing in Eq.~\eqref{eq:secondorder_05}
requires
\begin{equation}
  \label{eq:secondorder1_3}
  \f{1}{m^2}\Iqiv[\ffh] =O(1)\,.
\end{equation}
This request is made more precize by using Eq.~\eqref{eq:coeff_rel3}
to express the left-hand side of Eq.~\eqref{eq:secondorder_05} in
terms of susceptibilities. One finds
  \begin{equation}
  \label{eq:secondorder2}
  \begin{aligned}
    &   \lim_{m\to 0} \f{2}{m^2}\Iqiv[\ffh] =
      \f{1}{4}   \lim_{m\to 0} \left(\coeffi_{uu}- \coeffi_{ww}\right)\\
    &= \f{1}{16}   \lim_{m\to 0}  \left[ \chi\left((iP_a)^2(iP_b)^2\right)
      -\chi\left( S_a^2S_b^2\right)\right]|_{a\neq b}   \equiv \Delta_2\,,
  \end{aligned}
\end{equation}
and Eq.~\eqref{eq:secondorder1_3} requires $|\Delta_2|<\infty$.  This
quantity is an order parameter for $\mathrm{U}(1)_A$ that depends on
the correlation between zero and complex modes, as encoded in
$\rho_{Q^2\,c}^{(1)}$, Eq.~\eqref{eq:rho_def12_bis}.  This can be
written as
\begin{equation}
  \label{eq:topconst1_0}
  \begin{aligned}
    \rho_{Q^2\,c}^{(1)}(\lambda;m)
    &= \f{ \left\la Q^2   \rho_U(\lambda) \right\ra_c}{\lvol}
      = - \left.\de_\theta^2 \f{           \left\la \rho_U(\lambda)
      \right\ra_\theta}{\lvol}\right|_{\theta=0}    \\
    &=-  \left.  \de_\theta^2 \rho^{(1)}_c(\lambda;m;\theta)\right|_{\theta=0}\,,
  \end{aligned}
\end{equation}
where $\la\ldots\ra_\theta$ denotes the expectation value in the
presence of a $\theta$ term, defined by replacing
$-S_{\mathrm{eff}}\to -S_{\mathrm{eff}} + i\theta Q$ in
Eqs.~\eqref{eq:general_pf2} and \eqref{eq:general_pf3} [see
Eq.~\eqref{eq:topconst5_0} below], and
$\rho^{(1)}_c(\lambda;m;\theta) \equiv \f{1}{\lvol} \left\la
  \rho_U(\lambda) \right\ra_\theta $ is the normalized spectral
density (in a finite volume) of the GW Dirac operator in this case.
Then $\Iq[\ffh] = -\de_\theta^2 \Iu[\ffh;\theta]|_{\theta=0}$, where
\begin{equation}
  \label{eq:Iqder1_0_bis}
  \Iu[g;\theta] \equiv \int_0^2 d\lambda\, g(\lambda)  \rho^{(1)}_c(\lambda;m;\theta) \,,
\end{equation}
and so in the thermodynamic limit [that can be exchanged with the
derivatives with respect to $\theta$ at $m\neq 0$, see discussion
after Eq.~\eqref{eq:genfuncdef2_s}]
\begin{equation}
  \label{eq:Iqder2_0}
  \begin{aligned}
    \Iqiv[\ffh]
    &= - \f{1}{2}\left. \de_\theta^2 \left(\Iu_{\scriptscriptstyle\infty}[\ff;\theta]
      +\Iu_{\scriptscriptstyle\infty}[\fft;\theta]\right)\right|_{\theta=0} \\
    &   = - \f{1}{4}\left. \de_\theta^2 
      \left[\coeffi_u(\theta)+\coeffi_w(\theta)\right]\right|_{\theta=0} \\
    &=- \f{1}{8}\left.\de_\theta^2 \left[\chi_\pi(\theta)
      +\chi_\delta(\theta)\right]\right|_{\theta=0}\,,
  \end{aligned}
\end{equation}
where $\coeffi_{u,w}(\theta)$ denote the thermodynamic limit of the
expansion coefficients $\coeff_{u,w}(\theta)$ of $\coeff(\theta)$,
i.e., $\coeff$ of Eq.~\eqref{eq:def_C} evaluated in the presence of a
$\theta$ term; $\chi_\pi(\theta)$ and $\chi_\delta(\theta)$ are the
pion and delta susceptibilities in this case; and I have used
$2\ffh = \ff + \fft$ and Eq.~\eqref{eq:first_order_expl}, that holds
unchanged provided all quantities are evaluated at nonzero $\theta$.
Notice that although in this case the density of zero modes,
$b_{N_0}$, generally does not vanish in the thermodynamic limit, it
still exactly cancels out in $\coeff_u(\theta)+\coeff_w(\theta)$. From
Eq.~\eqref{eq:secondorder2} follows then that in the symmetric phase
the relation
\begin{equation}
  \label{eq:Iqder3_0}
  \Delta_2   =-   \lim_{m\to 0}\f{1}{4m^2}\left.\de_\theta^2
    \left(\chi_\pi(\theta)+ \chi_\delta(\theta)\right)\right|_{\theta=0}
\end{equation}
holds between $\Delta_2$ and the pion and delta susceptibilities.

Finally, finiteness of $\coeff_{\tu^2\tu^2}$,
Eq.~\eqref{eq:secondorder_06}, requires
\begin{equation}
  \label{eq:secondorder1_4}
  \f{\bbqq-\chit}{m^4} = 3\left(\left.\de_{m^2}\f{\chit}{m^2}\right|_{m= 0}
    - \Delta_2\vphantom{\f{2}{m^2}\Iqiv[\ffh]}\right)  + O(m^2)\,,
\end{equation}
where I used Eq.~\eqref{eq:secondorder2}.  If $\chit\propto m^2$, and
so $\mathrm{U}(1)_A$ remains effectively broken, this requires that
$\bbqq$ and $\chit$ be equal to leading order in $m$, and so that the
distribution of the topological charge be indistinguishable from that
of an ideal gas of instantons and anti-instantons, with identical and
vanishingly small density $\chit/2$, to lowest order in the fermion
mass and at the level of the first non-trivial cumulant. [If $\chit$
vanishes faster than $m^2$, and so is at least $O(m^4)$, this is not
necessarily the case, as Eq.~\eqref{eq:secondorder1_4} would generally
imply that $\bbqq$ and $\chit$ differ at leading order.] Leading
corrections to the ideal behavior are $O(m^4)$, and encoded in the
first term on the right-hand side of Eq.~\eqref{eq:secondorder1_4}. A
more general result was obtained in Ref.~\cite{Kanazawa:2014cua},
namely that an ideal instanton gas behavior holds to lowest order in
the fermion mass for all cumulants, under the assumption that in the
symmetric phase the free energy density at finite $\theta$ angle is
analytic in $m^2$ and $\mathrm{U}(1)_A$ remains effectively broken by
$\Delta\neq 0$. This conclusion can be obtained straightforwardly
using the formalism of the present paper, as I show below in
Sec.~\ref{sec:top_iig}. Moreover, the $m^2$-differentiability of
susceptibilities required in the symmetric phase (see
Sec.~\ref{sec:chisymrest_nsc}) justifies the expansion in powers of
$m^2$ used in Ref.~\cite{Kanazawa:2014cua}.

\subsection{Remarks} 
\label{sec:constr_remarks}

The constraints Eqs.~\eqref{eq:firstorder_again} and
\eqref{eq:firstorder_again1_2} [and Eq.~\eqref{eq:firstorder_again2}]
are not new, and have appeared in various forms (at least implicitly)
in the literature~\cite{Cohen:1997hz,Aoki:2012yj,Kanazawa:2015xna,
  Carabba:2021xmc,Azcoiti:2023xvu}. Here, however, they have been
fully justified from the theoretical point of view.  Moreover, the
present approach shows that these constraints, derived from the
first-order coefficients, are the only constraints that involve the
spectral density directly [see comments at the end of
Sec.~\ref{sec:constr_first}].

The constraints from the second-order coefficients,
Eqs.~\eqref{eq:secondorder1_1}, \eqref{eq:secondorder1_2},
\eqref{eq:secondorder1_3}, and \eqref{eq:secondorder1_4}, are instead
new,\footnote{An incomplete form of the first constraint in
  Eq.~\eqref{eq:secondorder1_1} appeared in Eq.~(4.28) of
  Ref.~\cite{Kanazawa:2015xna}. The constraint
  Eq.~\eqref{eq:secondorder1_4} follows from the results of
  Ref.~\cite{Kanazawa:2014cua}; here it is obtained from first
  principles.} and involve the thermodynamic limit of the two-point
correlation function of complex modes, $\rho_c^{(2)}$, and of the
correlation functions $\rho^{(1)}_{N_0\,c}$ and $\rho^{(1)}_{Q^2\,c}$
involving zero and complex modes.  These are only indirectly related
to the spectral density, through derivatives with respect to the mass
[see Eq.~\eqref{eq:mder8_nate}] or the $\theta$ angle [see
Eq.~\eqref{eq:topconst1_0}].

Without further assumptions, at this stage full restoration of
$\mathrm{SU}(2)_L\times \mathrm{SU}(2)_R$ symmetry is compatible both
with effective breaking and with effective restoration of
$\mathrm{U}(1)_A$, as the order parameters $\dcl$,
Eq.~\eqref{eq:firstorder_again2_0}, and $\Delta_2$,
Eq.~\eqref{eq:secondorder2}, can be nonzero without contradicting the
other requirements. In order to achieve effective breaking with
$\dcl\neq 0$, the topological susceptibility must be proportional to
$m^2$ (with a nonzero proportionality constant) in the chiral limit
[see Eq.~\eqref{eq:firstorder_again2}], in which case one finds an
ideal gas-like behavior for the topological charge [see
Eq.~\eqref{eq:secondorder1_4} and Sec.~\ref{sec:top_iig}], and
sufficiently strong repulsion between near-zero modes to satisfy
Eq.~\eqref{eq:secondorder4}. A nonzero $\Delta_2$ requires instead
that correlations between zero and complex modes do not vanish too
fast in the chiral limit [see Eq.~\eqref{eq:secondorder2}]. On the
other hand, restoration of $\mathrm{U}(1)_A$ requires
$\Delta=\Delta_2=0$, and therefore not only that $\chi_\pi$ and
$\chi_\delta$ become equal at $\theta=0$ in the chiral limit, but also
that their sum is independent of $\theta$ up to $O(\theta^2)$ and
$O(m^2)$ [see Eq.~\eqref{eq:Iqder3_0}].

\subsection{Comparison with
  Ref.~{\protect\ignorecitefornumbering{\cite{Aoki:2012yj}}}}
\label{sec:criticism0}

In Ref.~\cite{Aoki:2012yj} Aoki, Fukaya, and Taniguchi provided a
detailed discussion of spectral constraints resulting from chiral
symmetry restoration in the scalar and pseudoscalar sector. Their
requirements for symmetry restoration, however, differ from the one
used here, i.e., the finiteness of scalar and pseudoscalar
susceptibilities. Since the latter is a necessary and sufficient
condition for chiral symmetry restoration at the level of
susceptibilities (see Sec.~\ref{sec:chisymrest}), it should
automatically imply that the symmetry-restoration conditions of
Ref.~\cite{Aoki:2012yj} are satisfied, lest these are more restrictive
than necessary. If so, the approach of Ref.~\cite{Aoki:2012yj} should
not be able to provide more constraints on the spectrum than the ones
obtained here. I now show that this is indeed the case.

The requirements of Ref.~\cite{Aoki:2012yj} for symmetry restoration
are the following. (i.)\ For operators
$\Oc = \prod_{i \in I_{\Oc}} O_i$, with $I_{\Oc}=\{1,\ldots,N_{\Oc}\}$
and $O_i$ chosen from the set $ \{S,i\vec{P},iP,\vec{S}\}$, the
(suitably normalized) expectation values
$\lvol^{-n_{\delta_{Aa}\Oc}}\la \delta_{Aa}\Oc \ra$ of their
infinitesimal axial transformations $\delta_{Aa}\Oc$ [see
Eq.~\eqref{eq:deltaR_def}] vanish in the chiral limit (taken after the
thermodynamic limit). (ii.)\ Expectation values of $m$-independent
observables that depend only on gauge fields are analytic functions of
$m^2$. As I show below, condition (i.)\ follows from finiteness of
$\coeffi_{u,w,\tu^2}$ in the chiral limit.  Condition (ii.), relaxed
to the relevant quantities being $C^\infty$ functions of $m^2$ without
major practical effects on the approach of Ref.~\cite{Aoki:2012yj},
follows from the $m^2$-differentiability properties proved in
Secs.~\ref{sec:chisymrest_nsc} and \ref{sec:gauge_ext}, if also
nonlocal restoration or restoration in external fields is required.

In condition (i.), $n_{\delta_{Aa}\Oc}$ are appropriate powers of the
volume matching the leading volume dependence of
$\la\delta_{Aa}\Oc \ra$ as $\lvol\to\infty$. These powers are
determined using the cluster property of correlation functions,
encoded in the finiteness of
$\f{1}{\lvol}\ln \mathcal{Z} = \mathcal{W}$ in the thermodynamic
limit, which allows one to write for a generic observable $\Oc$ of the
type above
\begin{equation}
  \label{eq:aft_crit1}
  \la \Oc\ra  = \lvol^{n_{\Oc}}\sum_{\substack{\pi\in \Pi(I_\Oc)  \\|\pi| = n_{\Oc}}}
  \prod_{p\in \pi} \chi\left( \Oc(p)\right) + o\left(\lvol^{n_{\Oc}}\right)\,,
\end{equation}
where the sum is over the partitions of $I_{\Oc}$ with the maximal
amount $n_{\Oc}$ of parts $p$ such that $\chi(\Oc(p))\neq 0$, and
$\Oc(p)=\prod_{i\in p}O_{i}$.  These partitions involve only the
``irreducible'' correlation functions coinciding with their connected
part, i.e., those for which $\la \Oc\ra = \la \Oc \ra_c$. By
inspection of $\mathcal{W}$ as constrained by the symmetries of the
theory, Eq.~\eqref{eq:chiW0_2_fv_bis}, and taking into account $CP$
invariance, one sees that correlation functions not identically
vanishing must be even under $P,\vec{P}\to -P,-\vec{P}$, as well as
under $P_a,S_a\to -P_a,-S_a$ for each $a=1,2,3$ separately. This
allows one to obtain the following exhaustive list of irreducible
correlation functions,
\begin{equation}
  \label{eq:aft_crit3}
  \begin{aligned}
    \la S\ra    &=  -2m \coeffi_u \lvol +o(\lvol)\,,\\
    \la (iP_a)^2\ra   & =2\coeffi_u\lvol +o(\lvol)\,,\\
    \la (iP)^2\ra   & = \left(2\coeffi_w + 8m^2\coeffi_{\tilde{u}^2}\right)\!\lvol+o(\lvol)\,,\\
    \la (S_a)^2\ra  & = 2\coeffi_w\lvol +o(\lvol)\,,\\
    \la  (iP_a)(iP) S_a\ra  & = - 8m\coeffi_{\tilde{u}^2}\lvol +o(\lvol)\,,    \\
    \la(iP_a)(iP_b) S_a S_b\ra  & = 8\coeffi_{\tilde{u}^2}\lvol +o(\lvol) \qquad ( a\neq b)\,,
  \end{aligned}
\end{equation}
see
Eqs.~\eqref{eq:coeff_rel1}--\eqref{eq:coeff_rel1_bis}.\footnote{Invariance
  of
  $\left\la S^{n_S} (i\vec{P})^{\vec{n}_{P}} (iP)^{n_P}
    \vec{S}^{\vec{n}_{S}}\right\ra$ under $P,\vec{P}\to -P,-\vec{P}$
  and $P_a,S_a\to -P_a,-S_a$ requires
  \begin{equation*}
    \begin{aligned}
      (-1)^{n_{P_a}+ n_{S_a}}
      &= 1\,,\,\,\forall a\,,
      &&& (-1)^{n_P + \textstyle\sum_{a=1}^3 n_{P_a}}
      &= 1\,.
    \end{aligned}
  \end{equation*}
  These conditions are solved by
  \begin{equation*}
    \begin{aligned}
      n_P &= 2N_P + \tilde{n}_P\,,
      &&&
          n_{P_a} &= 2N_{P_a} + \tilde{o}_{a}\,, \\
      n_{S_a} &= 2N_{S_a} + \tilde{o}_{a}\,,
    \end{aligned}
  \end{equation*}
  with $N_P$, $N_{P_a}$, and $N_{S_a}$ non-negative integers,
  $\tilde{n}_P=0,1$, $\tilde{o}_{a}=0,1$, $a=1,2,3$, and with
  $\tilde{n}_P + \sum_{a=1}^3 \tilde{o}_{a}= 0,2,4$. The last
  condition has the solutions $\tilde{n}_P= 0$,
  $\tilde{o}_{a}= 0\,\forall a$, leading to the first four irreducible
  correlation functions in Eq.~\eqref{eq:aft_crit3}; $\tilde{n}_P=1$,
  $\sum_{a=1}^3 \tilde{o}_{a}=1$, leading to the fifth one;
  $\tilde{n}_P=0$, $\sum_{a=1}^3 \tilde{o}_{a}=2$, leading to the
  sixth one; and $\tilde{n}_P=1$, $\sum_{a=1}^3 \tilde{o}_{a}=3$,
  which does not lead to new irreducible correlation functions.}

Using the Ward-Takahashi identities Eq.~\eqref{eq:WT8}, 
condition (i.)\ is equivalent to requiring 
\begin{equation}
  \label{eq:aft_crit2}
  \lim_{m\to 0}m\lim_{\lvol\to\infty}  \lvol^{-n_{P_a\Oc}}\la P_{a}\Oc \ra = 0\,,  
\end{equation}
with $n_{P_a\Oc}=n_{\delta_{Aa}\Oc}$. The cluster property allows one
to write $\lvol^{-n_{P_a\Oc}}\la P_{a}\Oc \ra$ as a product of scalar
and pseudoscalar susceptibilities, that are necessarily finite in the
symmetric phase, as shown above in Sec.~\ref{sec:chisymrest_nsc}, and
so also this condition is automatically satisfied. Again, only the
irreducible correlation functions of Eq.~\eqref{eq:aft_crit3} appear
in this quantity, to leading order in the volume.  The approach of
Ref.~\cite{Aoki:2012yj} cannot then yield more constraints on the
Dirac spectrum than those following from the finiteness and
$m^2$-differentiability of $\coeffi_{u,w,\tilde{u}^2}$ (i.e.,
finiteness of $\susc_{n_u n_w n_{\tu}}$ with arbitrary $n_u$,
$n_w=0,1$, and $n_{\tu}=0,2$), which is only a subset of the
constraints following from finiteness of all the susceptibilities, and
is therefore a less general approach than the one used here.  In
particular, it can lead only to the same direct constraints
Eqs.~\eqref{eq:firstorder_again} and \eqref{eq:firstorder_again1_2} on
the spectral density obtained in this paper, that follow from
finiteness of $\coeffi_{u,w,\tilde{u}^2}$ in the chiral limit.

\section{Ideal instanton gas-like behavior}
\label{sec:top_iig}

Under the assumption that the free energy density at finite $\theta$
angle is analytic in $m^2$ in the symmetric phase, and that
$\mathrm{U}(1)_A$ remains effectively broken in the chiral limit
(which throughout this section will be synonymous with
$\Delta\neq 0$), Ref.~\cite{Kanazawa:2014cua} showed that to lowest
order in the fermion mass, the cumulants of the topological charge are
the same found for an ideal gas of instantons and anti-instantons,
with identical and vanishingly small densities $\chit/2\propto m^2$.
Here I rederive this conclusion directly from the assumption of chiral
symmetry restoration at the level of susceptibilities. As shown in
Sec.~\ref{sec:chisymrest_nsc}, this necessarily leads to
$m^2$-differentiability (rather than analyticity) of the free energy
density and of scalar and pseudoscalar susceptibilities at $\theta=0$;
I show below that this also implies the $m^2$-differentiability of the
derivatives of the free energy density with respect to $\theta$ at
$\theta=0$, i.e., of the cumulants of $Q$, as already argued [see
under Eq.~\eqref{eq:susc_condition_g2}].

Consider the usual partition function in the presence of a topological
term,
\begin{equation}
  \label{eq:topconst5_0}
  \begin{aligned}
    Z(\theta;m) \equiv
    \int DU\, e^{-S_{\mathrm{eff}}(U)+i\theta Q(U)}
    \!    \int D\Psi D\bar{\Psi} \,e^{- \bar{\Psi} D_m (U) \Psi} \,.
  \end{aligned}
\end{equation}
Using the transformation properties of the action and of the measure
under the $\mathrm{U}(1)_A$, flavor-singlet axial transformation
$\U_{A}^{(0)}\left(\tf{\theta}{4}\right)=
\U^{(0)}\left(-\tf{\theta}{4},\tf{\theta}{4}\right)$ [see
Eqs.~\eqref{eq:ch_transf3}, \eqref{eq:U1Ameas}, and
\eqref{eq:OVW_transf_singl}], one readily finds the identity
\begin{equation}
  \label{eq:topconst5_1}
  Z(\theta;m) = \tilde{\mathcal{Z}}\left(j_S(\theta;m),j_P(\theta;m);m\right) \,,
\end{equation}
with
$ \tilde{\mathcal{Z}}\left(j_S,j_P;m\right) \equiv
\mathcal{Z}\left(\sous,\soup;m\right)|_{\vec{\jmath}_{P,S} =\vec{0}}$
the generating function Eq.~\eqref{eq:partfunc} for vanishing
isotriplet sources, and
\begin{equation}
  \label{eq:sourcesth3_0}
  j_S(\theta;m) \equiv m \left(\cos\tf{\theta}{2}-1\right)\,,
  \quad j_P(\theta;m) \equiv m\sin\tf{\theta}{2}\,,
\end{equation}
are $m$- and $\theta$-dependent isoscalar
sources.\footnote{\label{foot:z2a}Setting $\theta=2\pi$ in
  Eqs.~\eqref{eq:topconst5_0} and \eqref{eq:topconst5_1} one finds the
  identity $Z(0;m)=Z(2\pi;m)=Z(0;-m)$, that implies an additional
  non-anomalous discrete $\mathbb{Z}_{2A}$ symmetry in the chiral
  limit.} The free energy density in the presence of a topological
term is then related to
$ \wtiv\equiv \wiv|_{\vec{\jmath}_{P,S} =\vec{0}}$ as
\begin{equation}
  \label{eq:topconst7_0}
  \begin{aligned}
    F(\theta;m) &\equiv -\lim_{\lvol\to\infty} \f{1}{\lvol}\ln Z(\theta;m) \\
                &  = -\wtiv\left(j_S(\theta;m),j_P(\theta;m);m\right)\,. 
  \end{aligned}
\end{equation}
Since $j_S(\theta;m)=O(\theta^2)$ and $j_P(\theta;m)=O(\theta)$ for
small $\theta$, the expansion of the right-hand side of
Eq.~\eqref{eq:topconst7_0} in powers of the sources corresponds to an
expansion for small $\theta$. More precisely, $\theta$ derivatives of
$F$ at $\theta=0$ equal finite linear combinations of derivatives of
$\wtiv$ with respect to scalar and pseudoscalar isosinglet sources at
zero sources (i.e., of isosinglet susceptibilities), with coefficients
the $\theta$-derivatives of powers of $j_{S,P}(\theta;m)$ at
$\theta=0$.\footnote{Since the expansion in powers of the sources is
  formal, so is the expansion in $\theta$, that may have zero radius
  of convergence, and may miss terms vanishing at zero with all their
  derivatives. This does not affect the correctness of the result for
  the $\theta$ derivatives of $F$ at $\theta=0$ [as long as they
  commute with the thermodynamic limit -- see discussion after
  Eq.~\eqref{eq:genfuncdef2_s} and footnote~\ref{foot:der}].}  Since
terms odd in $j_P$ vanish thanks to $CP$ invariance of the theory at
$\theta=0$ and $m\neq 0$, and since
$F(0;m) =-\wtiv(0,0;m) = -\wiv(0,0,m)$, one finds
\begin{equation}
  \label{eq:topconst11}
  \begin{aligned}
    F(\theta;m)&= F(0;m) -  j_S(\theta;m)\de_{j_S}\wiv(m)|_0 \\
               & \phantom{=}                  
                 - \tf{1}{2}j_P(\theta;m)^2\de_{j_P}^2\wiv(m)|_0\\
               & \phantom{=}                  
                 -  \tf{1}{2}j_S(\theta;m)^2\de_{j_S}^2\wiv(m)|_0 \\
               & \phantom{=}                  
                 - \tf{1}{2}j_S(\theta;m)j_P(\theta;m)^2\de_{j_S}\de_{j_P}^2\wiv(m)|_0    \\
               & \phantom{=} -\tf{1}{24}j_P(\theta;m)^4\de_{j_P}^4\wiv(m)|_0 + O(\theta^6)\,,
  \end{aligned}
\end{equation}
where the mass dependence of the derivatives of $\wiv$ at zero sources
is shown explicitly. This expression holds independently of the fate
of chiral symmetry in the chiral limit, and provides the correct mass
dependence of the $\theta$-derivatives of $F(\theta;m)$ and of its
mass-derivatives evaluated at $\theta=0$, including in the chiral
limit (i.e., with the chiral limit taken after setting $\theta=0$). In
particular, the topological susceptibility
$\chit = \de_\theta^2 F(\theta;m)|_{\theta=0}$ reads
\begin{equation}
  \label{eq:chitop_ward2}
  \chit   = \f{1}{4}  \left( m\Sigma - m^2\chi_\eta\right)
  = \f{m^2}{4}  \left( \chi_\pi - \chi_\eta\right)  \,,
\end{equation}
which is a well-known integrated Ward-Takahashi identity for the
anomalous $\mathrm{U}(1)_A$ symmetry~\cite{Giusti:2004qd}. Together
with Eq.~\eqref{eq:coeff_rel1_bis}, this relation implies
Eq.~\eqref{eq:coeff_real1_quater_bis}.

Using instead the functional form Eq.~\eqref{eq:chiW0_3} for the
generating function $\mathcal{W}_{\scriptscriptstyle\infty}$ one finds
\begin{equation}
  \label{eq:topconst8}
  F(\theta;m) =
  - \whiv\! \left(m^2 + u(\theta;m),w(\theta;m),\tilde{u}(\theta;m)\right)\,,
\end{equation}
where
\begin{equation}
  \label{eq:topconst_8bis}
  \begin{aligned}
    u(\theta;m)&= -w(\theta;m)\equiv  -m^2\left(\sin\tf{\theta}{2}\right)^2 \,,\\
    \tilde{u}(\theta;m)&\equiv m^2\sin\theta\,.
  \end{aligned}
\end{equation}
Independently of the fate of chiral symmetry as $m\to 0$, one can
expand $\whiv$ around zero sources in powers of $u(\theta;m)$, $w(\theta;m)$, and
$\tilde{u}(\theta;m)$, or rather
$\tilde{u}(\theta;m)^2 =
4m^4\left(\sin\tf{\theta}{2}\right)^2[1-\left(\sin\tf{\theta}{2}\right)^2]$
thanks to $CP$ invariance, finding [see Eqs.~\eqref{eq:def_C} and
\eqref{eq:cumexp_simpler9_simplified2}]
\begin{equation}
  \label{eq:topconst9}
  \begin{aligned}
    &  F(\theta;m) -      F(0;m)   \\
    &= m^2\left(\sin\tf{\theta}{2}\right)^2\left(  \coeffi_u(m^2)
      - \coeffi_w(m^2)   - 4m^2 \coeffi_{\tilde{u}^2}(m^2)\right)    \\
    &\phantom{=}      - \f{m^4}{2} \left(\sin\tf{\theta}{2}\right)^4 \left[
      \coeffi_{uu}(m^2) -2\coeffi_{uw}(m^2) + \coeffi_{ww}(m^2)      \right. \\
    &\phantom{=}   \left.      - 8 \coeffi_{\tilde{u}^2}(m^2)
      -8m^2\left(  \coeffi_{u\tilde{u}^2}(m^2)-\coeffi_{w\tilde{u}^2}(m^2) \right)
      \right. \\
    &\phantom{=}   \left.      + 16m^4 \coeffi_{\tilde{u}^2\tilde{u}^2}(m^2)    \right]
      + O\left(\left(\sin\tf{\theta}{2}\right)^6\right) \,.
  \end{aligned}
\end{equation}
This is generally not a systematic expansion in powers of
$m^{2n}\left(\sin\f{\theta}{2}\right)^2$, with $n\ge 1$: in the broken
phase $\coeffi_u= \f{\chi_\pi}{2} = \f{\Sigma}{2 m}$ diverges like
$1/m$ in the chiral limit, implying also that
$(m^2)^n\de_u^n \coeffi|_0 \sim m^{-1}$ for all $n\ge 1$. Identifying the
coefficient of a given power of $m$ requires then resumming infinitely
many terms.

In the symmetric phase, on the other hand, the coefficients of the
expansion Eq.~\eqref{eq:topconst9} are finite, and actually
$m^2$-differentiable in the chiral limit, so $F$ can be truly expanded
in powers of $m^2\left(\sin\tf{\theta}{2}\right)^2$ [or equivalently
in powers of $m^2\cos\theta$, or in a Fourier series in
$m^{2n}\cos(n\theta)$], with $O(1)$ coefficients that can be further
expanded in powers of $m^2$ around zero (possibly with zero radius of
convergence, and up to a function vanishing at $m=0$ with all its
derivatives). This shows that the $\theta$-derivatives of the free
energy are $m^2$-differentiable.  In particular, the omitted terms in
Eq.~\eqref{eq:topconst9} are then both $O(\theta^6)$ and $O(m^6)$.
Turning the argument around, one has that $F(\theta;m)$ can be
expanded in powers of $m^2$, with coefficients that are finite
polynomials in $\cos\theta$, and with the expansion being valid for
arbitrary $\theta$.

To lowest order in $m^2$, one finds in the symmetric phase
\begin{equation}
  \label{eq:topconst10}
  \begin{aligned}
    &  F(\theta;m)-F(0;m) \\
    &= (1-\cos\theta)\f{m^2}{2}\left( \coeffi_u(0) - \coeffi_w(0) \right) +O( m^4)\\
    &=   (1-\cos\theta)m^2\Delta    +O( m^4)   \\
    &= (1-\cos\theta)\chit +O( m^4)\,,
  \end{aligned}
\end{equation}
to all orders in $\theta$, and having used the constraint
Eq.~\eqref{eq:firstorder_again2} in the last passage. If
$\Delta\neq 0$, this is the free energy density of an ideal gas of
instantons and anti-instantons of equal densities
$\chit/2 = m^2 \Delta/2 + O(m^4)$. One concludes that an ideal
instanton gas-like behavior of the topological charge distribution in
the chiral limit is a necessary condition for chiral symmetry
restoration if $\mathrm{U}(1)_A$ remains effectively broken.

To find the corrections to the ideal gas behavior, one expands
Eq.~\eqref{eq:topconst9} up to order $O(\theta^4)$, obtaining
\begin{equation}
  \label{eq:topconst10_bis}
  F(\theta;m)-F(0;m) = \f{\theta^2}{2}\chit(m^2) -\f{\theta^4}{24}\bbqq(m^2)+ O(m^6\theta^6)\,,
\end{equation}
where to all orders in $m^2$ [see also
Eqs.~\eqref{eq:coeff_real1_quater_bis} and \eqref{eq:secondorder_06}]
\begin{equation}
  \label{eq:topconst10_ter}
  \begin{aligned}
    \f{ \chit(m^2)}{m^2}
    &=  \f{1}{2}\left(\coeffi_u(m^2) -\coeffi_w(m^2)\right)      - 2m^2 \coeffi_{\tilde{u}^2}(m^2)\,,\\
    \f{ \bbqq(m^2)}{m^4}
    &=  \f{\chit(m^2)}{m^4} + \f{3}{4}\left[     \coeffi_{uu}(m^2)-2\coeffi_{uw}(m^2)\right.\\
    &\phantom{=}\left. + \coeffi_{ww}(m^2)-8\coeffi_{\tilde{u}^2}(m^2)  \right. \\
    &\phantom{=}   \left. -8m^2\left(  \coeffi_{u\tilde{u}^2}(m^2)-\coeffi_{w\tilde{u}^2}(m^2) \right) \right. \\
    &\phantom{=}   \left. + 16m^4 \coeffi_{\tilde{u}^2\tilde{u}^2}(m^2)  \right]\,.
  \end{aligned}
\end{equation}
One finds for the coefficient $\bt$~\cite{DelDebbio:2004vxo} [see
Eq.~\eqref{eq:secondorder1_4}]
\begin{equation}
  \label{eq:topconst10_quinquies}
  \begin{aligned}
    \bt &\equiv -\f{1}{12}\f{\bbqq(m^2)}{\chit(m^2)}  \\
        &= -\f{1}{12}\left[1+ \f{3m^2}{\Delta
          + O(m^2)}\!\left(\left.\de_{m^2}\f{\chit}{m^2}\right|_{m= 0}
          - \Delta_2 \vphantom{\f{2}{m^2}\Iqiv[\ffh]}\right) \right. \\
        &\phantom{=} \left.\vphantom{\f{3m^2}{\Delta}}
          +  O\left(\f{m^4}{\Delta + O(m^2)}\right) \right] \,.
  \end{aligned}
\end{equation}
If $\Delta\neq 0$, to leading order one finds the ideal-gas result
$\bt = -\f{1}{12}$.  If $\Delta=0$ the second term in square brackets
is generally $O(1)$ and one does not find an ideal-gas behavior.

Since QCD at the physical point is rather close to the chiral limit,
the results above lead one to expect the onset of an instanton
gas-like behavior for the topological charge not far above the chiral
crossover temperature, $\mathrm{T}_c$, similarly to what has been
observed in the pure gauge case~\cite{Bonati:2013tt}, if
$\mathrm{U}(1)_A$ remains effectively broken.  The results of
Ref.~\cite{Bonati:2015vqz}, however, indicate the persistence of large
deviations of $\bt$ from $-1/12$ up to $\mathrm{T}\sim 2\mathrm{T}_c$
on fine lattices. It is possible, of course, that the large deviation
is actually physical, which would indicate that either
$\mathrm{U}(1)_A$ is effectively restored in the chiral limit, or
that, although $\mathrm{U}(1)_A$ is effectively broken, the
coefficient of the $O(m^2)$ correction is large. On the other hand, it
is known that taking the continuum limit of topology-related
observables is difficult, and algorithmic improvements have led to
significant revisions of some of the results of
Ref.~\cite{Bonati:2015vqz} (see
Refs.~\cite{Bonati:2018blm,Athenodorou:2022aay, Bonanno:2024zyn}), so
it seems safe to say that the situation is not yet settled.

To avoid misunderstandings, and as already pointed out in
Ref.~\cite{Kanazawa:2014cua}, it is worth stressing that the analysis
above by no means imply that if $\mathrm{U}(1)_A$ is effectively
broken in the symmetric phase, then the relevant topological degrees
of freedom in the chiral limit are the usual instantons and
anti-instantons, or more precisely their finite-temperature analogs,
i.e., calorons and
anti-calorons~\cite{Harrington:1978ve,Harrington:1978ua,Kraan:1998kp,
  Kraan:1998pm,Kraan:1998sn,Lee:1997vp,Lee:1998vu,Lee:1998bb,
  GarciaPerez:1999ux,Chernodub:1999wg,Diakonov:1995ea,Schafer:1996wv,
  Diakonov:2009jq}. This also means that the required instanton
gas-like behavior need not be the same behavior found in the usual
semiclassical dilute instanton gas~\cite{Gross:1980br,
  Boccaletti:2020mxu}. What is required is that the topological
properties of the system can be described (at least formally) in terms
of effective degrees of freedom corresponding to objects carrying unit
topological charge, of vanishingly small density, and fluctuating
independently of each other.

If localized objects of integer charge (of some sort) were indeed the
relevant topological degrees of freedom, thanks to the index theorem
they would support exact chiral zero modes if isolated from each
other. In the random matrix model for the low-lying Dirac spectrum of
Ref.~\cite{Kovacs:2023vzi}, based on a dilute instanton gas
interacting via the fermionic determinant, the mixing of these modes
leads to a singular peak in the spectral density. This, in turn,
affects the instanton density through the fermionic determinant
leading to $\chit \propto m^2$. Both these effects lead to effective
$\mathrm{U}(1)_A$ breaking in the chiral limit, fulfilling the
constraint Eq.~\eqref{eq:firstorder_again2_bis} in a nontrivial
way. The presence of a dilute gas of instanton-like objects in typical
gauge configurations is then likely to be a sufficient condition for
effective $\mathrm{U}(1)_A$ breaking in the symmetric phase.

For completeness, I conclude this section discussing briefly the
spontaneously broken phase. In this case it is convenient to stick to
the expansion of $F(\theta;m)$ in powers of $j_{S,P}(\theta;m)$,
Eq.~\eqref{eq:topconst11}. This expression contains only
susceptibilities involving the scalar and pseudoscalar singlet
operators, that are expected to remain finite in the chiral limit
since the corresponding particles (i.e., $\sigma$ and $\eta$) should
remain massive. To leading order in $m$ one finds then for small
$\theta$
\begin{equation}
  \label{eq:topconst12}
  \begin{aligned}
    F(\theta;m)-  F(0;m) &=  - j_S(\theta;m)\de_{j_S}\wiv(m)|_0  +o(m)\\
                         &= |m|\left(1-\cos\tf{\theta}{2}\right)\Sigma(0^+)+o(m)\,,
  \end{aligned}
\end{equation}
that provides the correct expression for the leading behavior in the
chiral limit of all the $\theta$ derivatives of $F(\theta;m)$ at
$\theta=0$. One has then $\chit= \f{1}{4}|m| \Sigma(0^+) + o(m)$ and
$\bt= -1/48 +o(m)$, up to corrections that vanish in the chiral limit,
matching the expectations obtained using chiral
Lagrangians~\cite{DiVecchia:1980yfw,Leutwyler:1992yt}. Amusingly,
Eq.~\eqref{eq:topconst12} is equal to the free energy density of an
ideal gas of topological objects of charge $\pm\f{1}{2}$, of equal
densities $\f{1}{2}|m|\Sigma(0^+)$. This effective description,
however, holds only in the vicinity of $\theta=0$, as
Eq.~\eqref{eq:topconst12} has no reason to be valid for larger
$\theta$, and even lacks the required $2\pi$-periodicity.  Enforcing
this property one finds
$F(\theta;m)- F(0;m) = |m|\left(1- |\cos\tf{\theta}{2}|\right)
\Sigma(0^+)$ to leading order in $m$, in agreement with the analysis
of Refs.~\cite{DiVecchia:1980yfw,Leutwyler:1992yt}, although the
present derivation guarantees the correctness of this expression only
in the vicinity of $\theta= 0 \mod 2\pi$.

\section{Conclusions}
\label{sec:concl1}

The nature of the chiral transition in the $N_f=2$ chiral limit of a
gauge theory and the fate of the anomalous $\mathrm{U}(1)_A$ symmetry
in the symmetric phase are still open problems in spite of extensive
investigations, both analytical~\cite{Pisarski:1983ms,
  Rajagopal:1992qz,Rajagopal:1993ah,Shuryak:1993ee,Butti:2003nu,
  Basile:2004wa,Pelissetto:2013hqa,Pisarski:2024esv,Giacosa:2024orp,
  Meggiolaro:1992wc,Marchi:2003wq,Meggiolaro:2013swa,
  Meggiolaro:2019rkl,Carabba:2021xmc,Fejos:2022mso,Fejos:2024bgl,
  Bernhardt:2023hpr,Cohen:1996ng,Evans:1996wf,Lee:1996zy,
  Cohen:1997hz,Aoki:2012yj,Kanazawa:2015xna,Kanazawa:2014cua,
  Azcoiti:2016zbi,Azcoiti:2017jsh,Azcoiti:2019moz, Azcoiti:2021gst,
  Azcoiti:2023xvu,Giordano:2020twm,Giordano:2021nat,Giordano:2022ghy,
  Giordano:2024jnc,Giordano:2024awb} and
numerical~\cite{HotQCD:2019xnw,
  Buchoff:2013nra,Bhattacharya:2014ara,Gavai:2024mcj,
  Chandrasekharan:1998yx,Dick:2015twa,Ding:2020xlj,Kaczmarek:2021ser,
  Kaczmarek:2023bxb,HotQCD:2012vvd,Aoki:2020noz,JLQCD:2024xey,
  Cossu:2013uua,Tomiya:2016jwr,Brandt:2016daq,Edwards:1999zm,
  Alexandru:2015fxa,Kovacs:2017uiz,Aoki:2020noz,Alexandru:2019gdm,
  Kaczmarek:2021ser,Vig:2021oyt,Kovacs:2021fwq,Meng:2023nxf,
  Kaczmarek:2023bxb,Alexandru:2024tel,Kovacs:2023vzi,Fodor:2025yuj}. In
this paper I have revisited the first-principles approach to this
problem based on the study of the Dirac
spectrum~\cite{Cohen:1997hz,Aoki:2012yj,Kanazawa:2015xna,
  Azcoiti:2023xvu,Giordano:2024jnc,Giordano:2024awb} using lattice
gauge theory and chiral (Ginsparg--Wilson)
fermions~\cite{Ginsparg:1981bj,Hasenfratz:1998ri,Hasenfratz:1998jp,
  Luscher:1998pqa,Kikukawa:1998py,Horvath:1999bk,Kaplan:1992bt,
  Shamir:1993zy,Furman:1994ky,Borici:1999zw,Chiu:2002ir,Brower:2005qw,
  Brower:2012vk,Narayanan:1993sk,Narayanan:1993ss,Neuberger:1997fp,
  Neuberger:1998wv,Hasenfratz:1993sp,Bietenholz:1995cy,DeGrand:1995ji,
  Hasenfratz:1998ri,Hasenfratz:1998jp,Chandrasekharan:1998wg,
  Niedermayer:1998bi,Azcoiti:2010ns,Giordano:2023spj,
  Hasenfratz:2002rp,Luscher:2004fu,Alexandrou:2000kj,Giusti:2004qd},
putting its foundations on solid ground and developing it in full
generality. The main results are the following.

(1.)\ I have proved that chiral symmetry is restored at the level of
scalar and pseudoscalar susceptibilities if and only if these are
finite (i.e., non-divergent) in the chiral limit; or equivalently, if
and only if susceptibilities involving an even number of isosinglet
scalar and pseudoscalar bilinears are $m^2$-differentiable (i.e.,
functions of $m^2$ infinitely differentiable at $m=0$), and $m$ times
an $m^2$-differentiable function if this number is odd
(Sec.~\ref{sec:chisymrest_nsc}). A symmetry being manifest at the
level of susceptibilities is a general property of a quantum field
theory within the symmetric phase, where the correlation length is
finite, and so the finiteness of scalar and pseudoscalar
susceptibilities in the chiral limit is a general property of the
chirally symmetric phase of a gauge theory.

(2.) Under the extended assumption that chiral symmetry is restored in
susceptibilities involving scalar and pseudoscalar bilinears and
general, possibly nonlocal operators containing only gauge fields
(nonlocal restoration), I have proved that also the spectral density
of the Dirac operator and similar spectral quantities are
$m^2$-differentiable in the symmetric phase
(Sec.~\ref{sec:gauge_ext}). This follows also if chiral symmetry is
restored in scalar and pseudoscalar susceptibilities involving
additional bilinears of external fermion fields
(Appendix~\ref{sec:rhochirest}).  These extended assumptions are still
based on the essential features of symmetry restoration in a quantum
field theory. Together with those in (1.), these results essentially
turn the analyticity assumptions of
Refs.~\cite{Cohen:1997hz,Aoki:2012yj,Kanazawa:2015xna,
  Azcoiti:2023xvu,Kanazawa:2014cua} into a necessary consequence of
symmetry restoration.

(3.)\ After making simplifying assumptions on the
Ginsparg--Wilson--Dirac operator that are met by its most common
realizations, I have obtained an explicit expression for the
generating function of scalar and pseudoscalar susceptibilities in
terms of the Dirac eigenvalues (Sec.~\ref{sec:det_sources}). Imposing
finiteness of these susceptibilities I have then obtained a set of
constraints on the Dirac spectrum (Sec.~\ref{sec:Dirac_constr}). In
particular, I have shown that the only constraints involving the
spectral density directly are those coming from finiteness in the
chiral limit of the pion susceptibility, $\chi_\pi$, that implies also
finiteness of the delta susceptibility, $\chi_\delta$; and from
$\f{\chi_\pi-\chi_\delta}{4}-\f{\chit}{m^2}=O(m^2)$, with $\chit$ the
topological susceptibility, implying
$\Delta=\lim_{m\to 0}\f{\chi_\pi-\chi_\delta}{4}=\lim_{m\to
  0}\f{\chit}{m^2}$ (Sec.~\ref{sec:constr_first}).  These constraints
are generally compatible with both effective breaking and effective
restoration of $\mathrm{U}(1)_A$.  Moreover, I have proved that
$\f{\chit}{m^2}$ is $m^2$-dif\-ferentiable in the symmetric phase, and
obtained a lower bound on $\chi_\pi-\chi_\delta$ that shows the
impossibility of effective $\mathrm{U}(1)_A$ restoration at nonzero
$m$ (Sec.~\ref{sec:constr_first}).

(4.)\ I have also obtained further constraints involving two-point
eigenvalue correlation functions, showing that the correlations among
near-zero complex modes, and between zero and complex modes, are
closely connected with topology and the fate of $\mathrm{U}(1)_A$. In
particular, an order parameter for $\mathrm{U}(1)_A$ is related to the
second derivative of $\chi_\pi+\chi_\delta$ with respect to the
``vacuum angle'' $\theta$ at $\theta=0$, which in turn is determined
by the correlation between zero and near-zero modes
(Sec.~\ref{sec:constr_second}).

(5.)\ I have shown that in the chiral limit the cumulants of the
topological charge must be identical to those found in an ideal gas of
instantons and anti-instantons of vanishingly small total density
$\chit\propto m^2$, to leading order in $m$, if $\mathrm{U}(1)_A$
remains effectively broken by $\Delta\neq 0$ (Sec.~\ref{sec:top_iig}).

The results in (1.)\ and (2.)\ are of very general nature, based only
on the properties expected of the symmetric phase of a quantum field
theory, and on the symmetry properties of a gauge theory with two
degenerate fermions. These results allow one to use the Dirac spectrum
to study chiral symmetry restoration and the fate of $\mathrm{U}(1)_A$
in a systematic and truly first-principles way. They also subsume and
extend the approach of Ref.~\cite{Aoki:2012yj}, allowing one to obtain
information not only on the spectral density, but also on the
correlations among eigenvalues (Sec.~\ref{sec:criticism0}).

The results discussed in (3.)\ and (4.)\ are also very general. In
particular, the representation of the generating function in terms of
spectral quantities applies independently of the status of chiral
symmetry. When translating the finiteness of susceptibilities in the
symmetric phase into constraints on the Dirac spectrum, no assumption
is made on how the various spectral quantities depend on the position
in the spectrum or on $m$; in particular, the assumption of nonlocal
restoration is not used.

Finally, (5.)\ is completely general, requiring only the finiteness
condition (1.)\ on susceptibilities and the known (anomalous) symmetry
properties of gauge theories under $\mathrm{U}(1)_A$ transformations,
besides the effective breaking of $\mathrm{U}(1)_A$ by $\Delta\neq 0$.
This result was already proved in Ref.~\cite{Kanazawa:2014cua} using
an effective-theory approach under the assumption of $m^2$-analyticity
of the free energy density in the presence of a $\theta$ term.  Here
the assumptions of Ref.~\cite{Kanazawa:2014cua} are put on firmer
ground, and the emergence of an ideal instanton-gas behavior in the
chiral limit if $\mathrm{U}(1)_A$ remains effectively broken is shown
rigorously, allowing also the study of corrections to the ideal gas
behavior.  Also in this case the assumption of nonlocal restoration
is not needed.

The results of this paper set the stage for a detailed study of the
Dirac spectrum in the symmetric phase of a gauge theory, once that
more detailed properties of the spectral density and eigenvalue
correlators are taken into account. This will be the purpose of the
second paper of this series.

\begin{acknowledgments}
  I thank V.~Azcoiti, C.~Bonanno, G.~Endr{\H o}di, I.~Horv\'ath,
  S.~D.~Katz, D.~N\'ogr\'adi, A.~Patella, A.~P\'asztor, Zs.~Sz\'ep,
  and especially T.~G.~Kov{\'a}cs for discussions. This work was
  partially supported by the NKFIH grants K-147396, NKKP Excellence
  151482, and TKP2021-NKTA-64.
\end{acknowledgments}

\appendix

\section{Connected correlation functions}
\label{sec:conncorrfunc}

\subsection{Recursive formula}
\label{sec:ccf_rec}

Consider families of observables of the general form
$A_{\vec{s}}^{(k)}(\Lambda_k)$, labeled by discrete indices
$\vec{s}=(s_1,\ldots,s_d)\in\mathbb{N}_0^d$ and $k\in\mathbb{N}_0$,
and by continuous variables
$\Lambda_k=\{\lambda_1,\ldots,\lambda_k\}$,
$\lambda_j\in I\subseteq \mathbb{R}$. Denote
$\vec{k}\equiv(\vec{s},k)$. Connected correlation functions for these
observables are defined recursively from the correlation functions
$\la A_{\vec{s}}^{(k)}(\Lambda_k) \ra$ via
\begin{equation}
  \label{eq:part_g1}
  \begin{aligned}
    \left\la A_{\vec{s}}^{(k)}(\Lambda_k)\right\ra_c
    &\equiv \left\la A_{\vec{s}}^{(k)}(\Lambda_k) \right\ra \\
    &\phantom{=} - \sum_{\substack{\pi\in\Pi(S_{\vec{k}})  \\ |\pi|>1}}
    \prod_{p\in \pi}   \left\la A_{\vec{s}(p)}^{(k(p))}\left(\Lambda_{k(p)}(p)\right) \right\ra_c\,,
  \end{aligned}
\end{equation}
where the sum runs over the partitions $\Pi(S_{\vec{k}})$ of a set
$S_{\vec{k}}$ of $s_1,s_2,\ldots, s_d, k$ elements of type
$1,2,\ldots, d,d+1$ into parts, $p$ (i.e., disjoint subsets whose
union is the whole set), containing $s(p)_j$ elements of type
$j=1,\ldots d$, and the $k(p)$ elements of type $d+1$ in the subset
$\Lambda_{k(p)}(p)\subseteq\Lambda_k$. Only nontrivial partitions are
included, i.e., the number of parts, $|\pi|$, obeys $|\pi|>1$. Clearly
$\sum_{p\in\pi} \vec{s}(p)=\vec{s}$, $\sum_{p\in\pi} k(p)=k$, and
$\cup_{p\in\pi} \Lambda_{k(p)}(p)=\Lambda_k$. Integrating over
$\Lambda_k$,
\begin{equation}
  \label{eq:intA}
  A_{\vec{k}}[g]  \equiv \int_I d\lambda_1\,g(\lambda_1)\ldots
  \int_I d\lambda_k\,g(\lambda_k)\, A_{\vec{s}}^{(k)}(\Lambda_k)\,,
\end{equation}
with $g$ continuous (and integrable over $I$), one obtains quantities
of the same general form, i.e.,
$A_{\vec{k}}[g]= A_{\vec{s}^{\,\prime}}^{\prime (k')}$ with
$\vec{s}^{\,\prime}=\vec{k}$ and $k'=0$ (and with trivial dependence
on $\Lambda_k$). The corresponding connected correlation functions
\begin{equation}
  \label{eq:intA_c}
  \left\la A_{\vec{k}}[g]\right\ra_c \equiv \left\la
    A_{\vec{k}}[g] \right\ra - \sum_{\substack{\pi\in\Pi(S_{\vec{k}})\\ |\pi|>1}}
  \prod_{p\in \pi} \left\la A_{\vec{k}(p)}[g] \right\ra_c\,,
\end{equation}
defined according to the general rule, Eq.~\eqref{eq:part_g1}, equal
for $k\ge 1$
\begin{equation}
  \label{eq:intA_c2}
  \left\la A_{\vec{k}}[g]\right\ra_c = \int_I d\lambda_1\,g(\lambda_1)\ldots
  \int_I d\lambda_k\,g(\lambda_k)\left\la
    A_{\vec{s}}^{(k)}(\Lambda_k)\right\ra_c \,.
\end{equation}
The proof by induction is straightforward.

\subsection{Partition function}
\label{sec:ccf_pf}

The partition function associated with the correlation functions
$\left\la A_{\vec{k}}[g]\right\ra$ is
\begin{equation}
  \label{eq:part_g2_bis}
  Z(t) = 1+ \sum_{\vec{k}\neq \vec{0}} \left(\prod_{j=1}^{d+1}\f{t_j^{k_j}}{k_j!}\right)
  \left\la A_{\vec{k}}[g]\right\ra\,,
\end{equation}
where $t$ denotes collectively the complex variables
$t_1,\ldots,t_{d+1}$.  Using the decomposition in connected
components,
\begin{equation}
  \label{eq:part_g2_bis1}
  Z(t) =1 + \sum_{\vec{k}\neq \vec{0}} \left(\prod_{j=1}^{d+1}\f{t_j^{k_j}}{k_j!}\right)
  \sum_{\pi \in\Pi(S_{\vec{k}})} \prod_{p\in \pi}\left\la A_{\vec{k}(p)}[g]\right\ra_c\,;
\end{equation}
taking into account that contributions to $Z$ are entirely
characterized by an integer-valued, non-identically vanishing function
$m(\vec{\sigma}) \in\mathbb{N}_0$, $m(\vec{\sigma})\not\equiv 0$,
counting the number of parts labeled by $\vec{\sigma}\neq\vec{0}$;
and taking into account that there are
$\prod_{j=1}^{d+1} k_j!\big/ \left[\prod_{\vec{\sigma}\neq\vec{0}}
  m(\vec{\sigma})!
  \left(\prod_{j=1}^{d+1}\sigma_j!\right)^{m(\vec{\sigma})}\right]$
partitions giving the same contribution, one finds
\begin{equation}
  \label{eq:part_g2_bis2}
  Z(t)  =\exp\left\{\sum_{\vec{\sigma}\neq 0}
    \left(\prod_{j=1}^{d+1}\f{t_j^{\sigma_j}}{\sigma_j!}\right)
    \left\la A_{\vec{\sigma}}[g]\right\ra_c\right\} \,.
\end{equation}
The connected correlation functions
$\left\la A_{\vec{k}}[g]\right\ra_c$ are then obtained by
differentiation with respect to the $t_j$ at $t_{1}=\ldots = t_{d+1}=0$,
and $\left\la A_{\vec{s}}^{(k)}(\Lambda_k)\right\ra_c$ are further
obtained by functional differentiation of these quantities with
respect to $g(\lambda_1),\ldots, g(\lambda_k)$.

\subsection{Relations between cumulants}
\label{sec:ccf_cr}

For the observables 
\begin{equation}
  \label{eq:ccf_cr_obs}
  \begin{aligned}
    A_{\vec{s}}^{(k)}(\Lambda_k)
    &=\left[\prod_{i=1}^d O_i^{s_i} \right]\gamma^{(k)}(\Lambda_k)\,,\\
    A_{\vec{s}}^{\prime (k)}(\Lambda_k)
    &=\left[\prod_{i=1}^d O_i^{\prime s_i}\right] \gamma^{(k)}(\Lambda_k)\,,
  \end{aligned}
\end{equation}
with $O_i=\sum_{j=1}^d\mathcal{C}_{ij}O_j'$, one has
$ A_{\vec{s}}^{(k)}= \sum_{\vec{s}^{\,\prime}}
C_{\vec{s}\vec{s}^{\,\prime}} A_{\vec{s}^{\,\prime}}^{\prime(k)} $ for
suitable $C_{\vec{s}\vec{s}^{\,\prime}}$.  One has for the
corresponding partition functions
\begin{equation}
  \label{eq:ccf_linrel}
  \begin{aligned}
    Z[t;g]
    &= \sum_{K,k=0}^\infty\f{1}{K!k!} \left\la \left(\sum_{i=1}^d t_i O_i  \right)^{K}
      \Gamma^{(k)}[g]\right\ra \\
    & = \sum_{K,k=0}^\infty \f{1}{K!k!} \left\la \left(\sum_{i=1}^d  t_i' O'_i  \right)^{K}
      \Gamma^{(k)}[g]\right\ra\\
    &= Z'[t';g]  \,,
  \end{aligned}
\end{equation}
with $t_i'=\sum_{j=1}^d \mathcal{C}_{ji} t_j$, $1\le i\le d$, and
\begin{equation}
  \label{eq:ccf_linrel2}
  \Gamma^{(k)}[g] \equiv \int_I d\lambda_1\,g(\lambda_1)\ldots
  \int_I d\lambda_k\,g(\lambda_k)\,\gamma^{(k)}(\lambda_1,\ldots,\lambda_k)\,.
\end{equation}
(Here I set $t_{d+1}= t_{d+1}'=1$ without loss of generality.) Full
correlation functions for the two sets of quantities are obtained by
applying the differential operator
$\mathrm{D}_{\vec{s}}\equiv \prod_{i=1}^d\de_{t_i}^{s_i}$ or
$\mathrm{D}'_{\vec{s}}\equiv \prod_{i=1}^d\de_{t'_i}^{s_i} $ to $Z$ or
$Z'$ and then setting $t_i=0$ or $t_i'=0$, $\forall i$,
respectively. Connected correlation functions are similarly obtained
by applying the same operators to $\ln Z$ or $\ln Z'$.  Since
$\mathrm{D}_{\vec{s}} =
\prod_{i=1}^d\left(\sum_{j=1}^d\mathcal{C}_{ji}\de_{t'_j}\right)^{s_i}
= \sum_{\vec{s}^{\,\prime}} C_{\vec{s}\vec{s}^{\,\prime}}
\mathrm{D}'_{\vec{s}^{\,\prime}}$, it follows that the same linear
relation holds between the connected correlation functions of
$A_{(\vec{s},k)}[g]=\left[\prod_{i=1}^d O_i^{s_i}\right]
\Gamma^{(k)}[g]$ and
$A_{(\vec{s},k)}'[g]= \left[\prod_{i=1}^d O_i^{\prime s_i}\right]
\Gamma^{(k)}[g]$ as between their full correlation functions and
between the observables themselves,
\begin{equation}
  \label{eq:ccf_linreal_final}
  \begin{aligned}
    \la A_{(\vec{s},k)}[g] \ra
    &= \sum_{\vec{s}^{\,\prime}} C_{\vec{s}\vec{s}^{\,\prime}}
      \la A_{(\vec{s}^{\,\prime}\!,k)}'[g]\ra\,, \\
    \la A_{(\vec{s},k)}[g] \ra_c
    &=  \sum_{\vec{s}^{\,\prime}} C_{\vec{s}\vec{s}^{\,\prime}}
      \la A_{(\vec{s}^{\,\prime}\!,k)}'[g]\ra_c\,.
  \end{aligned}
\end{equation} 
By functional differentiation one sees that the same holds for the
correlation functions of $ A_{\vec{s}}^{(k)}(\Lambda_k)$ and
$A_{\vec{s}}^{\prime(k)}(\Lambda_k) $.

If instead the following relation holds between the integrated
quantities $A_{(\vec{s},k)}[g]$ and $A_{(\vec{s},k)}'[g]$,
\begin{equation}
  \label{eq:ccf_sti1}
  \begin{aligned}
    A_{(\vec{s},k)}[g]
    &=  \sum_{\vec{s}^{\,\prime}}  C_{\vec{s}\vec{s}^{\,\prime}}A'_{(\vec{s}^{\,\prime}\!,k)}[g]  \,, \\
    C_{\vec{s}\vec{s}^{\,\prime}}
    &= \prod_{i=1}^d c^{(i)}_{ss'}\,,\qquad
      c^{(i)}_{ss'} = \left.\left(\f{d}{dx}\right)^{s}\f{\omega_i(x)^{s'}}{s'!}\right|_{x=0}\,,
  \end{aligned}
\end{equation}
for some functions $\omega_i(x)$ with $\omega_i(0)=0$, $i=1,\ldots,d$,
then
\begin{equation}
  \label{eq:ccf_sti2}
  \begin{aligned}
    Z[t;g] &=  \sum_{k=0}^\infty \sum_{\vec{s}}
             \left(\prod_{j=1}^{d}\f{t_j^{s_j}}{s_j!}\right)
             \left\la A_{(\vec{s},k)}[g]
             \right\ra  \\
           &= \sum_{k=0}^\infty \sum_{\vec{s}^{\,\prime}}
             \left(\prod_{j=1}^{d} \f{\omega_j(t_j)^{s_j'}}{s_j'!}\right)
             \left\la A'_{(\vec{s}^{\,\prime}\!,k)}[g] \right\ra  \\
           & = Z'[\omega(t);g]
             \,, 
  \end{aligned}
\end{equation}
and so one finds for any function $\Omega(Z)$ 
\begin{equation}
  \label{eq:ccf_sti3}
  \begin{aligned}
    & \left. \mathrm{D}_{\vec{s}}\,
      \Omega\left(Z[t;g]\right)\right|_{t=0}  
      = \left.\mathrm{D}_{\vec{s}}\,
      \Omega\left(Z'[\omega(t);g]\right)\right|_{t=0}  \\
    &=   \sum_{\vec{s}^{\,\prime}}
      \left.       \mathrm{D}_{\vec{s}}\,  \left(\prod_{j=1}^{d}\f{\omega_j(t_j)^{s_j'}}{s_j'!}\right)\right|_{t=0}
      \left.\mathrm{D}'_{\vec{s}^{\,\prime}} \Omega\left(Z'[t';g]\right)\right|_{t'=0}\\
    &=     \sum_{\vec{s}^{\,\prime}}C_{\vec{s}\vec{s}^{\,\prime}}
      \left. \mathrm{D}'_{\vec{s}^{\,\prime}}
      \Omega\left(Z'[t';g]\right)\right|_{t'=0}  \,,
  \end{aligned}
\end{equation}
with $\mathrm{D}_{\vec{s}}$ and $\mathrm{D}'_{\vec{s}}$ defined as above,  
so that both full and connected correlation functions of
$A_{(\vec{s},k)}$ and $A'_{(\vec{s}^{\,\prime}\!,k)}$ are linearly
related by the same $C_{\vec{s}\vec{s}^{\,\prime}}$, Eq.~\eqref{eq:ccf_sti1}. By
functional differentiation, one shows that the same holds for
$A_{\vec{s}}^{(k)}(\Lambda_k)$ and
$A_{\vec{s}}^{\prime(k)}(\Lambda_k)$ and their full and connected
correlation functions.

\subsection{Relevant observables}
\label{sec:ccf_relob}

The correlation functions considered in this paper involve observables
of the form [see Eqs.~\eqref{eq:densities} and \eqref{eq:rho_def4},
and under Eq.~\eqref{eq:largeV_cu4} for notation]
\begin{equation}
  \label{eq:obs0}
    A_{\vec{s}}^{(0)}= S^{n_S}  (i\vec{P})^{\vec{n}_{P}}  (iP)^{n_P} \vec{S}^{\vec{n}_{S}} \,, 
\end{equation}
with $\vec{s}=(n_S,\vec{n}_P,n_P,\vec{n}_S)$, where additional
functionals $G_i$ of gauge fields only may be included; and
observables of the form
\begin{equation}
  \label{eq:obs1}
    A_{\vec{s}}^{(k)}    (\lambda_1,\ldots,\lambda_k)
    = N_+^{n_+} N_-^{n_-}\rho_U^{(k)}(\lambda_1,\ldots,\lambda_k)\,, 
\end{equation}
with $\vec{s}=(n_+,n_-)$, including the case $k=0$ where
$\rho_U^{(0)}\equiv 1$, or their linear combinations
\begin{equation}
  \label{eq:obs2}
    A_{\vec{s}}^{(k)} (\lambda_1,\ldots,\lambda_k)  
    =  N_0^{n_0} Q^{n_1}\rho_U^{(k)}(\lambda_1,\ldots,\lambda_k)\,, 
\end{equation}
with $\vec{s}=(n_0,n_1)$, or the nonlinear combinations
\begin{equation}
  \label{eq:obs1bis}
    A_{\vec{s}}^{(k)} (\lambda_1,\ldots,\lambda_k)
    =  s_{n_1}(N_+) s_{n_2}(N_-)\rho_U^{(k)}(\lambda_1,\ldots,\lambda_k)\,, 
\end{equation}
with $\vec{s}=(n_1,n_2)$ and $s_n(t)$ defined in Eq.~\eqref{eq:spol},
and the integrals of the quantities in
Eqs.~\eqref{eq:obs1}--\eqref{eq:obs1bis} over
$\lambda_1,\ldots,\lambda_k$. Finally, in the discussion in
Appendix~\ref{sec:rhochirest}, observables of the following more
general form are involved, namely
\begin{equation}
  \label{eq:obs3}
  \begin{aligned}
    A_{\vec{s}}^{(k)}(\lambda_1,\ldots,\lambda_k)  
    &= S^{n_S}  (i\vec{P})^{\vec{n}_{P}}  (iP)^{n_P} \vec{S}^{\vec{n}_{S}} \\
    &\phantom{=}\times N_+^{n_+} N_-^{n_-}\rho_U^{(k)}(\lambda_1,\ldots,\lambda_k)\,, 
  \end{aligned}
\end{equation}
with $\vec{s}=(n_S,\vec{n}_P,n_P,\vec{n}_S,n_+,n_-)$,
or their linear combinations
\begin{equation}
  \label{eq:obs3_1}
  \begin{aligned}
    A_{\vec{s}}^{(k)}(\lambda_1,\ldots,\lambda_k)  
    &= S^{n_S}  (i\vec{P})^{\vec{n}_{P}}  (iP)^{n_P} \vec{S}^{\vec{n}_{S}}  \\
    &\phantom{=}\times  N_0^{n_0} Q^{n_1}\rho_U^{(k)}(\lambda_1,\ldots,\lambda_k)\,,
  \end{aligned}
\end{equation}
with $\vec{s}=(n_S,\vec{n}_P,n_P,\vec{n}_S,n_0,n_1)$. These
observables are precisely of the form discussed above in
Appendix~\ref{sec:ccf_rec}, and the scalar and pseudoscalar connected
correlation functions, Eq.~\eqref{eq:genfuncdef2}, and the spectral
correlators in Eqs.~\eqref{eq:rho_def5_0} and \eqref{eq:rho_def8_gen}
for $O=N_+^{n_+}N_-^{n_-}$, $O=N_0^{n_0}Q^{n_1}$, or
$O=s_{n_1}(N_+) s_{n_2}(N_-)$, are then defined by
Eq.~\eqref{eq:part_g1}.  Moreover, full and connected correlation
functions of the quantities in Eqs.~\eqref{eq:obs1} and
\eqref{eq:obs2} [and those of the quantities in Eqs.~\eqref{eq:obs3}
and \eqref{eq:obs3_1}] are related by the same linear transformation.
The same applies to the correlation functions of their integrals over
$\lambda_1,\ldots, \lambda_k$,
$I^{(n_{\smash{3}})}_{N_{\smash{+}}^{\smash{k_{\smash{1}}}}N_{\smash{-}}^{\smash{k_{\smash{2}}}}}$
and
$I^{(n_{\smash{3}})}_{N_{\smash{0}}^{\smash{k_{\smash{1}}}}Q^{\smash{k_{\smash{2}}}}}$
[see Eq.~\eqref{eq:cum_sys_3_0_again}]. In particular,
\begin{equation}
  \label{eq:inpnminq}
  \begin{aligned}
    I^{(n_3)}_{N_+^{k_1}N_-^{k_2}}
    &= \f{1}{2^{k_1+k_2}}  \sum_{j_1=0}^{k_1}\sum_{j_2=0}^{k_2}
      \begin{pmatrix}
        k_1 \\ j_1
      \end{pmatrix}
      \begin{pmatrix}
        k_2 \\ j_2
      \end{pmatrix}
      (-1)^{j_2}\\
    &\phantom{=}\times I_{N_0^{k_1+k_2-j_1-j_2} Q^{j_1+j_2} }^{(n_3)}\,.
  \end{aligned}
\end{equation}
Finally, $A_{\vec{s}}^{(k)}= s_{n_1}(N_+) s_{n_2}(N_-) \rho_U^{(k)}$,
$\vec{s}=(n_1,n_2)$, and
$A_{\vec{s}^{\,\prime}}^{\prime\,(k)} = N_+^{n_+}
N_-^{n_-}\rho_U^{(k)}$, $\vec{s}^{\,\prime}=(n_+,n_-)$, are related by
[recall $s(n,k)=0$ if $k>n$]
\begin{equation}
  \label{eq:ccf_sti4}
  \begin{aligned}
    A_{\vec{s}}^{(k)}(\lambda_1,\ldots,\lambda_k)
    &  =  \sum_{n_+=0}^\infty\sum_{n_-=0}^\infty s(n_1,n_+)s(n_2,n_-)\\
    &\phantom{=}\times A_{\vec{s}^{\,\prime}}^{\prime\,(k)}(\lambda_1,\ldots,\lambda_k)\,.
  \end{aligned}
\end{equation}
Since [Eq.~\eqref{eq:u_func}]
\begin{equation}
  \label{eq:ccf_sti5}
  \sum_{n=0}^\infty s(n,n')\f{x^{n}}{n!} = \f{[\ln (1+x)]^{n'}}{n'!}  = \f{\sti(x)^{n'}}{n'!}\,,
\end{equation}
the relation Eq.~\eqref{eq:ccf_sti4} is of the form
Eq.~\eqref{eq:ccf_sti1} with $\omega_{1}(x) = \omega_{2}(x) = \sti(x)$
[and of course with $A_{(\vec{s},k)}$ and
$A_{(\vec{s}^{\,\prime}\!,k)}'$ replaced by $A_{\vec{s}}^{(k)}$ and
$A_{\vec{s}}^{\prime(k)}$], and Eq.~\eqref{eq:sti_cu1} follows.

The generating function $\mathcal{Z}/\mathcal{Z}|_0$ of full scalar
and pseudoscalar correlation functions, normalized by the partition
function and expressed in terms of Dirac eigenvalues,
Eq.~\eqref{eq:fdet_final3}, is of the form Eq.~\eqref{eq:part_g2_bis}
with $A_{\vec{k}}[g]= N_+^{n_+} N_-^{n_-}Y_k$, $\vec{k}=(n_+,n_-,k)$,
where
\begin{equation}
  \label{eq:Yk_app}
  Y_k = \int_0^2  d\lambda_1\, X(\lambda_1)\ldots \int_0^2  d\lambda_k\, X(\lambda_k)\,
  \rho_U^{(k)}(\lambda_1,\ldots,\lambda_{k})\,,
\end{equation}
so $g=X$ and $I=[0,2]$, and $t_1=S(X_0)$, $t_2=S(X_0^*)$,
$t_3=1$. From Eq.~\eqref{eq:part_g2_bis2} follows then
Eq.~\eqref{eq:cumexp_simpler2}; taking the logarithm to obtain the
generating function $\mathcal{W}-\mathcal{W}|_0$,
Eq.~\eqref{eq:cumexp_simpler5} follows. Alternatively,
$\mathcal{Z}/\mathcal{Z}|_0$ can be put in the form
Eq.~\eqref{eq:part_g2_bis} with
$A_{\vec{k}}[g]= s_{n_1}(N_+) s_{n_2}(N_-)Y_k$, $\vec{k}=(n_1,n_2,k)$,
$t_1=X_0$, $t_2=X_0^*$, $t_3=1$, from which Eq.~\eqref{eq:cumexp_S}
follows.

\section{Reality of the partition function}
\label{sec:Trefl} 

For $\gamma_5$-Hermitean GW operators with $2R=\mathbf{1}$, the
transformation properties Eq.~\eqref{eq:sources3} imply the reality of
$\mathcal{Z}$, and so of the derivatives of $\mathcal{W}$ at zero
sources.  Setting
$\mathcal{K}(V,W)\equiv j_S\mathbf{1}_{\mathrm{f}} + i
\vec{\jmath}_P\cdot \vec{\sigma} \gamma_5 + i
j_P\mathbf{1}_{\mathrm{f}}\gamma_5 - \vec{\jmath}_S\cdot
\vec{\sigma}$, since
$\mathcal{K}(\mathcal{C}V,-\mathcal{C}W)=\mathcal{K}(V,W)^\dag$ and
$[\gamma_5,\mathcal{K}(V,W)]=0$, from Eq.~\eqref{eq:sources4} and
$\det \gamma_5=1$ one readily finds
\begin{equation}
  \label{eq:Trefl_real1}
  \begin{aligned}
    \mathcal{Z}(\sous,\soup;m)^*
    &= \int DU\, e^{-S_{\mathrm{eff}}(U)}
      \det \left[\vphantom{\tf{1}{2}}D_m(U)\mathbf{1}_{\mathrm{f}} \right.\\
    &\phantom{=}\left. +  \mathcal{K}(V,W)\left(\mathbf{1}-\tf{1}{2}D(U)\right)\right]\,.
  \end{aligned}
\end{equation}
Since $D_m$ is invertible ($\det D_m> 0$), one can exchange the order
of factors in the second term in square brackets by virtue of
Sylvester's theorem, $\det[\mathbf{1} + XY] =\det[\mathbf{1} + YX]$,
and one concludes that
\begin{equation}
  \label{eq:Trefl_real_new3}
  \mathcal{Z}(\sous,\soup;m)^*  =  \mathcal{Z}(\sous,\soup;m) \,.
\end{equation}
Since
$\mathcal{Z}(0,0;m) =\int DU\, e^{-S_{\mathrm{eff}}(U)} \left[\det
  D_m(U)\right]^2 > 0$, it also follows that
$\mathcal{W}(0,0;m)\in \mathbb{R}$.

\section{Proof of $m^2$-differentiability of spectral quantities using
  partially quenched theories}
\label{sec:rhochirest}

In this Appendix I argue that spectral quantities such as the spectral
density $\rho(\lambda;m)$ are $m^2$-differentiable, i.e., finite with
finite $m^2$-derivatives in the chiral limit, under the assumption
that in this limit chiral symmetry becomes manifest in scalar and
pseudoscalar susceptibilities involving not only bilinears built with
the physical fermionic fields, but also their analogs built with
external fermionic fields.  This is done in a partially quenched (PQ)
setup, including both fermion and suitable scalar fields in the
partition function in order for the corresponding determinants to
cancel out exactly.

\subsection{Spectral density}
\label{sec:rhochirest_rho}

Define the following quantities,
\begin{equation}
  \label{eq:specPQ1}
  \Upsilon^{(k)}(z^2;\sous,\soup;m)  \equiv  \lim_{\lvol\to\infty}\lvol^{-1}
  \left\la \left( iP_{a}^{\mathrm{PQ}}\right)^2 \right \ra_{k;\sous,\soup}\,, 
\end{equation}
with $k=0,\pm 1$, where for observables independent of the
light-fermion fields
\begin{equation}
  \label{eq:specPQ1_2}
\begin{aligned}
  \left\la \Oc \right\ra_{k;\sous,\soup}
  &\equiv \f{1}{Z^{(k)}}   \int dU \, e^{-S_{\mathrm{eff}}(U)} \det \Mc(U;\sous,\soup;m)   \\
  &\phantom{=} \times  \int d\omega d\bar{\omega}  \int d\varphi d\varphi^*
    e^{-S^{(k)}(\omega,\bar{\omega}, \varphi,\varphi^*,U;z)}  \\
  &\phantom{=}\times \Oc(\omega,\bar{\omega},\varphi,\varphi^*,U)\,, \\
  Z^{(k)}
  &\equiv  \int dU \,e^{-S_{\mathrm{eff}}(U)} \det\Mc(U;\sous,\soup;m) \\
  &\phantom{\equiv} \times  \int d\omega d\bar{\omega}\int d\varphi d\varphi^*\,
    e^{-S^{(k)}(\omega,\bar{\omega}, \varphi,\varphi^*,U;z)}\,,
\end{aligned}
\end{equation}
with $\Mc$ the fermionic matrix defined in Eq.~\eqref{eq:fdet1}, and
\begin{equation}
  \label{eq:specPQ1_3}
  iP_{a}^{\mathrm{PQ}}(\omega,\bar{\omega},U)
  \equiv i\bar{\omega}\left(\mathbf{1}-\tf{D(U)}{2}\vphantom{D^\dag}\right)\gamma_5\sigma_a\omega\,.
\end{equation}
Here $\omega_{x\alpha c f}$ and $\bar{\omega}_{x\alpha c f}$, $f=1,2$,
are a ``flavor'' doublet of Grassmann field variables, and
$\varphi_{x \alpha c}$ are $c$-number complex field variables,
carrying spacetime coordinate $x$, Dirac index $\alpha=1,\ldots,4$,
and color index $c=1,\ldots,N_c$. Both $\omega$ and $\phi$ transform
in the same representation of the gauge group as the physical
light-fermion fields, and $\bar{\omega}$ (and $\phi^*$) in its complex
conjugate. All indices are suppressed in Eqs.~\eqref{eq:specPQ1_2} and
\eqref{eq:specPQ1_3} and in the following. Moreover, $S^{(k)}$,
$k=0,\pm 1$, are the partially quenched actions
\begin{equation}
  \label{eq:specPQ2}
\begin{aligned}
  &  S^{(k)}(\omega,\bar{\omega},\varphi,\varphi^*,U;z)  \\
  &\equiv  \bar{\omega}\left[D(U) +z\left(\mathbf{1}-\tf{D(U)}{2}
    \vphantom{D^\dag}\right)\right]\mathbf{1}_{\mathrm{f}}\,\omega  \\
  &\phantom{\equiv} +(-i)^k  \varphi^\dag \left[D(U)D(U)^\dag
    + z^2 H(U) \right]\varphi\,,
\end{aligned}
\end{equation}
with $D(U)$ a $\gamma_5$-Hermitean GW Dirac operator with
$2R=\mathbf{1}$, and $H(U)
\equiv\mathbf{1}-\tf{D(U)D(U)^\dag}{4}$. The independence of
$\Upsilon^{(k)}$ of the index $a=1,2,3$ follows from the vector flavor
symmetry of the partially quenched actions. The quantities
$\Upsilon^{(k)}$ provide three representations of the following
resolvent,
\begin{equation}
  \label{eq:specPQ3}
  \begin{aligned}
    G(z^2;\sous,\soup;m)
    & \equiv \lim_{\lvol\to\infty} \lvol^{-1}
      \left\la \mathcal{B}(U;z^2) 
      \right\ra_{\sous,\soup}\,,\\
    \mathcal{B}(U;z^2)
    &\equiv \tr \f{ H(U)}{D(U)D(U)^\dag  + z^2 H(U)}\,,
  \end{aligned}
\end{equation}
where the trace runs over coordinate, Dirac, and color indices, in
terms of expectation values in a partially quenched theory, that are
valid in three different domains of the complex mass $z$ where the
corresponding path integrals are convergent. In Eq.~\eqref{eq:specPQ3}
$\la\ldots\ra_{\sous,\soup}$ denotes the expectation value in the
presence of source terms,
\begin{equation}
  \label{eq:specPQ3_bis}
  \left\la \Oc(U) \right\ra_{\sous,\soup}  \equiv
  \f{\left\la \Oc(U)  \f{\det  \Mc(U;\sous,\soup;m)}{\left[\det D_m(U)\right]^2}  \right\ra }{
    \left\la\f{ \det  \Mc(U;\sous,\soup;m)}{ \left[\det  D_m(U)\right]^2}\right\ra} \,,
\end{equation}
with $\la\ldots\ra$ defined in Eq.~\eqref{eq:general_pf3}. The
relation between $G$ and $\Upsilon^{(k)}$ reads
\begin{equation}
  \label{eq:specPQ4}
  2 G(z^2;\sous,\soup;m)  = \left\{
    \begin{aligned}
      &\Upsilon^{(+1)}(z^2;\sous,\soup;m), &&& \Im z^2 &> 0\,,\\
      & \Upsilon^{(-1)}(z^2;\sous,\soup;m), &&& \Im z^2 &< 0\,,\\
      &\Upsilon^{(0)}(z^2;\sous,\soup;m), &&& \Re z^2 &> 0\,.
    \end{aligned}
  \right.
\end{equation}
The proof of this statement is straightforward, and uses only the
properties of $D$, and the fact that
$\det \left[(-i)^k\mathbf{1}\right] =(-i)^{ 4N_c \lvol k} =1$, for any
$k$. Since $G$ is analytic in the cut complex plane,
$\mathbb{C}\setminus\{\Re z^2\le 0, \Im z^2=0\}$, this shows that each
$\Upsilon^{(k)}$ is analytic in its domain of definition.  The domains
of definition (and analyticity) of $\Upsilon^{(\pm 1)}$ overlap with
that of $\Upsilon^{(0)}$, and the union of the three domains covers
the whole cut complex plane. Writing the trace explicitly in terms of
the eigenvalues of $D$, one finds
\begin{equation}
  \label{eq:specPQ8}
  \begin{aligned}
    G(z^2;\sous,\soup;m)  
    &=  \f{n_0(V,\soup;m)}{z^2}\\
    &\phantom{=} + 2\int_0^2d\lambda \,\rho(\lambda;\sous,\soup;m)\ff(\lambda;z) \,,
  \end{aligned}
\end{equation}
where $\ff$ is defined in Eq.~\eqref{eq:fdet8_shorter_bis},
and\footnote{At $\soup=0$, the symmetry-breaking term
  $mS + \sous\cdot O_\sous=(m+j_S)S+i\vec{\jmath}_P\cdot \vec{P}$ can
  be rotated by a suitable chiral transformation to $m_\sous S$, with
  $m_\sous^2 \equiv (m+j_S)^2 + \vec{\jmath}_P^{\,2}=m^2+u$. In this
  case $\det\Mc$ is then positive-definite, as can also be seen
  directly from its explicit expression,
  Eqs.~\eqref{eq:fdet_final}--\eqref{eq:fdet8_shorter_bis}.  For
  $j_S=0$ and $\vec{\jmath}_P=m_t\hat{\alpha}$, this shows the
  equivalence of a theory including also a twisted-mass term,
  $im_t\hat{\alpha}\cdot \vec{P}$, besides the usual term $mS$, with a
  theory including only the usual mass term but with mass
  $(m^2+m_t^2)^{1/2}$. More generally, since the Dirac spectrum is
  unaffected by the transformation as it depends only on $U$, under
  the usual assumption $N_+N_-=0$ a.e.\ one finds
  $\lim_{\lvol\to\infty}\lvol^{-1}\la N_0\ra_{\sous,0} =0$ also at
  $\sous\neq 0$.}
\begin{equation}
  \label{eq:specPQ10}
  \begin{aligned}
    n_0(\sous,\soup;m)
    &\equiv\lim_{\lvol\to\infty}\lvol^{-1}\la N_0\ra_{\sous,\soup} \,,\\
    \rho(\lambda;\sous,\soup;m)
    &\equiv\lim_{\lvol\to\infty}\lvol^{-1}\left\la
      \rho_U(\lambda)\right\ra_{\sous,\soup}\,, 
  \end{aligned}
\end{equation}
see Eqs.~\eqref{eq:rho_def3_th} and \eqref{eq:bdef}. Instead of $\rho$
it is convenient to use the spectral density $r$ associated with
$x_n \equiv \lambda_n/\sqrt{1-\lambda_n^2/4}$ [see under
Eq.~\eqref{eq:rho_def1} for notation],
\begin{equation}
  \label{eq:specPQ12}
  \begin{aligned}
    r(x;\sous,\soup;m)
    &\equiv\lim_{\lvol\to\infty}\lvol^{-1}\left\la r_U(x)\right\ra_{\sous,\soup}\,,\\
    r_U(x)
    &\equiv \sum_n\delta(x-x_n)\,,
  \end{aligned}
\end{equation}
related to $\rho$ as $\rho = \f{dx}{d\lambda}\, r $. One has
\begin{equation}
  \label{eq:specPQ13}
  \begin{aligned}
    G(z^2;\sous,\soup;m)
    & = \f{n_0(\sous,\soup;m)}{z^2}\\
    &\phantom{=}  + 2\int_0^\infty dx \,\f{r(x;\sous,\soup;m)}{x^2 + z^2} \,.
\end{aligned}
\end{equation}
For $z^2=w +i\epsilon$ with $\epsilon>0 $ and $w\in \mathbb{R}$ one
finds
\begin{equation}
  \label{eq:specPQ14}
  \begin{aligned}
    &  \lim_{\epsilon\to 0} \left[G(w+i\epsilon;\sous,\soup;m)-G(w-i\epsilon;\sous,\soup;m)\right]   \\
    &  = -2\pi i\theta(-w)\f{r(\sqrt{- w};\sous,\soup;m)}{\sqrt{-w}}\,.
\end{aligned}
\end{equation}
The discontinuity of $G$ along the cut yields then the spectral
density.

One makes now the symmetry-restoration assumption that the
susceptibilities involving $i\vec{P}^{\,\mathrm{PQ}}$ as well as
physical scalar and pseudoscalar bilinears are symmetric in the chiral
limit, for arbitrary complex mass $z$ in the domains of definition of
the three partially quenched theories defined by $S^{(k)}$, as well as
on the boundaries of these domains. In the various domains of $z^2$
these susceptibilities are obtained from
$\Upsilon^{(k)}(z^2;\sous,\soup;m)$ by taking derivatives with respect
to the sources at zero sources, and so the symmetry requirement reads
\begin{equation}
  \label{eq:specPQ15}
  \lim_{m\to 0} \left[ \Upsilon^{(k)}(z^2;\sous,\soup;m)-\Upsilon^{(k)}(z^2;R\sous,R\soup;m)\right] =0\,,
\end{equation}
for $ k=0,\pm 1$ and $\forall R\in\mathrm{SO}(4)$. This extends by
analytic continuation, patching the three $\Upsilon^{(k)}$ together,
to
\begin{equation}
  \label{eq:specPQ16}
  \lim_{m\to 0} \left[ G(z^2;\sous,\soup;m)-G(z^2;R\sous,R\soup;m)\right] =0\,,
\end{equation}
$\forall R\in\mathrm{SO}(4)$, with the susceptibilities obtained from
$G(z^2;\sous,\soup;m) $ defined in the whole cut complex
$z^2$-plane. By the same argument as in
Appendix~\ref{sec:chisymrest_nsc}, Eq.~\eqref{eq:specPQ15}, or
equivalently Eq.~\eqref{eq:specPQ16}, imply that these
susceptibilities are finite in the chiral limit, with finite
$m^2$-derivatives of arbitrary order if they contain an even number of
isoscalar bilinears; the same applies to the susceptibilities divided
by $m$ if this number is odd. From Eq.~\eqref{eq:specPQ16} one obtains
a similar relation for the discontinuity of $G$, and so for the
spectral density one has
\begin{equation}
  \label{eq:specPQ17}
  \lim_{m\to 0} \left[ r(x;\sous,\soup;m)-r(x;R\sous,R\soup;m)\right] =0\,,
\end{equation}
$\forall R\in\mathrm{SO}(4)$. By the same argument as above one has
that the derivatives of $r$ with respect to the sources at zero
sources are finite quantities (possibly distributions) in the chiral
limit, $m^2$-differentiable if they contain an even number of
isoscalar bilinear and $m$ times an $m^2$-differentiable quantity if
this number is odd.\footnote{To proceed rigorously, one should
  consider the discontinuity of $G$ integrated over an interval
  $[x_0,x]$ of nonzero length, that yields the following generating
  function (see footnote~\ref{foot:rho})
  $$
  \mdens^{(r)}(x_0,x;\sous,\soup;m) = \lim_{\lvol\to\infty} \lvol^{-1}
  \left\la \int_{x_0}^x dx'\,r_U(x')\right\ra_{\sous,\soup} \,.
  $$
  
  \noindent For any interval this has the same symmetry property as in
  Eq.~\eqref{eq:specPQ17}, and so it is an ordinary generating
  function with finite $m^2$-derivatives at $m=0$. The statement for
  $r(\sous,\soup;m)$ follows by taking the derivative of
  $\mdens^{(r)}$ with respect to $x$.
  \mbox{}

} This applies in particular to
$r(x;m)\equiv r(x;V,W;m)|_0$, which is just the spectral density for
the variable $x$, from which one recovers the usual spectral density
$\rho(\lambda;m) = \rho(\lambda;\sous,\soup;m)|_0$ as
\begin{equation}
  \label{eq:specPQ17_bis}
  \rho(\lambda;m) =  \left(1-\tf{\lambda^2}{4}\right)^{-\f{3}{2}}
  r\left(\lambda\left(1-\tf{\lambda^2}{4}\right)^{-\f{1}{2}};m\right)\,.
\end{equation}
The validity of Eq.~\eqref{eq:specPQ17} relies upon the assumption
that symmetry of the susceptibilities in the chiral limit holds all
the way to the boundary of the domain of definition of the partially
quenched theories defined by the actions $S^{(\pm 1)}$. One could
relax this assumption to that of symmetry restoration only in the
interior of these analyticity domains, provided one also assumes that
the chiral limit $m\to 0$ and the limit $\epsilon\to 0$ defining the
discontinuity can be exchanged. Alternatively, one can assume that $G$
can be further analytically extended beyond the cut onto some suitable
Riemann surface, and that the symmetry restoration condition holds
there as well.

\newpage \mbox{}

\subsection{Higher-order correlation functions}
\label{sec:sp_rho2}

One can prove the $m^2$-differentiability of higher-order eigenvalue
correlation functions using a similar construction, adding to the
theory several partially quenched flavor doublets of fermion fields
$\omega=(\omega_1,\ldots,\omega_{N_{\mathrm{PQ}}})$,
$\bar{\omega}=(\bar{\omega}_1,\ldots,\bar{\omega}_{N_{\mathrm{PQ}}})$,
and complex scalar fields
$\varphi=(\varphi_1,\ldots,\varphi_{N_{\mathrm{PQ}}})$, of masses
$z=(z_1,\ldots, z_{N_{\mathrm{PQ}}})$, and defining the partially
quenched actions
\begin{equation}
  \label{eq:specPQ18}
  S^{(\vec{k})}(\omega,\bar{\omega},\varphi,\varphi^*,U;z)
  \equiv \sum_{j=1}^{N_\mathrm{PQ}}
  S^{(k_j)}(\omega_j,\bar{\omega}_j,\varphi_j,\varphi^*_j,U;z_j)\,,
\end{equation}
where $\vec{k}=(k_1,\ldots,k_{N_{\mathrm{PQ}}})$, $k_j=0,\pm 1$.  One
then makes the symmetry-restoration assumption on the correlators
\begin{equation}
  \label{eq:specPQ19}
  \Upsilon^{(\vec{k})}(z^2;\sous,\soup;m)
  =\left\la \prod_{j=1}^{N_\mathrm{PQ}} \left(iP_{j\,a_j}\right)^2 \right\ra_{\vec{k};\sous,\soup}\,, 
\end{equation}
where $z^2=(z_1^2,\ldots , z_{N_{\mathrm{PQ}}}^2)$,
\begin{equation}
  \label{eq:specPQ19bis}
    iP_{j\,a}
    = i\bar{\omega}_j\left(\mathbf{1}-\f{D(U)}{2}\vphantom{D^\dag}\right)\gamma_5\sigma_{a}\omega_j\,,
\end{equation}
and the expectation values $\la\ldots\ra_{\smash{\vec{k};\sous,\soup}}$ are
defined as in Eq.~\eqref{eq:specPQ1_2}, replacing $S^{(k)}$ with
$S^{(\vec{k})}$. These quantities are again independent of the indices
$a_j=1,2,3$ thanks to the vector flavor symmetry of the partially
quenched fermion doublets. For two-point spectral correlation
functions, one sets $N_{\mathrm{PQ}}=2$ and defines the following
resolvent,
\begin{widetext}
  \begin{equation}
    \label{eq:specPQ20}
    \begin{aligned}
      G(z_1^2,z_2^2;\sous,\soup;m)
      &\equiv\lim_{\lvol\to\infty}\lvol^{-1} 
        \left[\left\la  \mathcal{B}(U;z_1^2)\mathcal{B}(U;z_2^2)\right\ra_{\sous,\soup}
        -\left\la \mathcal{B}(U;z_1^2)\right\ra_{\sous,\soup}
        \left\la\mathcal{B}(U;z_2^2) \right\ra_{\sous,\soup}\right]\,,\\
      4 G(z_1^2,z_2^2;\sous,\soup;m)
      &= \lim_{\lvol\to\infty}\lvol^{-1}\left\la (iP_{1\,a}^{\mathrm{PQ}})^2
        (iP_{2\,b}^{\mathrm{PQ}})^2   \right\ra_{\vec{k};\sous,\soup;c}  \,,
    \end{aligned}
  \end{equation}
\end{widetext}
where $c$ denotes the connected part defined in the usual way (see
Appendix~\ref{sec:conncorrfunc}), and with the representation in terms
of the partially quenched theory labeled by $\vec{k}=(k_1,k_2)$
holding within the convergence domains discussed above in
Appendix~\ref{sec:rhochirest_rho} ($k_i=\pm 1$ if
$\Im z_i^2\gtrless 0$, $k_i=0$ if $\Re z_i^2> 0$). Writing the traces
in terms of the eigenvalues of $D$ one finds
\begin{widetext}
  \begin{equation}
    \label{eq:specPQ21}
    \begin{aligned}
      G(z_1^2,z_2^2;\sous,\soup;m)
      &= \f{\bbnn(\sous,\soup;m)}{ z_1^2 z_2^2}   
        + 2 \int_0^\infty dx\, \left[\f{1}{z_1^2(x^2 + z_2^2)}
        + \f{1}{(x^2 + z_1^2)z_2^2}\right]   r_{N_0\,c\,{\scriptscriptstyle\infty}}(x;\sous,\soup;m)\\
      &\phantom{=}  + 4 \int_0^\infty dx \int_0^\infty dx' \,
        \f{r^{(2)}_{c\,{\scriptscriptstyle\infty}}(x,x';\sous;m)
        + \left(\delta(x-x')+\delta(x+x')\right)r(x;\sous,\soup;m)}{(x^2 + z_1^2)(x^{\prime\,2} + z_2^2)}\,,
    \end{aligned}
  \end{equation}
\end{widetext}
where [see Eqs.~\eqref{eq:Ncum_def} and \eqref{eq:mder_ext0},
\eqref{eq:rho_def12_bis} and \eqref{eq:rho_def8_gen_th}, and
\eqref{eq:rho_def7}]
\begin{equation}
  \label{eq:specPQ21_2}
  \begin{aligned}
    \bbnn(\sous,\soup;m)
    &=\lim_{\lvol\to\infty}  \lvol^{-1}\left\la N_0^2 \right\ra_{\sous,\soup\,c}\,, \\
    r_{N_0\,c\,{\scriptscriptstyle\infty}}(x;\sous,\soup;m)
    &    =\lim_{\lvol\to\infty}\lvol^{-1}\left\la N_0 r_U(x) \right\ra_{\sous,\soup\,c}\,, \\
    r^{(2)}_{c\,{\scriptscriptstyle\infty}}(x,x';\sous,\soup;m)
    &=\lim_{\lvol\to\infty}\lvol^{-1}\left\la r_U(x) r_U(x')\right\ra_{\sous,\soup\,c}\\
    &\hphantom{=} - \left(\delta(x-x')+\delta(x+x')\right) \\
    & \hphantom{=}\times r(x;\sous,\soup;m)\,,
  \end{aligned}
\end{equation}  
understood as generating functions. Approaching $z_{1,2}^2=0$ from
positive real values, $z_{1,2}^2=w_{1,2}^2\in \mathbb{R}^+$, one finds
\begin{equation}
  \label{eq:specPQ22}
  \lim_{w_{1,2}\to 0}  w_1^2 w_2^2  G(w_1^2,w_2^2;\sous,\soup;m) =\bbnn(\sous,\soup;m)\,,
\end{equation}
while approaching $z_{2}^2=0$ from positive real values,
$z_{2}^2= w_{2}^2\in \mathbb{R}^+ $, and computing the discontinuity
on the negative real axis of $z_1^2$, one gets
\begin{equation}
  \label{eq:specPQ23}
  \begin{aligned}
    &     \lim_{\epsilon\to 0}\lim_{w_{2}\to 0}   w_2^2  \left[G(-w_1^2+i\epsilon,w_2^2;\sous,\soup;m)  \right. \\
    & \left.\hphantom{\lim_{\epsilon\to 0}\lim_{w_{2}\to 0} w_2^2} -G(-w_1^2-i\epsilon,w_2^2;\sous,\soup;m) \right]\\
    & = -2\pi i \f{   r_{N_0\,c\,{\scriptscriptstyle\infty}}(|w_1|;\sous,\soup;m)}{|w_1|}\,.
  \end{aligned}
\end{equation}
Finally, from the double discontinuity on the negative real axes of
$z_{1,2}^2$ one gets
\begin{equation}
  \label{eq:specPQ24}
  \begin{aligned}
    &\lim_{\epsilon\to 0}\lim_{\epsilon'\to 0} \left[G(-w_1^2+i\epsilon,-w_2^2+i\epsilon';\sous,\soup;m)   \right. \\
    & \phantom{\lim_{\epsilon\to 0}\lim_{\epsilon'\to 0}} \left.     -G(-w_1^2-i\epsilon,-w_2^2+i\epsilon';\sous,\soup;m) \right.\\
    &\phantom{\lim_{\epsilon\to 0}\lim_{\epsilon'\to 0}}
      \left. -G(-w_1^2+i\epsilon,-w_2^2-i\epsilon';\sous,\soup;m)   \right. \\
    &  \phantom{\lim_{\epsilon\to 0}\lim_{\epsilon'\to 0}}\left.  +G(-w_1^2-i\epsilon,-w_2^2-i\epsilon';\sous,\soup;m) \right]\\
    &  = \f{4\pi^2}{|w_1| |w_2|} \left[r^{(2)}_c(|w_1|,|w_2|;\sous,\soup;m)  \right. \\
    & \hphantom{=\f{4\pi^2}{|w_1| |w_2|}[]}  \left. + \delta(|w_1|-|w_2|)r(|w_1|;\sous,\soup;m)\vphantom{r^{(2)}_c(|w_1|,|w_2|;\sous,\soup;m)}\right] \,.
  \end{aligned}
\end{equation}
Assuming as above in Appendix~\ref{sec:rhochirest_rho} that chiral
symmetry is realized in the chiral limit in the partially quenched
theories labeled by $\vec{k}$, one proves that $G$ has the same
$m^2$-differentiability properties at $m=0$ as its counterpart in the
previous subsection, in the whole domain of analyticity and at its
boundary, and this property is inherited by $\bbnn(0,0;m)$,
$r_{N_0\,c\,{\scriptscriptstyle\infty}}(x;0,0;m)$ and
$r^{(2)}_{c\,{\scriptscriptstyle\infty}}(x,x';0,0;m)$, and so by
$\rho_{N_0\,c\,{\scriptscriptstyle\infty}}(\lambda;m)$ and
$\rho^{(2)}_{c\,{\scriptscriptstyle\infty}}(\lambda,\lambda';m)$, that
are obtained by a mass-independent change of variables.

\section{$\susc_n$ and susceptibilities}
\label{sec:ansusc}

Consider the generating function for vanishing isosinglet sources,
$j_{S,P}=0$, Eq.~\eqref{eq:expansion2}. I consider for notational
simplicity the $CP$-invariant case, in which
$\susc_{n_u n_w n_{\tu}}(m^2)\neq 0$ only for $n_{\tu} =
2n_{\bu}$. For $(n_u,n_w,n_{\tu})\neq (0,0,0)$, setting
$ B_{n_u n_w n_{\tu}}(m^2) \equiv 4^{n_{\bu}}\susc_{n_u n_w
  2n_{\bu}}(m^2)$, and further choosing $j_{P3}=j_{S2}=j_{S3}=0$, one
finds after simple algebra
\begin{equation}
  \label{eq:suscA8}
  \begin{aligned}
    &  \wiv(\sous,\soup;m)|_{j_S=j_P=j_{P3}=j_{S2}= j_{S3}=0} \\
    &  = \sum_{a,b,c=0}^\infty \f{\left(j_{P1}^2\right)^{a}  \left(j_{P2}^2\right)^{b}
      \left(j_{S1}^2\right)^{c}}{(2a)!(2b)!(2c)!}\chi_{abc}(m^2)\,,
  \end{aligned}
\end{equation}
where for $(a, b, c) \neq (0,0,0)$
\begin{equation}
  \label{eq:suscA8_bis}
  \begin{aligned}
    \chi_{abc}(m)&\equiv      \sum_{d=0}^{\min(a,c)} C_{abcd} B_{a-d+b\,\, c-d\,\, d}(m^2)\,,    \\
    C_{abcd}&\equiv \f{(2a)!(2b)!(2c)!}{(a-d)!b!(c-d)!(2d)!}\,,
  \end{aligned}
\end{equation}
are just the usual susceptibilities in the restricted subset
\begin{equation}
  \label{eq:suscA9}
  \chi_{abc}(m) =\chi\left((iP_1)^{2a}(iP_2)^{2b}S_1^{2c}\right)\,,
\end{equation}
while $\chi_{000} = \susc_{000} = \wiv|_0$ is minus the free energy density.
For $a=z$, $b=x$, $c=y+z$ one has from Eq.~\eqref{eq:suscA8_bis}
\begin{equation}
  \label{eq:suscA14_bis}
  \begin{aligned}
    B_{x y z}
    &=     \f{x!y!\chi_{z\,x\,y+z}}{(2x)![2(y+z)]!} \\
    &\phantom{=} - \sum_{k=1}^{z}\f{y!(2z)!B_{x+k\, y+k\, z-k}}{k!(y+k)![2(z-k)]!}
      \,, &&& z&\ge 1\,,\\
    B_{xy0}
    &=\f{x!y!\chi_{0xy}}{(2x)!(2y)!}\,, &&& x+y&>0\,.
  \end{aligned}
\end{equation}
By recursion on $z$, one shows that $B_{xyz}$, $x,y,z\ge 0$,
$x+y+z>0$, are finite linear combinations of $\chi_{abc}$ with $c=y+z$
and $a+b=x+z$, with mass-independent coefficients.  To see this,
notice that for the term $B_{x' y' z'}=B_{x+k\, y+k\, z-k}$ inside the
summation in Eq.~\eqref{eq:suscA14_bis} one has $0\le z'< z$, i.e.,
the maximal $z'$ decreases at every step of the recursion, while
$y'+z' = y+ z$ and $x'+z' = x+ z$ remain constant, and
$x'+y'+z'>x+y+z>0$. The extension to the non-$CP$-invariant case
presents no difficulty.

The same result applies to the coefficients of the generating function
of susceptibilities involving also functionals of gauge fields only,
$G_i(U)$, obtained from
\begin{equation}
  \label{eq:partfunc_G}
  \begin{aligned}
    \mathcal{Z}_G(\sous,\soup;J_G;m)
    &\equiv \int DU    \int D\Psi D\bar{\Psi} \, e^{-S_{\mathrm{eff}}(U)}\\
    &\phantom{=}\times e^{-\bar{\Psi} D_m (U)\mathbf{1}_{\mathrm{f}}  \Psi - \des( \Psi,\bar{\Psi},U; \sous,\soup)}\\
    &\phantom{=}\times e^{ \sum_{i=1}^N J_{G_i} G_i(U)} \,,\\
    \mathcal{W}_G(\sous,\soup;J_G;m)
    &\equiv \f{1}{\lvol} \ln\mathcal{Z}_G(\sous,\soup;J_G;m) \,,\\
    \wivG(\sous,\soup;J_G;m)
    &\equiv \lim_{\lvol\to\infty}\mathcal{W}_G(\sous,\soup;J_G;m)\,.
  \end{aligned}
\end{equation}
Since gauge fields are unaffected by chiral transformations one has again
\begin{equation}
  \label{eq:chiW0_3_G}
  \begin{aligned}
    & \wivG(\sous,\soup;J_G;m) = \hat{\Wc}_{G{\scriptscriptstyle\infty}}(m^2 + u,w,\tilde{u};J_G)
    \\ & = \sum_{n_u,n_w,n_{\tu}=0}^\infty  \f{u^{n_u}w^{n_w}\tu^{n_{\tu}}}{n_u! n_w! n_{\tu}!}
         \susc_{n_u n_w n_{\tu}}(m^2;J_G)\\
    &= \sum_{n_u,n_w,n_{\bu}=0}^\infty  \f{u^{n_u}w^{n_w}\left(\f{\tu}{2}\right)^{2 n_{\bu}}}{n_u! n_w! (2n_{\bu})!}
      B_{n_u n_w n_{\bu}}(m^2;J_G) \,,
  \end{aligned}
\end{equation}
having used $CP$ invariance in the last passage, and having set
$B_{n_u n_w n_{\bu}}(m^2;J_G)= 4^{n_{\bu}}\susc_{n_u n_w
  2n_{\bu}}(m^2;J_G)$.  One finds again
\begin{equation}
  \label{eq:suscA8_G}
  \begin{aligned}
    & \wivG(\sous,\soup;J_G;m)|_{j_S= j_P=j_{P3}=j_{S2}=j_{S3}=0} \\
    & =    \sum_{a,b,c=0}^\infty  \f{\left(j_{P1}^2\right)^{a}  \left(j_{P2}^2\right)^{b}
      \left(j_{S1}^2\right)^{c}}{(2a)!(2b)!(2c)!}\chi_{abc}(m^2;J_G)\,,
  \end{aligned}
\end{equation}
where now
\begin{equation}
  \label{eq:suscA8_bis_G}
  \chi_{abc}(m;J_G) \equiv  \sum_{d=0}^{\min(a,c)} C_{abcd} B_{a-d+b\, c-d\, d}(m^2;J_G)
\end{equation}
are the generating functions of susceptibilities involving scalar and
pseudoscalar bilinears and gauge operators,
\begin{equation}
  \label{eq:suscA8_bis_G2}
  \prod_{i=1}^N  \de_{J_{Gi}}^{n_i}\chi_{abc}(m;J_G)|_0
  =\chi\left((iP_1)^{2a}(iP_2)^{2b}S_1^{2c}\prod_{i=1}^N G_i^{n_i}\right)\,.
\end{equation}
Proceeding as above, one shows recursively that $ B_{x y z}(m;J_G)$
can be reconstructed from $\chi_{abc}(m;J_G)$, so they contain
equivalent information.

\section{Ward-Takahashi identities}
\label{sec:WTI_geo}

Using Eq.~\eqref{eq:mshift1} and the exact chiral invariance of the
massless theory, one finds after simple algebra
\begin{equation}
  \label{eq:chiZ0_5}
  \begin{aligned}
    & \mathcal{Z}(R^T\sous,R^T\soup;m)  = \mathcal{Z}(R^T\sous +me_0,R^T\soup;0) \\
    &=\mathcal{Z}(\sous +mR e_0,\soup;0)
      = \mathcal{Z}(\sous  +m(R-\mathbf{1}_4)e_0,\soup;m)\,,
  \end{aligned}
\end{equation}
where $e_0=(1,\vec{0})^T$ and $\mathbf{1}_4$ is the $4\times 4$
identity matrix, or equivalently
\begin{equation}
  \label{eq:chiZ0_6}
  \mathcal{Z}(\sous,\soup;m)  =    \mathcal{Z}(R\sous + m(R-\mathbf{1}_4)e_0,R\soup;m)  \,.
\end{equation}
The same relation holds replacing $\mathcal{Z}$ with $\mathcal{W}$.
The generating functions are therefore invariant under the
mass-dependent affine transformation Eq.~\eqref{eq:chiW0_7_0},
\begin{equation}
  \label{eq:chiW0_7}
    \sous \to R\sous + m(R-\mathbf{1}_4)e_0\,,\qquad
    \soup \to R\soup\,.
\end{equation}
Expanding Eq.~\eqref{eq:chiZ0_5} for small $R-\mathbf{1}_4$, one finds
to leading order
\begin{equation}
  \label{eq:WTquickproof1}
  \begin{aligned}
    &  \mathcal{Z}(R^T\sous,R^T\soup;m)-\mathcal{Z}(\sous,\soup;m)\\
    &    = m[(R-\mathbf{1}_4)e_0]\cdot \de_{\sous}\mathcal{Z}(\sous,\soup;m)  + \ldots  \,.
  \end{aligned}
\end{equation}
The quantity $\mathcal{Z}(R^T\sous,R^T\soup;m)$ is the generating
function of the correlators of the chirally transformed bilinears
$R O_{V,W}$ [see Eqs.~\eqref{eq:OVW_def} and \eqref{eq:O_VW_transf}].
Denoting with $O_i$ the components of the vector
\begin{equation}
  \label{eq:Odef}
  O \equiv    \left( O_{\sous} , O_{\soup}  \right)^T =  \left( S, i\vec{P}, iP, -\vec{S} \right)^T \,,
\end{equation}
after taking derivatives with respect to the sources, denoted
collectively as $J\equiv(\sous,\soup)$, and setting these to zero, one
finds for the left-hand side of Eq.~\eqref{eq:WTquickproof1}
\begin{equation}
  \label{eq:WTquickproof2}
  \begin{aligned}
    &      \left(  {\textstyle \prod_{s=1}^k} (-\de_{J_{i_s}})\right)\left[\mathcal{Z}(R^T\sous,R^T\soup;m)
      -\mathcal{Z}(\sous,\soup;m)\right]|_0 \\
    & = \left\la  \left({\textstyle \prod_{s=1}^k} \left(RO\right)_{i_s}\right)
      -\left({\textstyle \prod_{s=1}^k} O_{i_s}\right)\right\ra\,,
  \end{aligned}
\end{equation}
which is the expectation value of the variation of
$\prod_{s=1}^k O_{i_s}$ under a chiral transformation. Specializing
now to infinitesimal vector and axial nonsinglet transformations [see
under Eq.~\eqref{eq:ch_transf3}] this leads to the well-known
integrated Ward-Takahashi identities.  For general vector and axial
transformations $\U_{V,A}(\vec{\alpha})$, one correspondingly finds
the $\mathrm{SO}(4)$ matrices
$R_{V,A}(\vec{\alpha})\equiv \Rc(\U_{V,A}(\vec{\alpha}))$, with
\begin{equation}
  \label{eq:ch_transf7_bis}
  \begin{aligned}
    R_V(\vec{\alpha})
    &= \begin{pmatrix}
      1 & \vec{0}^{\,T}\\
      \vec{0} & \tilde{R}(\vec{\alpha})
    \end{pmatrix}\,, \\
    \tilde{R}(\vec{\alpha})\vec{v}
    &\equiv          \Pi_{\hat{\alpha}} \vec{v}
      + \cos|\vec{\alpha}| \, \Pi_{\hat{\alpha}_\perp} \vec{v}
      + \sin|\vec{\alpha}|\, \hat{\alpha}\wedge \vec{v} \,,
  \end{aligned}
\end{equation}
and
\begin{equation}
  \label{eq:ch_transf9_bis}
    R_A(\vec{\alpha})
     =\begin{pmatrix}
      \cos (|\vec{\alpha}|)& \sin (|\vec{\alpha}|) \hat{\alpha}^T \\
      -\sin (|\vec{\alpha} |) \hat{\alpha} & \Pi_{\hat{\alpha}_\perp}
                                             +  \cos (|\vec{\alpha}|) \Pi_{\hat{\alpha}}
    \end{pmatrix}\,.
\end{equation}
Here $\tilde{R}\in\mathrm{SO}(3)$,
$\Pi_{\hat{\alpha}}\equiv \hat{\alpha}\hat{\alpha}^T$ and
$\Pi_{\hat{\alpha}_\perp}\equiv\mathbf{1}_{3}-\Pi_{\hat{\alpha}}$,
with $\hat{\alpha}=\vec{\alpha}/|\vec{\alpha}|$ and $\mathbf{1}_{3}$
the $3\times 3$ identity matrix. Denoting for a generic function
$\mathcal{F}( O_{\sous}, O_{\soup})$
\begin{equation}
  \label{eq:deltaR_def}
  \begin{aligned}
    &  \mathcal{F}(R_{V,A}  O_{\sous}, R_{V,A} O_{\soup}) - \mathcal{F}(  O_{\sous},  O_{\soup})\\
    &\equiv i\vec{\alpha}\cdot\vec{\delta}_{V,A}  \mathcal{F}(  O_{\sous},  O_{\soup})
      +O(\vec{\alpha}^{\,2}) \,,
  \end{aligned}
\end{equation}
Eq.~\eqref{eq:WTquickproof1} reads for $R=R_{V,A}$
\begin{equation}
  \label{eq:WTquickproof2_bis}
  \begin{aligned}
    &  \left(  {\textstyle \prod_{s=1}^k} (-\de_{J_{i_s}})\right)
      \left[\mathcal{Z}(R_{V,A}^T\sous,R_{V,A}^T\soup;m)
      -\mathcal{Z}(\sous,\soup;m)\right]|_0 \\
    & = i\vec{\alpha}\cdot \left\la  \vec{\delta}_{V,A}\left({\textstyle \prod_{s=1}^k} O_{i_s}\right)\right\ra
      + O(\vec{\alpha}^{\,2}) \\
    &=    m[(R_{V,A}-\mathbf{1}_4)e_0]\cdot \de_V\left({\textstyle \prod_{s=1}^k}
      (-\de_{J_{i_s}})\right)\mathcal{Z}(\sous,\soup;m)|_0  \\
    &\phantom{=}+ O(\vec{\alpha}^{\,2}) \,.
  \end{aligned}
\end{equation}
Since $(R_V-\mathbf{1}_4)e_0=0$ and
$(R_A-\mathbf{1}_4)e_0= - (0,\vec{\alpha})^T$, one finds for the
right-hand side to order $|\vec{\alpha}|$
\begin{equation}
  \label{eq:WTquickproof3}
  m[(R_{V}-\mathbf{1}_4)e_0]\cdot \de_V\left({\textstyle \prod_{s=1}^k} (-\de_{J_{i_s}})\right)
  \mathcal{Z}(\sous,\soup;m)|_0 =   0\,,
\end{equation}
for a vector transformation, and
\begin{equation}
  \label{eq:WTquickproof4}
  \begin{aligned}
    &  m[(R_{A}-\mathbf{1}_4)e_0]\cdot \de_V \left({\textstyle \prod_{s=1}^k} (-\de_{J_{i_s}})\right)
      \mathcal{Z}(\sous,\soup;m)|_0 \\
    &= m  \vec{\alpha}\cdot    \left\la (i\vec{P}) \left({\textstyle \prod_{s=1}^k} O_{i_s}\right)\right\ra\,,
  \end{aligned}
\end{equation}
for an axial transformation, and one concludes
\begin{equation}
  \label{eq:WT8}
  \begin{aligned}
    \left\la  \delta_{Va} \left({\textstyle\prod_{s=1}^k}O_{i_s}\right)\right\ra
    &= 0\,, \\
    \left\la  \delta_{Aa} \left({\textstyle\prod_{s=1}^k}O_{i_s}\right)\right\ra
    &=m \left\la  P_a \left({\textstyle\prod_{s=1}^k} O_{i_s}\right)\right\ra\,.
  \end{aligned}
\end{equation}
Replacing $\mathcal{Z}$ with $\mathcal{W}$ in
Eq.~\eqref{eq:WTquickproof1} one shows that the same identities hold
for the connected correlation functions, i.e., for $\la\ldots\ra$
replaced with $\la\ldots\ra_c$ in Eq.~\eqref{eq:WT8}.

\section{Chiral limit of the generating function in the symmetric phase}
\label{sec:invchilim}

The symmetry-restoration condition Eq.~\eqref{eq:symrest1} formally
collects the symmetry-restoration conditions
Eq.~\eqref{eq:susc_condition} for the whole set of
susceptibilities. Denoting the susceptibilities with a fixed number
$n_\sous$ of $ O_\sous$-type bilinears and $n_\soup$ of
$ O_\soup$-bilinears compactly as
$\chi_{\vec{n}} \equiv \chi\left(\left(\otimes_{n_\sous}
    O_\sous\right)\otimes\left( \otimes_{n_\sous}
    O_\soup\right)\right)$, $\vec{n}=(n_\sous, n_\soup)$, one has
\begin{equation}
  \label{eq:chliminvalgsh1}
  \wiv(\sous,\soup;m)= \sum_{\vec{n}} 
  \f{J_{\vec{n}}\cdot \chi_{\vec{n}}(m)}{n_\sous! n_\soup!}   \,,
\end{equation}
where
$J_{\vec{n}}\equiv\left(\otimes_{n_\sous}\sous\right) \otimes
\left(\otimes_{n_\soup} \soup\right)$. Symmetry restoration requires
that for each $\vec{n}$ separately
\begin{equation}
  \label{eq:chliminvalgsh2}
  \lim_{m\to 0}\left[\mathbf{I} - \mathbf{R}(R)\right] \chi_{\vec{n}}(m)=0\,,
  \quad
  \forall R \in \mathrm{SO}(4)\,,
\end{equation}
where $\mathbf{I}\equiv \otimes_{\mathrm{n}}\mathbf{1}_4$ and
$ \mathbf{R}(R)\equiv \otimes_{\mathrm{n}}R$, with
$\mathrm{n} =n_\sous+n_\soup$. The representation space
$\otimes_{\mathrm{n}}\mathbb{R}^4$ can be decomposed into the
invariant subspace $\mathbb{I}$ formed by vectors invariant under any
transformation,
$\mathbb{I}\equiv\{x\in
\otimes_{\mathrm{n}}\mathbb{R}^4~|~\mathbf{R}(R)x=x~\forall
R\in\mathrm{SO}(4)\}$, and its orthogonal complement,
$\mathbb{I}_\perp$, which is also an invariant subspace. One has then
$\otimes_{\mathrm{n}}\mathbb{R}^4=\mathbb{I}\oplus \mathbb{I}_\perp$,
and one can write
$\chi_{\vec{n}}(m) = x_{\vec{n}}(m) + v_{\vec{n}}(m)$ with
$x_{\vec{n}}(m)\in\mathbb{I}$ and
$v_{\vec{n}}(m)\in \mathbb{I}_\perp$. It is easy to show that
$x\in \mathbb{I}$ if and only if $\mathbf{t}_a x=0$ for
$ a= 1,\ldots, d=\dim \mathrm{SO}(4)$, where the Hermitean matrices
$\mathbf{t}_a$ are the representatives of the group generators $t_a$
in the $\mathrm{n}$-fold tensor-product representation.\footnote{If
  $\mathbf{R}w=w ~\forall R\in\mathrm{SO}(4)$ then in particular
  $e^{i\epsilon \mathbf{t}_a}w=w~ \forall \epsilon$, and so
  $0= -i\de_\epsilon e^{i\epsilon \mathbf{t}_a}w|_{\epsilon=0}=
  \mathbf{t}_a w$, for $a=1,\ldots d$. Conversely, since
  $\mathrm{SO}(4)$ is a connected Lie group one can write any
  $R\in \mathrm{SO}(4)$ as $R=\prod_j R_j$ with
  $R_j=e^{i\sum_{a=1}^d\alpha_a^{(j)}t_a}$ for suitable
  $\alpha_a^{(j)}$.  If $\mathbf{t}_a w=0$ for $a=1,\ldots, d$ then
  $\mathbf{R}w = \left(\textstyle\prod_j \mathbf{R}(R_j)\right)w =
  \left(\textstyle\prod_j
    e^{i\sum_{a=1}^d\alpha_a^{(j)}\mathbf{t}_a}\right)w = w$.} Since
this is reducible for $\mathrm{n}>1$, $\mathbb{I}$ generally does not
have to be trivial.

Consider now
$\mathbf{R}_a = \mathbf{R}(e^{i\epsilon_a t_a}) = e^{i\epsilon_a
  \mathbf{t}_a}$ (no summation over $a$), $a=1,\ldots, d$. Choosing
$\epsilon_a$ so that $\epsilon_a \Vert\mathbf{t}_a\Vert <2\pi$, the
matrices $\mathbf{M}_a = \mathbf{I}- \mathbf{R}_a $ are positive
semidefinite, and $\mathbf{M}_a w = 0$ if and only if
$\mathbf{t}_a w =0$. Then for
$\mathbf{M} \equiv \sum_{a=1}^d \mathbf{M}_a$ one has $\mathbf{M}w =0$
if and only if $\mathbf{t}_a w =0$ for $a=1,\ldots, d$, so if and only
if $w\in \mathbb{I}$, and so the restriction of $\mathbf{M}$ to
$\mathbb{I}_\perp$ is invertible. Defining $\tilde{\mathbf{M}}w$ for
$w \in \otimes_{\mathrm{n}}\mathbb{R}^4$ by decomposing (uniquely)
$w=x+v$ with $x\in\mathbb{I}$ and $v\in \mathbb{I}_\perp$, and setting
$\tilde{\mathbf{M}}w\equiv x
+\left(\mathbf{M}|_{\mathbb{I}_\perp}\right)^{-1} v$, one has then
\begin{equation}
  \label{eq:chliminvalgsh3}
\begin{aligned}
  0 &= \tilde{\mathbf{M}}\sum_{a=1}^d \left(\lim_{m\to 0}\mathbf{M}_a \chi_{\vec{n}}(m)\right)
      =  \lim_{m\to 0}\tilde{\mathbf{M}}\!\left(\sum_{a=1}^d\mathbf{M}_a\right) v_{\vec{n}}(m)  \\
    &=  \lim_{m\to 0}\tilde{\mathbf{M}}\mathbf{M} v_{\vec{n}}(m) =  \lim_{m\to 0} v_{\vec{n}}(m)\,,
\end{aligned}
\end{equation}
as a consequence of Eq.~\eqref{eq:chliminvalgsh2}.  Setting
\begin{equation}
  \label{eq:chliminvalgsh4}
    \wiv^{\mathrm{inv}}(\sous,\soup;m) 
    \equiv \sum_{\vec{n}}\f{J_{\vec{n}} \cdot  x_{\vec{n}}(m)}{n_\sous! n_\soup!}\,, 
\end{equation}
with
$\wiv^{\mathrm{inv}}(R\sous,R\soup;m)=\wiv^{\mathrm{inv}}(\sous,\soup;m)$,
one has then
\begin{equation}
  \label{eq:chliminvalgsh5}
  \lim_{m\to 0}\left[\wiv(\sous,\soup;m)-\wiv^{\mathrm{inv}}(\sous,\soup;m)\right]  =0\,,
\end{equation}
and in particular
\begin{equation}
  \label{eq:chliminvalgsh6}
  \begin{aligned}
    &  \lim_{m\to 0}\left[\de_{j_S}-2\left(j_S\de_{\sous^2}+j_P\de_{2\sous\cdot\soup}\right)\right]
      \wiv(\sous,\soup;m)\\
    &= \lim_{m\to 0}\left[\de_{j_S}-2\left(j_S\de_{\sous^2}+j_P\de_{2\sous\cdot\soup}\right)\right]
      \wiv^{\mathrm{inv}}(\sous,\soup;m)      =0\,.
  \end{aligned}
\end{equation}
In Refs.~\cite{Giordano:2024jnc,Giordano:2024awb} it is implicitly
made use of the relation
$\lim_{m\to 0}(\mathbf{I}-\mathbf{R})\chi_{\vec{n}}(m)=
(\mathbf{I}-\mathbf{R})\lim_{m\to 0}\chi_{\vec{n}}(m)$ [see Eqs.~(S7)
and (S8) of the Supplemental Material of Ref.~\cite{Giordano:2024jnc},
and Eq.~(9) of Ref.~\cite{Giordano:2024awb}], that is true if one
assumes the existence of $\lim_{m\to 0}\chi_{\vec{n}}(m)$, but is not
warranted otherwise. The argument above shows that this assumption is
not necessary to prove Eq.~\eqref{eq:chliminvalgsh5}, upon which the
arguments of Refs.~\cite{Giordano:2024jnc,Giordano:2024awb} rely.

\section{Contribution of complex modes to the fermionic determinant}
\label{sec:compl_det}

For the contribution $\det\mnz(\mu)$ to the fermionic determinant in
the presence of sources of a pair of complex modes $\mu,\mu^*$, with
$|\mu|^2 = 2\Re\mu$ and $0< |\mu|^2 < 4$, see Eqs.~\eqref{eq:fdet3}
and \eqref{eq:fdet2_2_0}, one finds
\begin{widetext}
\begin{equation}
  \label{eq:fdet5}
  \begin{aligned}
    \det \mnz(\mu)
    &=\det
    \begin{pmatrix}
      \left(1-\tf{\mu}{2}\right)\mathbf{1}_{\mathrm{f}} & 0 \\
      0 & \left(1-\tf{\mu^*}{2}\right)\mathbf{1}_{\mathrm{f}}
    \end{pmatrix}
      \det \begin{pmatrix}
        \tf{\mu}{1-\tf{\mu}{2}}\mathbf{1}_{\mathrm{f}} + A & iB \\
        iB & \tf{\mu^*}{1-\tf{\mu^*}{2}}\mathbf{1}_{\mathrm{f}} + A
      \end{pmatrix}
      = \left|1-\tf{\mu}{2}\right|^{4}
      \det \begin{pmatrix}
        A +iB  & i \tf{\Im\mu}{1-\tf{|\mu|^2}{4}}\mathbf{1}_{\mathrm{f}} \\
        i \tf{\Im\mu}{1-\tf{|\mu|^2}{4}}\mathbf{1}_{\mathrm{f}} & A -iB 
      \end{pmatrix}\\
    &= \left(1-\tf{|\mu|^2}{4}\right)^{2}  \det\left((A+iB)(A-iB) +
      \tf{|\mu|^2}{1-\tf{|\mu|^2}{4}}\right)
      = \det\left(|\mu|^2 +
      \left(1-\tf{|\mu|^2}{4}\right)(A+iB)(A-iB) \right) \,,
  \end{aligned}
\end{equation}
\end{widetext}
having used the following identity for determinants of block square
matrices with square blocks of equal size,
\begin{equation}
  \label{eq:fdet4}
  \det \begin{pmatrix}
    a & b \\ c & d
  \end{pmatrix}
  =\det \begin{pmatrix}
    \f{a+b+c+d }{2} &     \f{a-b+c-d }{2}\\
    \f{a+b-c-d }{2} &     \f{a-b-c+d }{2}
  \end{pmatrix}\,,
\end{equation}
the following identities satisfied by the complex eigenvalues,
\begin{equation}
  \label{eq:fdet6}
  \begin{aligned}
    \left|1-\tf{\mu}{2}\right|^2 &= 1-\tf{|\mu|^2}{4}\,,
    &&& \Re\tf{\mu}{1-\tf{\mu}{2}} &= 0\,, \\
    \Im\tf{\mu}{1-\tf{\mu}{2}} &= \f{\Im\mu}{1-\tf{|\mu|^2}{4}}\,,
    &&& (\Im\mu)^2 &= |\mu|^2\left(1-\tf{|\mu|^2}{4}\right)\,,
  \end{aligned}
\end{equation}
and the properties of determinants of block matrices with commuting
blocks~\cite{silvester2000determinants}. Setting now
$|\mu|^2=\lambda^2$ [see above Eq.~\eqref{eq:rho_def1}] and
$h(\lambda)= 1-\f{\lambda^2}{4}$ [see Eq.~\eqref{eq:fdet_hdef}], since
\begin{equation}
  \label{eq:fdet7_shorter}
  \begin{aligned}
    (A+iB)(A-iB)
    &= (\tilde{\sous}^2 + \soup^2)\mathbf{1}_{\mathrm{f}} +   2\vec{L}\cdot \vec{\sigma} \\
    &= (m^2 + u + w)\mathbf{1}_{\mathrm{f}} +  2\vec{L}\cdot \vec{\sigma} \,,
  \end{aligned}
\end{equation}
with $\tilde{\sous}$ defined in Eq.~\eqref{eq:mshift1}, and
\begin{equation}
  \label{eq:fdet7_shorte_bis}
  \begin{aligned}
    \vec{L} &\equiv j_P\vec{\jmath}_P-(m+j_S)\vec{\jmath}_S + \vec{\jmath}_P\wedge \vec{\jmath}_S\,, \\
    \vec{L}^{\,2}&= \tilde{V}^2W^2 - (\tilde{V}\cdot W)^2 = ( m^2+u) w -\tf{1}{4}\tilde{u}^2\,,
\end{aligned}
\end{equation}
one has
\begin{widetext}
  \begin{equation}
    \label{eq:fdet8_shorter}
    \begin{aligned}
      \det \mnz(\mu)
      &= \det\left( \left(\lambda^2 + h(\lambda)(m^2+u+w)\right)\mathbf{1}_{\mathrm{f}}
        + 2 h(\lambda) \vec{L}\cdot \vec{\sigma} \right)
        = \left(\lambda^2 + h(\lambda)(m^2 + u+w)\right)^2 -  4h(\lambda)^2 \vec{L}^{\,2}\\
      &= \left(\lambda^2 + h(\lambda)(m^2 + u +w )\right)^2 -
        4h(\lambda)^2( ( m^2+u) w -\tf{1}{4}\tilde{u}^2 )\\
      &= \lambda^4 + 2\lambda^2h(\lambda)(m^2 + u + w ) +  h(\lambda)^2\left( (m^2 + u -w )^2
        + \tilde{u}^2 \right)\\
      &= [\lambda^2 + m^2 h(\lambda)]^2 + 2h(\lambda)[ (\lambda^2+ m^2
        h(\lambda))u + (\lambda^2- m^2h(\lambda))w ]
        + h(\lambda)^2\left[ (u -w )^2 + \tilde{u}^2 \right]\,.
    \end{aligned}
  \end{equation}
\end{widetext}

\bibliographystyle{../apsrev4-2_mod}
\bibliography{../references_chi_PRD}

\end{document}